\documentclass[10pt]{article}
\usepackage{ifpdf}

\usepackage{latexsym}                   
\usepackage{epsfig,color}                     
\usepackage{graphicx}
\usepackage{amssymb, amsmath, amsthm}   
\usepackage{enumerate}                  
\usepackage{mathrsfs}                   
\usepackage{geometry}                                                                          
\usepackage{verbatim}                                                                           
 \usepackage{url}
\usepackage{xr}
\graphicspath{{./Figures/}}
\providecommand\Z[1]{\boldsymbol{Z_{#1}}}

\def\R{\mathbb{R}}
\def\E{\mathbb{E}}

\def\C{\mathbb{C}}

\def\Z{\mathbb{Z}}
\def\ind{{\mathchoice {\rm 1\mskip-4mu l} {\rm 1\mskip-4mu l}
{\rm 1\mskip-4.5mu l} {\rm 1\mskip-5mu l}}}
\newtheorem{theorem}{Theorem}[section]

\newtheorem{proposition}[theorem]{Proposition}
\newtheorem{lemma}[theorem]{Lemma}

\newtheorem{remark}[theorem]{Remark}

\newtheorem*{theorem*}{Theorem}
\usepackage{multirow}

\begin{document}
\thispagestyle{plain}
  \title{A Spectral Koopman Approximation Framework for Stochastic Reaction Networks}
  \author{Ankit Gupta and Mustafa Khammash\\
    \date{\today}
   {\small Department of Biosystems Science and Engineering, ETH Z\"{u}rich, Basel, Switzerland. } 
  }

  \maketitle

\begin{abstract}
Stochastic reaction networks (SRNs) are a general class of continuous-time Markov jump processes used to model a wide range of systems, including biochemical dynamics in single cells, ecological and epidemiological populations, and queueing or communication networks. Yet analyzing their dynamics remains challenging because these processes are high-dimensional and their transient behavior can vary substantially across different initial molecular or population states. Here we introduce a spectral framework for the stochastic Koopman operator that provides a tractable, low-dimensional representation of SRN dynamics over continuous time, together with computable error estimates. By exploiting the compactness of the Koopman operator, we recover dominant spectral modes directly from simulated or experimental data, enabling efficient prediction of moments, event probabilities, and other summary statistics across all initial states. We further derive continuous-time parameter sensitivities and cross-spectral densities, offering new tools for probing noise structure and frequency-domain behavior. We demonstrate the approach on biologically relevant systems, including synthetic intracellular feedback controllers, stochastic oscillators, and inference of initial-state distributions from high-temporal-resolution flow cytometry. Together, these results establish spectral Koopman analysis as a powerful and general framework for studying stochastic dynamical systems across the biological, ecological, and computational sciences.
\end{abstract}

\section{Introduction}\label{sec:intro}

The internal dynamics of biological cells—the fundamental units of life—are governed by complex, nonlinear reaction networks \cite{loskot2019comprehensive}. These networks often involve biomolecular species present in low copy numbers, which leads to reaction events that occur intermittently rather than continuously \cite{thattai2001intrinsic,elowitz2002stochastic}. As a result, intracellular reactions exhibit randomness in their timing, a phenomenon referred to as intrinsic noise. This dynamical randomness can profoundly influence cellular behavior \cite{raj2008nature,eldar2010functional}. To capture such effects accurately, mathematical models of intracellular network dynamics must adopt a stochastic framework. A widely used approach is to model these dynamics as a continuous-time Markov chain (CTMC), where each state is a vector of non-negative integers representing the copy numbers of all molecular species involved \cite{anderson2015stochastic}. This modeling framework is known as a stochastic reaction network (SRN), and it has been extensively applied to explore the impact of randomness and to construct predictive models across a range of disciplines, including systems and synthetic biology \cite{wilkinson2018stochastic,briat2016antithetic,gupta2019universal}, ecology \cite{black2012stochastic}, epidemiology \cite{allen2008introduction, stadler2013birth}, oncology \cite{foo2011stochastic}, and pharmacokinetics \cite{irurzun2020beyond}.

Let $(X_x(t))_{t \geq 0}$ denote the CTMC that captures the dynamics of an SRN, with initial state $x$ belonging to the state space $\mathcal{E}$, a subset of the non-negative integer lattice. To analyze the behavior of the SRN and calibrate its parameters using experimental data, a key objective is to estimate the time evolution of the probability distribution of the random state $X_x(t)$ at time $t$, given by
\begin{align}
\label{prob_distr_random_state}
p_t(y \vert x) = \mathbb{P}(X_x(t) = y), \quad y \in \mathcal{E}, \quad t \geq 0.
\end{align}
This probability distribution evolves according to the forward Kolmogorov equation, commonly referred to as the \emph{Chemical Master Equation} (CME) \cite{anderson2015stochastic}. The CME constitutes a system of coupled linear ordinary differential equations (ODEs) that account for the probabilistic inflow and outflow at each state \(y \in \mathcal{E}\) (see \eqref{defn_cme}). Since the cardinality of \(\mathcal{E}\) is countably infinite in most examples of interest, the CME typically results in an infinite-dimensional linear system and is analytically intractable except in some special cases \cite{jahnke2007solving,iyer2009stochasticity}. By enumerating the state space as \(\mathcal{E} = \{y_1, y_2, y_3, \dots\}\), the CME can be formally written as
\begin{align}
\label{cme_as_linear_system}
\frac{d\mathbf{p}^{(x)}_t}{dt} = \mathbf{A} \mathbf{p}^{(x)}_t,
\end{align}
where \(\mathbf{p}^{(x)}_t = [p_t(y_1 \vert x), p_t(y_2 \vert x), \dots]\) is the time-dependent probability (column-)vector and \(\mathbf{A}\) is a (bi-)infinite matrix describing the transition structure of the CTMC. 

A widely used numerical method for approximating the solution of the CME is the \emph{Finite State Projection} (FSP) algorithm \cite{munsky2006finite}. FSP circumvents the infinite dimensionality by projecting the CTMC onto a finite subset of \(\mathcal{E}\), thereby yielding a truncated system that can be solved with standard ODE techniques. The appeal of FSP lies in its conceptual simplicity and its ability to provide computable error bounds for the approximation it delivers. However, the principal limitation of FSP is its susceptibility to the \emph{curse of dimensionality} (CoD) in the CME \cite{fang2024advanced}: the size of the truncated state space required for an accurate approximation increases exponentially with the number of species in the SRN. Consequently, FSP becomes computationally infeasible for high-dimensional SRNs and remains practical only for systems with a small number of species. Moreover, even in such low-dimensional settings, FSP can only yield reliable solutions over finite time horizons; it does not support accurate long-time approximation over the entire interval \([0, \infty)\) (see \cite{gupta2017finite}). 

These limitations of the FSP method are fundamentally rooted in the mathematical structure of the CME~\eqref{cme_as_linear_system}. As a linear operator, the matrix \(\mathbf{A}\) is typically \emph{non-compact} (see the Appendix, Section \ref{supp:compactness_of_the_koopman_operator}) and hence it cannot be accurately approximated by finite-dimensional operators \cite{conway1994course}—an assumption implicitly made by the FSP method.

To develop a comprehensive understanding of the dynamics of a SRN, it is often necessary to compute the solution of the CME for multiple initial states \(x\) (e.g. see \cite{andreychenko2012approximate}). This task is the stochastic analogue to constructing the \emph{flow map} in deterministic dynamical systems \cite{katok1995introduction}, as it reveals how the system responds to variations in initial conditions and supports tasks such as initial state inference from experimental data. However, given that solving the CME for a single initial state is already computationally demanding, extending this to a large set of initial conditions significantly compounds the challenge. Therefore, there is a critical need for efficient computational methods that can estimate the CME solution at multiple initial conditions without duplicating the full computational effort for each one. The method we develop in this paper is designed to address this critical need, as we elaborate later.

In recent years, numerous machine learning (ML)-based approaches have been proposed to address the computational challenges associated with estimating solutions to the Chemical Master Equation (CME). These methods have shown significant promise in circumventing the curse of dimensionality and providing efficient approximations to the CME solutions. For instance, the \textit{DeepCME} method~\cite{gupta2021deepcme} employs a reinforcement learning framework grounded in an almost sure relationship derived from Kolmogorov’s backward equation, which must be satisfied by each CTMC trajectory. Another notable method, \textit{Nessie}~\cite{sukys2022approximating}, assumes that the marginal distribution of a particular output species can be represented as a mixture of negative binomial distributions. By leveraging simulated trajectories across multiple parameter values, it trains a neural network to estimate the time-dependent marginal distributions over the entire parameter space. Similarly, the \textit{NNCME} method~\cite{tang2023neural} focuses on estimating the full joint probability distribution of the CME by assuming a finite state space and factorizing the distribution as a product of conditional distributions. Each conditional distribution is modeled using a Variational Autoregressive Network (VAN), trained by aligning its predictions with the CME's transition kernel over small time increments. Building on this, the \textit{MET} method~\cite{liu2024distilling} extends NNCME by applying reinforcement learning-based knowledge distillation. MET enhances generalization capabilities, enabling accurate predictions of CME solutions across previously unseen parameter values, initial conditions, and time points. Despite their demonstrated effectiveness, ML-based approaches come with inherent limitations. First, their accuracy is difficult to rigorously assess without access to the exact solution, and the resulting predictions often lack interpretability. Furthermore, methods like NNCME and MET rely on the assumption of a finite state space—a significant restriction for many practical systems. Since their training depends on time discretizations, accumulated errors can become non-negligible over long time horizons. Moreover, approaches such as DeepCME, Nessie, and NNCME assume a fixed initial state and do not readily generalize to scenarios requiring solutions across varying initial conditions.

In contrast, the method proposed in this paper is not based on machine learning. It is specifically designed to produce interpretable estimates of the temporal dynamics of user-defined summary statistics of the CME solution, valid over the entire time interval $[0, \infty)$ and for any given initial state. Importantly, our approach also provides an \textit{approximate error bound} that can be computed without requiring access to the true solution, offering a significant advantage in both theoretical insight and practical reliability.

We now formally describe the central goal of our study. Let $\mathcal{E}$ denote the state space of the stochastic reaction network (SRN), and let $\mathcal{F}$ be a finite collection of real-valued \emph{observable functions} defined on $\mathcal{E}$. For each $f \in \mathcal{F}$, we define a corresponding \emph{summary statistic} of the CME solution $p_t(y \vert x)$ as
\begin{align}
\label{defn_koopman_operater}
\mathcal{K}_t f(x) := \mathbb{E}[f(X_x(t))] = \sum_{y \in \mathcal{E}} f(y) \, p_t(y \vert x),
\end{align}
where $\E$ denotes the expectation operator. These summary statistics encode specific features of the distribution $p_t(\cdot \vert x)$ as numerical quantities. By computing them for each $f \in \mathcal{F}$ over all times $t \in [0, \infty)$, we obtain a concise yet informative description of how the CME solution evolves over time. To illustrate, consider an SRN with $D$ species, where the state at any time is represented by a vector $y = (y_1, \dots, y_D)$, denoting the copy-numbers of each species. Let $k \in \{1, \dots, D\}$, and define $\mathcal{F} = \{f_1, f_2\}$ with $f_j(y) = y_k^j$ for $j = 1, 2$. Then $\mathcal{K}_t f_1(x)$ and $\mathcal{K}_t f_2(x)$ represent, respectively, the first and second moments of the copy-number of the $k$-th species at time $t$, given that the system starts from initial state $x = (x_1, \dots, x_D)$. Depending on the application, $\mathcal{F}$ can be expanded to include observable functions capturing higher-order moments, cross-moments between multiple species, or probabilities of specific events by incorporating indicator functions.

This framework is particularly relevant in biological applications, where single-cell measurement techniques—such as Flow Cytometry or single-cell RNA sequencing—yield empirical estimates of marginal distributions for a subset of species at various time points \cite{munsky2009listening,gorin2023studying}. By selecting an appropriate set of observable functions $\mathcal{F}$, one can compute the corresponding summary statistics from experimental data and compare them with the predicted values of $\mathcal{K}_t f(x)$ for each \( f \in \mathcal{F} \). This enables a principled approach to assessing the consistency of a given stochastic reaction network (SRN) model with empirical observations.

Note that $\mathcal{K}_t$, as defined in \eqref{defn_koopman_operater}, is essentially the transition semigroup associated with the underlying Markov process~\cite{ethier2009markov}, and can also be interpreted as the \emph{Koopman operator} corresponding to the stochastic dynamics \cite{mauroy2020introduction}. This operator, introduced by Bernard Koopman in 1931~\cite{koopman1931hamiltonian}, acts linearly on observable functions \( f \), thereby enabling the analysis of nonlinear dynamical systems through the lens of linear systems theory. In recent years, Koopman operators have received considerable attention in the context of data-driven modeling of complex dynamical systems~\cite{williams2015data}. They have been applied to the development of control strategies grounded in linear control theory~\cite{korda2018linear}, the identification of dominant spatiotemporal patterns via Koopman mode decomposition~\cite{rowley2009spectral}, and the spectral analysis of system dynamics. In the latter case, the eigenvalues and eigenfunctions of the Koopman operator yield insights into the long-term behavior of the system, including properties such as stability, periodicity, and coherence~\cite{rowley2009spectral,mauroy2016global}.

Our objective in this paper is related to the latter application. However, rather than using the Koopman operator for spectral analysis of the system dynamics, we leverage spectral analysis techniques to construct a low-dimensional representation of the Koopman operator for a general nonlinear SRN. Specifically, our aim is to solve the following estimation problem:
\begin{align}
\label{estimation_task1}
\textbf{Task 1:} \qquad \textnormal{for any initial state} \ x\in \mathcal{E}, \ \textnormal{estimate} \ \mathcal{K}_t f(x) \ \textnormal{for each} \ f\in \mathcal{F} \ \textnormal{and} \ t \geq 0.
\end{align}
As mentioned above, if we succeed in this goal then it would help assess fidelity of the SRN model by comparing the Koopman operator estimated from it with the Koopman operator derived from empirical single-cell data (e.g., obtained via Flow Cytometry, single-cell RNA sequencing, etc.). Furthermore, as we discuss later, our estimation framework naturally extends to compute the sensitivities of the Koopman operator with respect to the SRN model parameters \cite{gupta2013unbiased,gupta2014efficient}. These sensitivities can be used to construct gradient-based optimization procedures that adjust model parameters to align more closely with the data-derived Koopman operator, thus enabling effective parameter inference. In addition, our Koopman operator estimation framework provides a pathway for computing cross-spectral densities associated with the underlying Markov chain trajectories. These spectral densities quantify signal strength across frequencies and offer valuable insights into the dynamical structure of the system \cite{gupta2022frequency}. They can also serve as diagnostic tools for evaluating model consistency against experimental time-lapse microscopy data.

We now explain why the Koopman operator estimation task \eqref{estimation_task1} is computationally challenging. Even if we assume mass–action kinetics (see \eqref{main:massactionkinetics}) for the SRN and restrict the class of observable functions \(\mathcal{F}\) to monomials (e.g. first- and second-order monomials corresponding to the first two moments of a subset of species), the resulting system of ODEs describing the time evolution of \(\mathcal{K}_t f(x)\) is not closed whenever the SRN is nonlinear (e.g. contains bimolecular reactions). This leads to the familiar moment-closure problem~\cite{schnoerr2015comparison}, for which no universally satisfactory remedy exists~\cite{schnoerr2014validity}. A viable alternative is to simulate trajectories of the underlying CTMC using, for example, Gillespie’s stochastic simulation algorithm (SSA) \cite{gillespie1977exact} and to estimate the action of the Koopman operator at multiple time points via a Monte Carlo (MC) estimator for each \(f \in \mathcal{F}\) (see the Appendix, Section \ref{supp_sec:monte_carlo_estimators}). This strategy, however, has several drawbacks. First, the computational cost of SSA-based sampling grows linearly with the sample size $N$, while the MC standard error decreases only as \(1/\sqrt{N}\); hence, halving the error requires quadrupling the number of samples. Second, whereas the true dynamics evolve continuously over \([0,\infty)\), this approach yields only a time-discretized approximation of the Koopman operator over a finite time horizon, hampering the identification of long-term trends. Third—and most importantly—for every new initial state the MC estimator must be constructed afresh by simulating a new batch of CTMC trajectories; trajectories generated for previous initial states cannot be directly reused to reduce the variance of the MC estimator for the new state. Taken together, these issues motivate us to develop a new approach for estimating the Koopman operator that leverages spectral analysis to produce a simple analytical expression for its continuous-time evolution. Our method applies to SRN models with arbitrary kinetics and to arbitrary classes of observable functions \(\mathcal{F}\). Furthermore, while our method is not exact, it yields a computable \emph{a posteriori} error bound that does not require access to the true Koopman operator.

We now elaborate on the core idea underlying our approach. We assume that the continuous-time Markov chain (CTMC) governing the dynamics of the stochastic reaction network (SRN) is \emph{exponentially ergodic}, meaning that the probability distribution \eqref{prob_distr_random_state} solving the CME converges at an exponential rate to a unique stationary distribution $\pi$ on the state space $\mathcal{E}$. This property can be verified using the techniques of \cite{gupta2014scalable,gupta2018computational}, and it implies that the CTMC generator $\mathbb{A}$ (the adjoint of the CME operator $\mathbf{A}$; see \eqref{main:defn_genA}) has a simple eigenvalue at $0$ (i.e., geometric multiplicity one), while all other eigenvalues have strictly negative real parts. Typically, these eigenvalues are also simple and distinct and tend to $-\infty$, allowing us to order them by decreasing real part as
\begin{align}
\label{eigen_val_seq}
0 = -\sigma_0,\; -\sigma_1,\; -\sigma_2,\; \dots \;\longrightarrow\; -\infty,
\end{align}
where $-\sigma_j$ denotes the $j$th eigenvalue, and $\sigma_j$ is a complex number with positive real part that we refer to as the $j$th \emph{decay mode} of the SRN. The eigenvalues of the Koopman operator $\mathcal{K}_t$ are $\{ e^{-\sigma_j t} : j = 0,1,\dots \}$, and for each $t>0$ they converge to $0$ as $j \to \infty$, indicating the compactness of operator (see the Appendix, Section \ref{supp:compactness_of_the_koopman_operator}). The eigenvalue decay ensures that higher (larger-index $j$) modes carry progressively smaller weight, enabling an accurate finite-dimensional approximation of $\mathcal{K}_t$. This observation is the cornerstone of our methodology.

Let $\phi_j(x)$ denote the eigenfunction of the CTMC generator $\mathbb{A}$ associated with eigenvalue $-\sigma_j$. For $j = 0$ (with $\sigma_0 = 0$), the eigenfunction is the constant function $\phi_0(x) \equiv \mathbf{1}$. Furthermore, each $\phi_j(x)$ is also an eigenfunction of the Koopman operator $\mathcal{K}_t$, corresponding to eigenvalue $e^{-\sigma_j t}$. Thus, for any function $f \in \mathcal{F}$, we may express its spectral expansion as
\begin{align*}
f(x) = \sum_{j=0}^\infty \alpha_j(f) \phi_j(x) = \alpha_0(f) + \sum_{j=1}^\infty \alpha_j(f) \phi_j(x).
\end{align*}
Letting $\alpha_j(f,x) := \alpha_j(f) \phi_j(x)$, the action of the Koopman operator on $f$ becomes
\begin{align*}
\mathcal{K}_t f(x) = \alpha_0(f) + \sum_{j=1}^\infty \alpha_j(f,x) e^{-\sigma_j t},
\end{align*}
and it is immediate that $\alpha_0(f)$ is equal to the stationary expectation
\begin{align*}
\E_\pi(f) = \sum_{x \in \mathcal{E}} f(x) \pi(x).
\end{align*}
Our approach involves estimating a truncated version of this spectral series, namely,
\begin{align}
\label{spectral_expansion}
\mathcal{K}_t f(x) \approx \E_\pi(f) + \sum_{j=1}^J \alpha_j(f,x) e^{-\sigma_j t},
\end{align}
using a specialized procedure detailed below. In the first stage, we estimate the components that are independent of the initial state $x$, which include the dominant decay rates $\sigma_1,\dots,\sigma_J$ and the stationary expectation $\E_\pi(f)$ for each $f \in \mathcal{F}$. These state-independent quantities are estimated only once per SRN. Subsequently, for any initial state $x$, the approximation in \eqref{spectral_expansion} becomes linear in the state-dependent coefficients $\alpha_1(f,x), \dots, \alpha_J(f,x)$, which can then be efficiently estimated using a linear regression procedure based on a small number of CTMC simulations.

We demonstrate that this two-tiered approach yields highly accurate estimates of the Koopman operator's continuous-time evolution. In particular, the approximation error decreases exponentially as the number of decay modes \(J\) increases, a trend consistent with existing results on the use of spectral expansions in solving partial differential equations (PDEs) \cite{davies1995spectral,gross2024sparse}. We refer to our methodology as SKA, an abbreviation for {\bf S}tochastic {\bf K}oopman {\bf A}pproximation.

Through a range of illustrative examples, we show that SKA is significantly more computationally efficient than standard methods that rely on Monte Carlo estimators based on stochastic simulation algorithm (SSA) trajectories—particularly when estimating the Koopman operator across multiple initial states, which is important for obtaining the flow map for the stochastic dynamical system or tasks such as inferring the initial state distribution from Flow Cytometry data (see Section \ref{main:self_ge_network}). This efficiency gain becomes even more pronounced in tasks such as computing parameter sensitivities of the operator or estimating spectral densities of the underlying stochastic processes.

The main computational challenge lies in accurately estimating the dominant decay rates \(\sigma_1, \dots, \sigma_J\). To address this, we work in the \emph{frequency domain}, focusing not on the Koopman operator itself but on its \emph{Laplace transform}, which corresponds to the \emph{resolvent} of the Markov transition semigroup. By leveraging the complex-analytic nature and certain other properties of the resolvent, we design an efficient estimation algorithm based on convex optimization. This procedure recovers the dominant decay modes from quantities estimated via carefully designed, highly parallelized large-scale simulations of the CTMC, ideally performed on a graphical processing unit (GPU) using code we provide.

We now define and motivate the problem of estimating the parameter sensitivities and spectral densities. Suppose that our SRN model depends of a finite set of parameters $\Theta$. These could include the rate constants, enzyme concentrations, temperature, etc. that may influence the reaction kinetics. Letting $\mathcal{K}^{ (\Theta)}_t$ be the parameter-dependent Koopman operator, we define the sensitivity with respect to parameter $\theta$
\begin{align}
\label{theta_sensitivity}
\mathcal{S}_{t,\theta}^{ (\Theta)} f(x):= \frac{\partial}{\partial \theta}\mathcal{K}^{ (\Theta)}_t f(x).
\end{align}
Then our sensitivity estimation task can be stated as
\begin{align}
\label{estimation_task_sensitivity}
\textbf{Task 2:} \qquad & \textnormal{for any initial state} \ x\in \mathcal{E}, \notag  \\ & \ \textnormal{estimate} \ \mathcal{S}_{t,\theta}^{ (\Theta)} f(x)\ \ \textnormal{for each} \ f\in \mathcal{F}, \ \theta \in \Theta \ \  \textnormal{and} \ \ t \geq 0.
\end{align}
As mentioned before, having access to parameter sensitivities can help in parameter inference. It can also help examine the robustness properties in networks, and we shall show in Section \ref{main:aif} how the explicit time-evolution of the sensitivities that our approach provides can help assess competing synthetic \emph{Cybergenetic} controllers \cite{briat2016antithetic} that impute the property of Robust Perfect Adaptation (RPA) to any intracellular SRN. 

Dropping the explicit dependence on parameters $\Theta$, let $(X_x(t))_{t \geq 0}$ be a CTMC trajectory for the SRN starting with initial state $x \in \mathcal{E}$. From this trajectory, suppose for any two observable functions $f_1, f_2 \in \mathcal{F}$, we obtain two real-valued time-series $(f_j(X_x(t)))_{t\in [0,T]}$ for $j=1,2$ in the time interval $[0,T]$. We compute their one-sided Fourier Transform as
\begin{align*}
\mathcal{F}_{f_j} (\omega, x, T) = \frac{1}{\sqrt{T}} \int_0^T f_j(X_x(t)) e^{ - i \omega t}dt,
\end{align*}
where $i = \sqrt{-1}$ and $\omega$ is the frequency. We can define the \emph{cross-spectral Density (CSD)} between the two time-series as
\begin{align}
\label{csd_definition}
\textnormal{CSD}_{f_1, f_2} (\omega, x, T) = \E\left( \mathcal{F}_{f_1} (\omega, x, T) \overline{\mathcal{F}_{f_2} (\omega, x, T)}\right),
\end{align}
where $\bar{z}$ denotes the complex conjugate of $z$. With this definition in place, our spectral density estimation task can be stated as
\begin{align}
\label{estimation_task_csd}
\textbf{Task 3:} \qquad & \textnormal{for any initial state} \ x\in \mathcal{E} \ \textnormal{and terminal time} \  T > 0, \notag \\ & \textnormal{estimate} \ \textnormal{CSD}_{f_1, f_2} (\omega, x, T) \ \ \textnormal{for each} \ f_1,f_2\in \mathcal{F} \  \textnormal{and} \ \ \omega \geq 0.
\end{align}

Note that the magnitude of CSD represents the portion of the two signals that oscillates in synchrony at a given frequency, while the argument of CSD estimates the phase difference. In the special case, $f_1=f_2=f$, this CSD becomes the \emph{Power Spectral Density (PSD)} for the signal $(f(X_x(t)))_{t\in [0,T]}$ and we write it as $\textnormal{PSD}_{f} (\omega, x, T)$. Quantities like CSD or PSD help us analyze SRNs corresponding to synthetic oscillators, like the famous \emph{Repressilator} \cite{elowitz2000synthetic}, and help us identify ways to tune the parameters so as to make the spectral density peaks sharper, enhancing the oscillatory power (see Section \ref{main:repressilator_network}). While comparing RPA-achieving synthetic controllers, it is desirable to reduce oscillations and spectral analysis assists in that effort (see Section \ref{main:aif}).

We now outline the core idea behind extending our Koopman operator estimation framework (see \textbf{Task 1} given \eqref{estimation_task1}) to two additional tasks: estimating parameter sensitivities (\textbf{Task 2} given by \eqref{estimation_task_sensitivity}) and spectral densities (\textbf{Task 3} given by \eqref{estimation_task_csd}). Leveraging a result from~\cite{gupta2018estimation}, we express the parameter sensitivity $\mathcal{S}_{t,\theta}^{(\Theta)} f(x)$ as a time integral whose integrand involves the Koopman operator applied to a new class of functions $\mathcal{G} \neq \mathcal{F}$. Since we already estimate the Koopman operator’s action on a known set of functions $\mathcal{F} = \{f_1, \dots, f_F\}$, we also know its action on their linear span $\mathcal{L}(\mathcal{F})$ (see \eqref{defn_linear_span_bar_F}). This allows us to project any function $g \in \mathcal{G}$ onto $\mathcal{L}(\mathcal{F})$, enabling an accurate approximation of its Koopman image—and hence, an explicit expression for the parameter sensitivity. The same idea extends to estimating cross-spectral densities $\textnormal{CSD}_{f_1, f_2} (\omega, x, T)$, by leveraging the Markov property to relate them to the Koopman operator.

A key advantage of our method is that it yields an explicit, continuous mapping from time $t$ to the sensitivity $\mathcal{S}_{t,\theta}^{(\Theta)} f(x)$, allowing us to track how sensitivity evolves from $t = 0$ to $\infty$. In contrast, existing methods provide only time-discretized estimates over finite intervals \cite{Rathinam2010, anderson2012efficient, gupta2013unbiased,gupta2014efficient,gupta2018estimation}. These approaches often rely on finite-difference approximations and simulations of coupled trajectories—methods that introduce bias and become unreliable over long time horizons due to the decaying strength of coupling, which leads to high variance in the estimates. While our method is also biased, it avoids this long-term variance issue, offering a more robust estimate across the entire time interval $[0, \infty)$.

Similarly, for spectral density estimation, our approach yields an explicit continuous map from both the frequency $\omega$ and the terminal time $T$ to the quantity $\textnormal{CSD}_{f_1, f_2} (\omega, x, T)$. This enables a complete examination of how signal strength and phase relationships are distributed across frequencies, and how this distribution evolves with the trajectory length $T$ (see Supplementary Movies 1-4). In contrast, standard spectral estimation methods typically rely on discretizing the trajectory in time and then applying the Discrete Fourier Transform (DFT). This restricts CSD estimates to a finite, $T$-dependent set of sampled frequencies. Each change in $T$ requires a full recomputation of the DFT, making it difficult to track the evolution of CSD with time. Moreover, discretizing continuous trajectories can introduce aliasing effects, which may distort the spectral density \cite{Engelberg2008}. By avoiding time discretization altogether, our method sidesteps these aliasing effects and provides a more reliable estimate.

\section{Mathematical Preliminaries} \label{sec:prelim}

In this section, we introduce key mathematical concepts that form the foundation of our approach. Throughout, we use the notation \( \mathbb{R} \) (resp.\ \( \mathbb{R}_+ \)) to denote the set of all real numbers (resp.\ strictly positive real numbers), \( \mathbb{Z} \) (resp.\ \( \mathbb{Z}_{\geq 0} \)) for the set of all integers (resp.\ nonnegative integers), and \( \mathbb{C} \) (resp.\ \( \mathbb{C}_+ \)) for the set of all complex numbers (resp.\ those with strictly positive real part). 

\subsection{The stochastic model of a reaction network}

Consider a reaction network with $D$ species, denoted $\mathbf{X}_1,\dots,\mathbf{X}_D$, and $K$ reactions. In the classical stochastic formulation, the dynamics of the system are described by a \emph{continuous-time Markov chain} (CTMC) \cite{anderson2015stochastic}, whose states represent the copy numbers of the $D$ species. If the system is in state $x = (x_1,\dots,x_D) \in \Z^D_{ \geq 0}$ and reaction $k$ occurs, the state changes by the stoichiometric vector $\zeta_k \in \mathbb{Z}^D$. The rate at which reaction $k$ fires in state $x$ is determined by the propensity function $\lambda_k(x)$. Under the mass-action kinetics assumption \cite{anderson2015stochastic}, this function is given by
\begin{align}
\label{main:massactionkinetics}
\lambda_k(x_1,\dots,x_D) = \theta_k \prod_{j =1}^D \frac{ x_j(x_j-1)\dots (x_j - \nu_{jk} +1 ) }{  \nu_{jk} ! },
\end{align}
where $\theta_k$ is the reaction rate constant, and $\nu_{jk}$ denotes the number of molecules of $\mathbf{X}_j$ consumed in the $k$-th reaction.

The CTMC can be formally defined via its infinitesimal generator $\mathbb{A}$, which encodes the instantaneous rate of change of expectations of bounded real-valued functions $f$ defined on the state space $\mathcal{E} \subseteq \mathbb{Z}_{\geq 0}^D$, the set of all accessible states. The generator of the CTMC is defined by
\begin{align}
\label{main:defn_genA}
(\mathbb{A} f)(x)  = \sum_{k=1}^K \lambda_k(x) \big( f(x+\zeta_k) - f(x) \big),
\end{align}
for all bounded functions $f : \mathcal{E} \to \mathbb{R}$. Under suitable growth conditions on the propensities $\lambda_k$, the same expression defines the \emph{extended (Dynkin) generator} on classes of polynomially growing functions $f$; see 
\cite{anderson2015stochastic}.

Let $(X_x(t))_{t \geq 0}$ denote the CTMC with generator $\mathbb{A}$ and initial state $x \in \mathcal{E}$, and let $p_t(y \vert x)$ denote the probability distribution of the state $X_x(t)$. Then, the evolution of this distribution is governed by the \emph{Chemical Master Equation (CME)}, a system of linear ordinary differential equations given by
\begin{align}
\label{defn_cme}
\frac{ d p_t(y \vert x) }{dt} = \sum_{k=1}^K  p_t(y -\zeta_k \vert x) \lambda_k(y - \zeta_k) -p_t(y \vert x) \sum_{k=1}^K \lambda_k(y), \quad \textnormal{for each } y \in \mathcal{E}.
\end{align}
Upon enumerating the countable state space $\mathcal{E}$, the probability distribution $p_t(y \vert x)$ may be represented as a vector, in which case the CME can be written as the linear system \eqref{cme_as_linear_system} with $\mathbf{A} = \mathbb{A}^\dagger$, the adjoint of the generator $\mathbb{A}$.

As stated in the introduction, we assume that the CTMC is exponentially ergodic with stationary distribution $\pi$ over $\mathcal{E}$. To develop our approach, we shall work over the Hilbert space
\begin{align*}
\mathcal{L}_2(\pi) = \left\{f : \mathcal{E} \to \mathbb{R} :  \left\| f \right\|^2_{\mathcal{L}_2(\pi)}  <\infty  \right\},
\end{align*}
where the associated norm is
\begin{align}
\label{defn_l2_norm}
\left\| f \right\|^2_{\mathcal{L}_2(\pi)}:= \sum_{y \in \mathcal{E} } (f(y))^2 \pi(y).
\end{align}

The Koopman operator $\mathcal{K}_t$, defined by \eqref{defn_koopman_operater}, can be viewed as an operator from $\mathcal{L}_2(\pi)$ to $\mathcal{L}_2(\pi)$, and it corresponds to the transition semigroup of the CTMC. Formally, it can be expressed as the operator exponential
\begin{align*}
\mathcal{K}_t f(x) = \left[e^{t \mathbb{A}} \right]f (x).
\end{align*}
This implies that $-\sigma_j$ is an eigenvalue of $\mathbb{A}$ with corresponding eigenfunction $\phi_j(x)$ if and only if $e^{-\sigma_j t}$ is an eigenvalue of $\mathcal{K}_t$ with the same eigenfunction $\phi_j(x)$. As discussed in Section \ref{supp:compactness_of_the_koopman_operator} of the Appendix, for most SRN models of interest, the operator $\mathcal{K}_t$ can be expected to be compact for all $t > 0$. This compactness implies that $e^{-\sigma_j t} \to 0$ as $j \to \infty$, or equivalently, $-\sigma_j \to -\infty$, as indicated in \eqref{eigen_val_seq}. We shall assume that all the eigenfunctions belong to $\mathcal{L}_2(\pi)$.

\subsection{The Resolvent Operator}

To develop our approach for estimating the Koopman operator, we shift our perspective from the time domain to the frequency domain by employing the \emph{resolvent operator}~\cite{ethier2009markov}. For any frequency \( s \in \mathbb{C}_+ \), the resolvent operator $\mathcal{R}_s:\mathcal{L}_2(\pi) \to \mathcal{L}_2(\pi)$ is defined as
\begin{align}
\label{defn_resolvent_operator}
\mathcal{R}_s f(x) = \int_{0}^{\infty} s e^{- s t } \mathcal{K}_t f(x) \, dt.
\end{align}
The compactness of \( \mathcal{R}_s \) is equivalent to the compactness of the Koopman operator \( \mathcal{K}_t \). Moreover, the spectral properties of the two operators are closely related: \( \mathcal{K}_t \) admits eigenvalue \( e^{-\sigma_j t} \) with eigenfunction \( \phi_j(x) \) if and only if \( \mathcal{R}_s \) has eigenvalue \( \frac{s}{s + \sigma_j} \) with the same eigenfunction. This spectral correspondence underpins our strategy for constructing a frequency-domain approximation of the Koopman operator.

The resolvent operator possesses several properties that are instrumental in our methodology. These include the complex-analytic nature of the map \( s \mapsto \mathcal{R}_s f(x) \), its representation via a Taylor series expansion, the initial and final value theorems, and a relation between its iterates\footnote{The \( m \)-th iterate of the resolvent operator is defined by composition: 
\[
\mathcal{R}^{m}_s = \underbrace{\mathcal{R}_s \circ \mathcal{R}_s \circ \dots \circ \mathcal{R}_s}_{m \textnormal{ times}}.
\]
By convention, \( \mathcal{R}^{0}_s \equiv \mathbf{I} \), the identity operator.}
and its derivatives with respect to \( s \) (see the Appendix, Proposition \ref{properties_of_the_resolvent_operator}). Of particular importance for us is the following expression for the iterates of the resolvent map:
\begin{align}
\label{resolvent_iterates_formula}
\mathcal{R}^{m}_s f(x) = 
\begin{cases}
f(x), & \text{if } m = 0, \\[1ex]
\displaystyle\frac{s^m}{(m-1)!} \int_0^\infty t^{m-1} e^{- s t} \mathcal{K}_{t} f(x) \, dt, & \text{for } m = 1,2,\dots.
\end{cases}
\end{align}

\subsection{Expression for Parameter Sensitivity}
\label{prelim:expr_param_sens}
Suppose that the propensities in the stochastic reaction network (SRN) model depend on a finite set of parameters \( \Theta \), so that the propensity function for the $k$th reaction is written as \( \lambda_k(x, \Theta) \) rather than \( \lambda_k(x) \). Let \( \mathcal{K}^{(\Theta)}_t \) denote the corresponding parameter-dependent Koopman operator. For each $\theta \in \Theta$, we define the \emph{parameter sensitivity} of the Koopman operator as the partial derivative with respect to $\theta$, as in \eqref{theta_sensitivity}.

Let $\left(X^{(\Theta)}_x(t)\right)_{t \geq 0}$ denote the parameter-dependent CTMC with initial state \( x \in \mathcal{E} \). Theorem 3.3 in~\cite{gupta2018estimation} establishes that the sensitivity of the Koopman operator admits the following integral representation:
\begin{align}
\label{sensitivity_koopman_repr}
\mathcal{S}^{(\Theta)}_{t,\theta} f(x) = \frac{\partial}{\partial \theta} \mathcal{K}^{(\Theta)}_t f(x) = \sum_{k=1}^K \int_0^t \mathbb{E} \left[ \frac{\partial \lambda_k}{\partial \theta}\left(X^{(\Theta)}_x(s), \Theta \right) \Delta_k \mathcal{K}^{(\Theta)}_{t-s} f \left( X^{(\Theta)}_x(s) \right) \right] ds,
\end{align}
where \( \Delta_k \) is the difference operator defined by \( \Delta_k f(x) := f(x + \zeta_k) - f(x) \).

As shown in~\cite{gupta2018estimation}, this formulation enables the construction of efficient Monte Carlo estimators for parameter sensitivities. However, the computational cost remains substantial due to the need to evaluate
\begin{align}
\label{koopman_differences_path}
\Delta_k \mathcal{K}^{(\Theta)}_{t-s} f \left( X^{(\Theta)}_x(s) \right) = \mathcal{K}^{(\Theta)}_{t-s} f \left( X^{(\Theta)}_x(s) + \zeta_k \right) - \mathcal{K}^{(\Theta)}_{t-s} f \left( X^{(\Theta)}_x(s) \right),
\end{align}
along the trajectory \( \left(X^{(\Theta)}_x(t)\right)_{t \geq 0} \), which involves simulating \emph{auxiliary paths} that branch from the main trajectory (see~\cite{gupta2018estimation, gupta2013unbiased, gupta2014efficient}). Consequently, computing the full time-integral becomes particularly challenging.

In this work, we show that our spectral approximation of the Koopman operator simplifies the time-integral in \eqref{sensitivity_koopman_repr} and yields an explicit expression for parameter sensitivity valid for all $t > 0$, thereby addressing \textbf{Task 2} (see~\eqref{estimation_task_sensitivity}). Moreover, we demonstrate that certain quantities can be pre-computed, allowing rapid evaluation of the sensitivity for any new initial state \( x \) with minimal additional computation.

\subsection{Expression for Cross-Spectral Density (CSD)}

Dropping parameter dependence for now, suppose the SRN dynamics are governed by the CTMC \( \left(X_x(t)\right)_{t \geq 0} \) with initial state \( x \in \mathcal{E} \). Let \( f_1, f_2 \in \mathcal{F} \) be two observable functions, generating the real-valued stochastic processes \( \left(f_1(X_x(t))\right)_{t \geq 0} \) and \( \left(f_2(X_x(t))\right)_{t \geq 0} \), respectively. The CSD over the time interval \([0, T]\) is defined in equation~\eqref{csd_definition}. Substituting the expressions for the one-sided Fourier transforms, we obtain
\begin{align*}
\textnormal{CSD}_{f_1, f_2} (\omega, x, T) &= \frac{1}{T} \int_0^T \int_0^T 
\mathbb{E}\left[ f_1(X_x(t)) f_2(X_x(u)) \right] e^{ - i \omega (t-u)} \, dt \, du \\
&= \frac{1}{T} \int_0^T \left[ \int_u^T 
\mathbb{E}\left[ f_1(X_x(t)) f_2(X_x(u)) \right] e^{ - i \omega (t-u)} \, dt \right] du \\
&\quad + \frac{1}{T} \int_0^T \left[\int_u^T 
\mathbb{E}\left[ f_1(X_x(u)) f_2(X_x(t)) \right] e^{  i \omega (t-u)} \, dt \right] du.
\end{align*}

Using the Markov property and the law of iterated expectations, we can simplify the joint expectation for $s>t$ as:
\begin{align*}
\mathbb{E}\left[ f_1(X_x(t)) f_2(X_x(s)) \right] = \mathbb{E}\left[ f_1(X_x(t)) \, \mathbb{E} \left[f_2(X_x(s)) \mid X_x(t) \right] \right] =  \mathbb{E}\left[ f_1(X_x(t)) \, \mathcal{K}_{s-t} f_2(X_x(t)) \right].
\end{align*}

This yields the following Koopman operator representation of the CSD:
\begin{align}
\label{csd_in_terms_Koopman_operator}
\textnormal{CSD}_{f_1, f_2} (\omega, x, T) 
&= \frac{1}{T} \int_0^T \mathbb{E} \left[ f_2(X_x(u)) \int_u^T 
\mathcal{K}_{t-u} f_1(X_x(u)) \, e^{ - i \omega (t-u)} \, dt \right] du \\
&\quad + \frac{1}{T} \int_0^T \mathbb{E} \left[ f_1(X_x(u)) \int_u^T \mathcal{K}_{t-u}
f_2(X_x(u)) \, e^{  i \omega (t-u)} \, dt \right] du. \notag
\end{align}

This representation involves a double time-integral over the action of the Koopman operator, making its evaluation particularly demanding at each frequency \( \omega \). We show that our spectral approximation once again simplifies this integral expression and leads to an explicit analytical form for the CSD as a function of both frequency \( \omega \) and terminal time \( T \), thereby addressing \textbf{Task 3} (see~\eqref{estimation_task_csd}). Furthermore, as in the case of parameter sensitivity, many quantities involved in the CSD computation can be pre-computed, enabling fast and efficient evaluation for any new initial condition \( x \).

We note that this approach generalises the spectral method developed in~\cite{gupta2022frequency} for estimating the steady-state power spectral density (PSD), where the observable functions \( f_1 \) and \( f_2 \) were required to be identical. In contrast, the method proposed here allows for estimation of the full CSD, where \( f_1 \) and \( f_2 \) may differ, and it applies not only in the steady-state regime  $T \to \infty$ but also for any finite terminal time $T$, making it more reflective of real-world experimental settings.

\section{Main Results}

In this section, we present the core mathematical results that form the foundation of our spectral approximation of the stochastic Koopman operator, as expressed in equation \eqref{spectral_expansion}.

\subsection{Frequency Domain Characterization of the Koopman operator}

The basis of our approach is the following theorem which addresses the case in which the finite spectral expansion \eqref{spectral_expansion} is exact, providing a characterization of this scenario in the frequency domain. Before we state the result we recall that the $J \times J$ Lagrange matrix $\mathbf{L}$ with interpolation points $x_1,\dots,x_J$ is defined as the matrix whose $j$-th row is the vector of coefficients of the Lagrange interpolation polynomial, i.e.\
\begin{align}
\label{lagrange_interpolation_poly}
\sum_{k=1}^J \mathbf{L}_{jk} x^{k-1} = \prod_{k=1, k\neq j }^J \left( \frac{x-x_k}{x_j-x_k}\right).
\end{align}

\begin{theorem}[Frequency Domain Characterization of the Exact Finite Representation of the Koopman operator]
\label{thm_1}
Let $\sigma_1,\dots, \sigma_J  \in \C_+$ and suppose $f \in \mathcal{F}$ is an observable function with stationary expectation $\E_\pi (f)$. Then \eqref{spectral_expansion} holds exactly for all $t \geq 0$ and all $x \in \mathcal{E}$, if and only if for \textbf{some} $s \in \C_+$
\begin{align}
\label{resol_cond1}
\left( \frac{ \prod_{j=1}^J \left( s( \mathcal{R}_s -\mathbf{I}) +\sigma_j \mathcal{R}_s \right) }{ \prod_{j=1}^J \sigma_j }\right) f(x) = \E_\pi(f) \quad \textnormal{for all} \quad x \in \mathcal{E}.
\end{align}
Moreover, if \eqref{resol_cond1} holds for one $s \in \C_+$ then it will hold for each $s \in  \C_+$, and the coefficients $\alpha_1(f,x),\dots, \alpha_J(f,x)$ in \eqref{spectral_expansion} are given by
\begin{align}
\label{coeff_alphas}
\alpha_j(f,x) = \left( \frac{s+\sigma_j}{\sigma_j} \right) \sum_{k=1}^J \mathbf{L}_{jk} (s) \left(  \mathcal{R}^{k-1}_s -   \mathcal{R}^{k}_s \right) f(x)  \qquad  \textnormal{\textbf{for any}} \qquad s \in \C_+,
\end{align}
where $\mathbf{L}_{jk} (s)$ is the $J \times J$ Lagrange matrix with interpolation points $s/(s+\sigma_j)$ for $j=1,\dots, J$. In fact, if $\alpha_j(f,x) \neq 0$ then it an eigenfunction for the resolvent operator $\mathcal{R}_s$ corresponding to the eigenvalue $s/(s+\sigma_j)$, i.e.\
\begin{align}
\label{eigenfunction_reln_phi_j}
\mathcal{R}_s \alpha_j(f,x) = \left( \frac{s}{s+\sigma_j} \right) \alpha_j(f,x).
\end{align}
\end{theorem}

A complete proof of this theorem is given in the Appendix, Section~\ref{supp:sec_frequency_domain_characterization}. Intuitively, \eqref{resol_cond1} holds because the operator involving the resolvent on the left-hand side is designed such that it only removes the contributions of all $J$ eigenmodes from the dynamics of $\mathcal{K}_t f(x) = \E(f(X_x(t)))$, leaving just the stationary expectation $\E_\pi(f)$ of the observable function.

A few remarks are in order about this theorem and its implications.
\begin{remark}
\label{thm_rem1}
While proving the result we shall show that if \eqref{spectral_expansion} holds exactly for all $t \geq 0$ and all $x \in \mathcal{E}$ then \eqref{resol_cond1} will actually hold \textbf{for each and every} $s \in \C_+$. However for the reverse implication, it is sufficient to check that \eqref{resol_cond1} holds \textbf{for just one} $s\in \C_+$, in order to ensure that \eqref{spectral_expansion} holds exactly for all $t \geq 0$ and all $x \in \mathcal{E}$. In other words, checking a time-domain relation for all values of time, becomes equivalent to a frequency-domain relation at a single frequency. This occurs due to the complex-analytic nature of the resolvent map and forms a crucial piece in our method.
\end{remark}
\begin{remark}
\label{thm_rem2}
Note that condition~\eqref{resol_cond1} is independent of the coefficients $\alpha_1(f,x), \dots, \alpha_J(f,x)$ appearing in the spectral expansion \eqref{spectral_expansion}. This independence enables us to exploit the relation to estimate the decay modes $\sigma_1, \dots, \sigma_J$ without needing to disentangle the influence of the initial state $x$ or the choice of the observable function $f$. Secondly, although \eqref{coeff_alphas} suggests that the coefficients depend on the specific choice of frequency $s \in \mathbb{C}_+$ at which the right-hand side is evaluated, this theorem ensures that these coefficients are, in fact, independent of this choice.
\end{remark}
\begin{remark}
\label{thm_rem3}
This theorem proves that if \eqref{resol_cond1} holds, then $\alpha_j(f,x)$ defined by \eqref{coeff_alphas} is in fact an eigenfunction for the resolvent operator $\mathcal{R}_s$ corresponding to the eigenvalue $s/(s+\sigma_j)$. This is equivalent to saying that $\alpha_j(f.x)$ is an eigenfunction for the CTMC generator $\mathbb{A}$ corresponding to the eigenvalue $-\sigma_j$, i.e.\
\begin{align}
\label{eigenfunction_reln_phi_j_2}
\mathbb{A} \alpha_j(f,x) = -\sigma_j \alpha_j(f,x).
\end{align}
\end{remark}

Let us illustrate Theorem \ref{thm_1} with a simple example.

\subsubsection{Constitutive gene-expression network}
\label{ex:cons_gene_ex}

This network involves two species, namely the mRNA ($\mathbf{X}_1$) and the protein ($\mathbf{X}_2$), and the following four reactions:
\begin{align*}
& \emptyset  \stackrel{k_r}{\longrightarrow} \mathbf{X}_1  \quad \textnormal{(mRNA transcription)} \\
& \mathbf{X}_1  \stackrel{k_p}{\longrightarrow} \mathbf{X}_1  + \mathbf{X}_2  \quad \textnormal{(protein translation)}  \\
& \mathbf{X}_1   \stackrel{\gamma_r}{\longrightarrow}  \emptyset  \quad \textnormal{(mRNA degradation)}  \\
& \mathbf{X}_2  \stackrel{\gamma_p}{\longrightarrow}   \emptyset  \quad \textnormal{(protein degradation)}.
\end{align*}
We shall assume mass-action kinetics \eqref{main:massactionkinetics} with the reaction rate constants displayed above the reaction arrows. Considering two observable functions $f_1(x_1,x_2) = x_1$ and $f_2(x_1,x_2) =x_2$, $\mathcal{K}_{t} f_1(x_1,x_2)$ and $\mathcal{K}_{t} f_2(x_1,x_2)$ denote the expected number of mRNA and protein in the system at time $t$, given that we have $x_1$ mRNA and $x_2$ protein at the start. As the propensity functions are linear in the state variables, we can compute these quantities exactly as
\begin{align}
\label{koopman_gene_ex_mrna}
\mathcal{K}_{t} f_1(x_1,x_2) &= \frac{k_r}{\gamma_r} + \left( x_1 -  \frac{k_r}{\gamma_r}  \right) e^{ -\gamma_r t} \\
\label{koopman_gene_ex_protein}
\mathcal{K}_{t} f_2(x_1,x_2) &=  \frac{k_r k_p}{\gamma_r \gamma_p} + \frac{k_p}{\gamma_p - \gamma_r}\left( x_1 -  \frac{k_r}{\gamma_r}  \right) e^{ -\gamma_r t} \notag \\
&+ \left[  x_2  - \frac{k_r k_p}{\gamma_r \gamma_p} - \frac{k_p}{\gamma_p - \gamma_r}\left( x_1 -  \frac{k_r}{\gamma_r}  \right)\right] e^{ -\gamma_p t}.
\end{align}
It is immediate that $\E_{\pi} (f_1) =  \frac{k_r}{\gamma_r} $ and $\E_{\pi} (f_2) =  \frac{k_r k_p}{\gamma_r \gamma_p}$, and the finite spectral expansion \eqref{spectral_expansion} is exact, with two decay modes $\sigma_1 = \gamma_r$ and $\sigma_2 = \gamma_p$, and with nonzero coefficients $\alpha_j(f_\ell, x)$ given by
\begin{align}
\label{non_zero_alpha_gene_ex}
\alpha_1(f_1,x) &= \left( x_1 - \frac{k_r}{\gamma_r} \right) \\
\alpha_1(f_2,x) &= \frac{k_p}{\gamma_p - \gamma_r}\left( x_1 -  \frac{k_r}{\gamma_r}  \right) \quad \textnormal{and} 
\quad
\alpha_2(f_2,x) =  x_2  - \frac{k_r k_p}{\gamma_r \gamma_p} - \frac{k_p}{\gamma_p - \gamma_r}\left( x_1 -  \frac{k_r}{\gamma_r}  \right). \notag
\end{align}
Let us see how the two decay modes and the nonzero coefficients, can be identified in the frequency domain, as per Theorem \ref{thm_1}.

Applying formula \eqref{resolvent_iterates_formula}, for any $m=0,1,\dots$, we obtain
\begin{align}
\label{gene_ex_resol_1}
 \mathcal{R}^m_s f_1(x) & = \frac{k_r}{\gamma_r} + \left( x_1 -  \frac{k_r}{\gamma_r}  \right) \left( \frac{s}{s+\gamma_r}\right)^m \\
\label{gene_ex_resol_2}
\textnormal{and} \quad \mathcal{R}^m_s f_2(x) & = \frac{k_r k_p}{\gamma_r \gamma_p} + \frac{k_p}{\gamma_p - \gamma_r}\left( x_1 -  \frac{k_r}{\gamma_r}  \right) \left( \frac{s}{s+\gamma_r}\right)^m \\
 & + \left[  x_2  - \frac{k_r k_p}{\gamma_r \gamma_p} - \frac{k_p}{\gamma_p - \gamma_r}\left( x_1 -  \frac{k_r}{\gamma_r}  \right)\right] \left( \frac{s}{s+\gamma_p}\right)^m. \notag
\end{align} 
For the observable function $f_1(x)$, we see that \eqref{resol_cond1} holds with $J=1$ and $\sigma_1 = \gamma_r$ because
\begin{align*}
\left( \frac{ \prod_{j=1}^J \left( s( \mathcal{R}_s -\mathbf{I}) +\sigma_j \mathcal{R}_s \right) }{ \prod_{j=1}^J \sigma_j }\right) f_1(x) & = \left( \frac{ \left( s( \mathcal{R}_s -\mathbf{I}) +\gamma_r \mathcal{R}_s \right) }{ \gamma_r }\right) f_1(x) \\ 
&= \left( \frac{s+\gamma_r}{\gamma_r}\right) \mathcal{R}_s f_1(x) - \frac{s}{\gamma_r} x_1 \\
& =  \frac{k_r}{\gamma_r} \left( \frac{s+\gamma_r}{ \gamma_r}\right) + \frac{s}{\gamma_r}  \left( x_1 -  \frac{k_r}{\gamma_r}  \right)  - \frac{s}{\gamma_r} x_1 \\
& = \frac{k_r}{\gamma_r} = \E_\pi(f_1).
\end{align*}
The $1\times 1$ Lagrange matrix is just the scalar $1$ and so from formula \eqref{coeff_alphas} we get
\begin{align*}
\alpha_1(f_1, x) & = \left( \frac{s+\sigma_1}{\sigma_1} \right) \sum_{k=1}^J \mathbf{L}_{jk} (s) \left(  \mathcal{R}^{k-1}_s -   \mathcal{R}^{k}_s \right) f_1(x) = \left( \frac{s+\gamma_r}{\gamma_r} \right) \left( x_1 -   \mathcal{R}_s f_1(x) \right) =  \left( x_1 -  \frac{k_r}{\gamma_r}  \right),
\end{align*}
which agrees with \eqref{non_zero_alpha_gene_ex}.

Similarly for the observation function $f_2(x)=x_2$, with $J=2$, $\sigma_1 = \gamma_r$ and $\sigma_2 = \gamma_p$, one can verify that
\begin{align*}
\left( \frac{ \prod_{j=1}^J \left( s( \mathcal{R}_s -\mathbf{I}) +\sigma_j \mathcal{R}_s \right) }{ \prod_{j=1}^J \sigma_j }\right) f_2(x) & = \left( \frac{ \left( s( \mathcal{R}_s -\mathbf{I}) +\gamma_r \mathcal{R}_s \right) \left( s( \mathcal{R}_s -\mathbf{I}) +\gamma_p \mathcal{R}_s \right) }{ \gamma_r \gamma_p }\right) f_2(x) \\ 
& = \frac{k_r k_p}{\gamma_r \gamma_p} = \E_\pi(f_2).
\end{align*}
The $2\times 2$ Lagrange matrix is given by
\begin{align*}
\mathbf{L}(s) =\frac{1}{\gamma_p- \gamma_r} \left[
\begin{array}{cc}
-(s+\gamma_r) & \frac{(s+\gamma_r) (s+\gamma_p)}{s} \\
(s+\gamma_p) & -\frac{(s+\gamma_r) (s+\gamma_p)}{s} \\
\end{array} \right].
\end{align*}
It can be checked that
\begin{align*}
\alpha_1(f_2,x) &= \left(\frac{s+\gamma_r}{\gamma_r} \right)\left[ \mathbf{L}_{11}(s) ( f_2(x) - \mathcal{R}^{1}_s f_2(x)  ) + \mathbf{L}_{12}(s)( \mathcal{R}^{1}_s f_2(x) - \mathcal{R}^{2}_s f_2(x)  )  \right] \\
& =\frac{k_p}{\gamma_p - \gamma_r}\left( x_1 -  \frac{k_r}{\gamma_r}  \right) 
\end{align*}
and similarly
\begin{align*}
\alpha_2(f_2,x) &= \left(\frac{s+\gamma_p}{\gamma_p} \right)\left[ \mathbf{L}_{21}(s) ( f_2(x) - \mathcal{R}^{1}_s f_2(x)  ) + \mathbf{L}_{22}(s)( \mathcal{R}^{1}_s f_2(x) - \mathcal{R}^{2}_s f_2(x)  )  \right] \\
& =x_2  - \frac{k_r k_p}{\gamma_r \gamma_p} - \frac{k_p}{\gamma_p - \gamma_r}\left( x_1 -  \frac{k_r}{\gamma_r}  \right),
\end{align*}
consistent with \eqref{non_zero_alpha_gene_ex}.

Now let us illustrate that having extra decay modes, does not alter the spectral expansion \eqref{spectral_expansion}. For this we return to the observable function $f_1(x)=x_1$ for which \eqref{spectral_expansion} holds with one decay mode (i.e.\ $J=1$) which is $\sigma_1 = \gamma_r$. Suppose we instead set $J=2$ with an extra decay mode $\sigma_2$. Then the $2\times 2$ Lagrange matrix is given by
\begin{align}
\label{lagrange_matrix_gene_exp}
\mathbf{L}(s) =\frac{1}{\sigma_2 - \gamma_r} \left[
\begin{array}{cc}
-(s+\gamma_r) & \frac{(s+\gamma_r) (s+\sigma_2)}{s} \\
(s+\sigma_2) & -\frac{(s+\gamma_r) (s+\sigma_2)}{s} \\
\end{array} \right].
\end{align}
Now applying formula \eqref{coeff_alphas} we get
\begin{align*}
\alpha_1(f_1,x) &= \left(\frac{s+\gamma_r}{\gamma_r} \right)\left[ \mathbf{L}_{11}(s) ( f_1(x) - \mathcal{R}^{1}_s f_1(x)) + \mathbf{L}_{12}(s) (\mathcal{R}^{1}_s f_1(x) -\mathcal{R}^{2}_s f_1(x) ) \right] \\
& = \left( x_1 -  \frac{k_r}{\gamma_r}  \right)  
\left[ -\frac{s+\gamma_r}{\sigma_2 - \gamma_r} + \left(\frac{s}{s+\gamma_r}\right) \frac{(s+\gamma_r) (s+\sigma_2)}{s(\sigma_2 - \gamma_r)}\right] \\
& = \left( x_1 -  \frac{k_r}{\gamma_r}  \right)  
\left[ -\frac{s+\gamma_r}{\sigma_2 - \gamma_r} +  \frac{ s+\sigma_2}{\sigma_2 - \gamma_r}\right]  = \left( x_1 -  \frac{k_r}{\gamma_r}  \right) 
\end{align*}
which is again correct. Now let us compute the coefficient $\alpha_2(f_1,x)$ for the extraneous decay mode $\sigma_2$
\begin{align*}
\alpha_2(f_1,x) &= \left(\frac{s+\gamma_p}{\gamma_p} \right)\left[ \mathbf{L}_{21}(s) (f_1(x) - \mathcal{R}^{1}_s f_1(x)) + \mathbf{L}_{22} ( \mathcal{R}^{1}_s f_1(x) -\mathcal{R}^{2}_s f_1(x))\right] \\
& =  \left( x_1 -  \frac{k_r}{\gamma_r}  \right) \left(\frac{s+\gamma_p}{\gamma_p} \right) \left(\frac{\gamma_r}{s+\gamma_r} \right) 
\left[ \frac{s+\lambda_2}{\lambda_2 - \gamma_r} - \left(\frac{s}{s+\gamma_r}\right) \frac{(s+\gamma_r) (s+\lambda_2)}{s(\lambda_2 - \gamma_r)}\right] \\
& =  \left( x_1 -  \frac{k_r}{\gamma_r}  \right) \left(\frac{s+\gamma_p}{\gamma_p} \right) \left(\frac{\gamma_r}{s+\gamma_r} \right) 
\left[ \frac{s+\lambda_2}{\lambda_2 - \gamma_r} - \frac{s+\lambda_2}{\lambda_2 - \gamma_r} \right] \\
& = 0.
\end{align*}
These calculations show that adding an extraneous decay modes does not change weight of the coefficient corresponding to the correct decay mode, and the coefficient for the extraneous decay mode automatically gets set to 0.

\subsection{Identification of the decay modes via convex optimization} \label{decay_mode_identification}

For a general nonlinear SRN, the finite spectral expansion \eqref{spectral_expansion} does not hold exactly, and our aim is to construct an approximation that is accurate in the sense of the induced operator norm on $\mathcal{L}_2(\pi)$. Theorem \ref{thm_rem1} suggests a \emph{two-tier} approach for this purpose. Using condition \eqref{resol_cond1}, we first estimate the relevant decay modes, which are independent of both the initial state $x$ and the observable function $f$. Then, for any observable function $f$ and any initial state $x$, the action of the Koopman operator $\mathcal{K}_t f(x)$ can be approximated by \eqref{spectral_expansion}, with the coefficients determined by \eqref{coeff_alphas}.

To this end, let us now discuss how the decay modes can be estimated with convex optimization. Fixing a frequency $s\in \C_+$, an observable function $f\in \mathcal{F}$, and the number of decay modes $J$, the ``error" in condition \eqref{resol_cond1} is given by the function
\begin{align}
\label{decay_mode_Estimation_error_fn1}
E_{f,J}(x,s):=\left( \frac{ \prod_{j=1}^J \left( s( \mathcal{R}_s -\mathbf{I}) +\sigma_j \mathcal{R}_s \right) }{ \prod_{j=1}^J \sigma_j}\right) f(x) - \E_\pi(f).
\end{align}
Let $\beta_1,\dots, \beta_J$ be the coefficients of the polynomial
\begin{align}
\label{beta_poly}
1 + \sum_{j=1}^J \beta_j s^j = \frac{\prod_{j=1}^{J} (s + \sigma_j)}{ \prod_{j=1}^J \sigma_j}. 
\end{align}
Then Lemma \ref{supp_lemm:error_function_form} in the Appendix shows that we can express the error function $E_{f,J}(x,s)$ in terms of these coefficients as
\begin{align}
\label{decay_mode_Estimation_error_fn2}
 E_{f,J}(x,s) = \mathcal{R}^J_sf(x) -  \E_\pi(f) + \sum_{j=1}^J \beta_j s^j C^{ (s) }_{j,J} (f,x), 
\end{align}
where for each $j=1,\dots,J$
\begin{align}
\label{error_fn_coefficient_defn}
 C^{ (s) }_{j,J} (f,x) =  \sum_{\ell=0}^{j} (-1)^{\ell} {j\choose \ell}  \mathcal{R}^{J-\ell}_s f(x)  . 
\end{align}

Recall the norm $\left\| f \right\|^2_{\mathcal{L}_2(\pi)}$ defined by \eqref{defn_l2_norm}. Under this norm, the normalized error is given by
\begin{align}
\label{const_fn_single_f_s}
\mathcal{C}_{f,s,J}(\beta):= \frac{\| E_{f,J}(\cdot,s)\|_{\mathcal{L}_2(\pi)} }{ \|f\|_{\mathcal{L}_2(\pi)} }   =\frac{1 }{ \|f\|_{\mathcal{L}_2(\pi)} } \sqrt{ \sum_{x \in \mathcal{E}} \left(\mathcal{R}^J_s f(x) -  \E_\pi(f) + \sum_{j=1}^J \beta_j s^j C^{ (s) }_{j,J} (f,x)   \right)^2 \pi(x)}.
\end{align} 
Note that this function is convex in the coefficient vector $\beta = (\beta_1,\dots,\beta_J)$ of the polynomial \eqref{beta_poly} whose roots are the eigenvalues (i.e.\ negative of the decay modes). This function also depends on the stationary expectation $\E_\pi(f)$, which we shall assume is known, as it can be accurately estimated with CTMC simulations (see Section \ref{SKA_details}).

In our approach we shall consider a finite set of frequency values $\mathbb{S}$ and a finite set of observable functions $\mathcal{F}$, and hence our cumulative cost becomes
\begin{align}
\label{cumulative_cost_fn}
\mathcal{C}_J(\beta) = \max_{s \in \mathbb{S}} \max_{f \in \mathcal{F}} \mathcal{C}_{f,s,J}(\beta),
\end{align}
which is also a convex function of $\beta$, because convexity is preserved by the maximum operation. Hence, we can reliably infer $\beta$ by solving the following convex optimization problem:
\begin{align}
\label{defn_convex_optimization}
\min_{ \beta \in \R^J_+} \mathcal{C}_J(\beta). 
\end{align}
We constrain the coefficient vector $\beta$ to lie in the strictly positive orthant, as each component must be positive for the inferred nonzero eigenvalues to have strictly negative real parts. This follows from the well-known Routh–Hurwitz criterion \cite{hurwitz1895ueber}.

Let $C^*_J$ denote the optimal value of the optimization problem \eqref{defn_convex_optimization}, and let $\beta^*_J = (\beta^*_{J1}, \dots, \beta^*_{JJ})$ denote an optimizer at which this value is attained. In our examples, $C^*_J$ decreases monotonically as the number of decay modes $J$ increases, and this decay is typically exponential. Using $\beta^*_J$ as the coefficient vector, we define the polynomial
\begin{align}
\label{defn_poly_beta}
\mathcal{P}_{\beta^*_J}(s) = 1 + \sum_{j=1}^J \beta^*_{jJ} s^j,
\end{align}
whose negated roots yield approximate values of the $J$ decay modes, denoted by $\bar{\sigma}_1, \dots, \bar{\sigma}_J$. These decay modes will be assumed to have positive real parts.

\subsection{Estimation of the linear coefficients} \label{num_coeff_identification}

Once the $J$ decay modes $\bar{\sigma}_1,\dots,\bar{\sigma}_J$ have been identified, we can fix some frequency $\bar{s} \in \C_+$ and identify the linear coefficients in \eqref{spectral_expansion} by applying formula \eqref{coeff_alphas}, i.e.\
\begin{align}
\label{coeff_alphas_2}
\bar{\alpha}_j(f,x) = \left( \frac{\bar{s}+\bar{\sigma}_j}{\bar{\sigma}_j} \right) \sum_{k=1}^J \mathbf{L}_{jk} (\bar{s}) \left(  \mathcal{R}^{k-1}_{\bar{s}} -   \mathcal{R}^{k}_{\bar{s}} \right) f(x)  \qquad  \textnormal{for each} \qquad j=1,\dots,J,
\end{align}
where $\mathbf{L}_{jk} (s)$ is the $J \times J$ Lagrange matrix with interpolation points $\bar{s}/(\bar{s}+\bar{\sigma}_j)$ for $j=1,\dots, J$. Note that since \eqref{resol_cond1} does not hold exactly, these coefficients will depend on the choice of frequency $\bar{s}$. Let $\bar{\mathcal{K}}_t f(x)$ be the approximate Koopman operator defined by the right-hand side of \eqref{spectral_expansion} with $\sigma_j$ replaced by $\bar{\sigma}_j$, and $\alpha_j(f,x) $ replaced by $\bar{\alpha}_j(f,x)$, i.e.\
\begin{align}
\label{approx_koopman_operator1}
\bar{\mathcal{K}}_t f(x) = \E_\pi(f) + \sum_{j=1}^J \bar{\alpha}_j(f,x) e^{ -\bar{\sigma}_j t } \quad \textnormal{for all} \quad t \geq 0.
\end{align}
Corresponding to this operator we define the approximate resolvent $\bar{\mathcal{R}}_s f(x)$ as 
\begin{align}
\label{defn_resolvent_operator_approximate}
\bar{\mathcal{R}}_s f(x) = \int_{0}^{\infty} s e^{- s t } \bar{\mathcal{K}}_t f(x) \, dt =  \E_\pi(f) + \sum_{j=1}^J \bar{\alpha}_j(f,x) \left( \frac{s}{s + \bar{\sigma}_j}\right).
\end{align}
We shall quantify the error between the true Koopman operator $\mathcal{K}_t f(x)$ and its approximation $\bar{\mathcal{K}}_t f(x)$, by estimating the error between the true resolvent $\mathcal{R}_s f(x)$ and its approximation $\bar{\mathcal{R}}_s f(x) $.

Lemma \ref{supp_lem:linear_coeff} in the Appendix shows that, at $s = \bar{s}$, 
the first $J-1$ derivatives of $\mathcal{R}_s f(x)$ and $\bar{\mathcal{R}}_s f(x)$ coincide:
\begin{align}
\label{derivatives_match_condition}
\frac{\partial^m}{\partial s^m} \mathcal{R}_s f(x) \Big|_{s = \bar{s}} 
= \frac{\partial^m}{\partial s^m} \bar{\mathcal{R}}_s f(x) \Big|_{s = \bar{s}}, 
\quad \text{for each } m = 1, \dots, J-1.
\end{align}
Let $\mathcal{D}_{\bar{s}} \subset \mathbb{C}$ denote the open disk of radius $|\bar{s}|$ 
centered at $\bar{s}$. Since the maps 
$s \mapsto \mathcal{R}_s f(x)$ and $s \mapsto \bar{\mathcal{R}}_s f(x)$ are complex-analytic, with singularities outside $\mathbb{C}_+$, their Taylor series expansions around $\bar{s}$ converge for any 
$s \in \mathcal{D}_{\bar{s}} \cap \mathbb{C}_+$. Approximating these maps by their respective Taylor expansions yields
\begin{align*}
\mathcal{R}_s f(x) &=\mathcal{R}_{\bar{s}} f(x) 
+ \sum_{m=1}^{J-1} \frac{(s-\bar{s})^m}{m!} 
\left.\frac{\partial^m}{\partial s^m} \mathcal{R}_s f(x) \right|_{s = \bar{s}}
+ O\big(|s - \bar{s}|^{J}\big), \\
\bar{\mathcal{R}}_s f(x) &= \bar{\mathcal{R}}_{\bar{s}} f(x) 
+ \sum_{m=1}^{J-1} \frac{(s-\bar{s})^m}{m!} 
\left.\frac{\partial^m}{\partial s^m} \bar{\mathcal{R}}_s f(x) \right|_{s = \bar{s}}
+ O\big(|s - \bar{s}|^{J}\big),
\end{align*}
where $O\big(|s - \bar{s}|^{J}\big)$ terms denote the Taylor remainder. Subtracting these expressions, ignoring the remainder terms and canceling the $(J-1)$ derivative terms using 
\eqref{derivatives_match_condition}, we obtain
\begin{align}
\label{resolvent_error_condition}
\mathcal{R}_s f(x) - \bar{\mathcal{R}}_s f(x) 
\approx \mathcal{R}_{\bar{s}} f(x) - \bar{\mathcal{R}}_{\bar{s}} f(x) 
:= \mathcal{E}_{\bar{s}} f(x),
\end{align}
which shows that, for any $s \in \mathcal{D}_{\bar{s}} \cap \mathbb{C}_+$, the error between the resolvents 
is well-approximated by the error at $s = \bar{s}$, denoted by $\mathcal{E}_{\bar{s}} f(x)$. 
The operator $\mathcal{E}_{\bar{s}}$ can also be viewed as a linear operator, 
whose action can be computed directly from the approximate linear coefficients estimated via 
\eqref{coeff_alphas_2}, as shown in Lemma \ref{supp_lem:linear_coeff} in the Appendix:
\begin{align}
\label{defn_e_f_x}
\mathcal{E}_{\bar{s}} f(x) = f(x) - \mathbb{E}_\pi(f) - \sum_{j=1}^J \bar{\alpha}_j(f,x).
\end{align}
In fact, this lemma also shows that if $C^*_J$ is the optimal value of the optimization problem 
\eqref{defn_convex_optimization}, and if $\bar{s} \in \mathbb{S}$ as well as $f \in \mathcal{F}$, 
then we must have
\begin{align*}
\frac{\| \mathcal{E}_{\bar{s}} f(\cdot)\|_{\mathcal{L}_2(\pi)} }{ \|f\|_{\mathcal{L}_2(\pi)} } 
\leq C^*_J.
\end{align*}
Therefore, if the number of decay modes $J$ is large enough to ensure that $C^*_J < \epsilon$, 
the relative error in our approximation, as measured by the norm 
$\left\| \cdot \right\|_{\mathcal{L}_2(\pi)}^2$, should be below $\epsilon$. 
Relation \eqref{resolvent_error_condition} further shows that, even for an individual state $x$, 
$\mathcal{E}_{\bar{s}} f(x)$ provides a tight estimate of the error for $s$ close to $\bar{s}$. Observe that, since $\mathcal{K}_t f(x) = f(x)$ at time $t = 0$, 
equation~\eqref{defn_e_f_x} represents the error in the approximate Koopman operator at $t = 0$. This error can be computed directly from the estimated linear coefficients via 
\eqref{defn_e_f_x}, without requiring the true value of either the Koopman operator 
$\mathcal{K}_t f(x)$ or its resolvent $\mathcal{R}_s f(x)$. 
The fact that the error can be evaluated at the initial time $t = 0$ is notable, 
as it contrasts with other numerical methods, such as FSP \cite{munsky2006finite}, where the error must be computed 
from the solution at the terminal time.

Let $\mathbf{R}_{\bar{s}}(f,x)$ and $\boldsymbol{\alpha}(f,x) $ be $J \times 1$ column vectors whose 
$j$-th components are $\mathcal{R}_{\bar{s}}^{j-1} f(x) - \mathcal{R}_{\bar{s}}^{j} f(x)$ and 
$\bar{\alpha}_j(f,x)$, respectively. 
Then, to compute the linear coefficients according to \eqref{coeff_alphas_2}, 
we can solve the linear system
\begin{align}
\label{linear_sys_num}
\mathbf{A}(\bar{s}) \, \boldsymbol{\alpha}(f,x)  = \mathbf{R}_{\bar{s}}(f,x),
\end{align}
where $\mathbf{A}(\bar{s})$ is the $J \times J$ matrix whose entry in row $j$ and column $k$ is given by
\begin{align*}
\mathbf{A}_{ jk}(\bar{s}) = \left( \frac{\bar{s}}{\bar{s}+\sigma_k} \right)^j 
\left( \frac{\bar{\sigma}_k}{\bar{s}+\bar{\sigma}_k} \right).
\end{align*}
To obtain a more robust approximation of the Koopman operator, it is preferable to select the linear coefficients so that the approximate resolvent operator is close to the true resolvent operator for multiple frequencies, rather than a single value \( \bar{s} \).  
Let \( \mathbb{S} = \{ s_1, \dots, s_L \} \) be a finite collection of \( L \) frequency values.  
To each \( s_\ell \) we assign an \emph{order} \( J_\ell \), construct the \( J_\ell \times J \) matrix \( \mathbf{A}^{(\ell)}(s_\ell) \), and estimate the \( J_\ell \times 1 \) vector \( \mathbf{R}^{(\ell)}_{s_\ell}(f,x) \).  
By vertically stacking these matrices and vectors, we obtain the \( J_{\textnormal{tot}} \times J \) matrix $\mathbf{A}_{\mathbb{S}}$ and the \( J_{\textnormal{tot}} \times 1 \) vector $\mathbf{R}_{\mathbb{S}}(f,x)$, given by
\begin{align}
\label{over_linear_system_single_f}
\mathbf{A}_{\mathbb{S}} =
\begin{bmatrix}
\mathbf{A}^{(1)}(s_1) \\
\vdots \\
\mathbf{A}^{(L)}(s_L)
\end{bmatrix}
\quad \textnormal{and} \quad 
\mathbf{R}_{\mathbb{S}}(f,x) =
\begin{bmatrix}
\mathbf{R}^{(1)}_{s_1}(f,x) \\
\vdots \\
\mathbf{R}^{(L)}_{s_L}(f,x)
\end{bmatrix},
\end{align}
where \( J_{\textnormal{tot}} = \sum_{\ell=1}^L J_\ell \). The vector of linear coefficients \( \boldsymbol{\alpha}(f,x) \) is then estimated in the least-squares sense by solving the aggregate linear system 
\begin{align}
\label{linear_sys_num_aggr}
\mathbf{A}_{\mathbb{S}} \, \boldsymbol{\alpha}(f,x) = \mathbf{R}_{\mathbb{S}}(f,x).
\end{align}
The approximate Koopman operator $\bar{\mathcal{K}}_t f(x)$ can then be constructed according to \eqref{approx_koopman_operator1}, thereby accomplishing \textbf{Task 1} defined by \eqref{estimation_task1}. In Section \ref{SKA_details}, we discuss how the components of \( \mathbf{R}_{\mathbb{S}}(f,x) \) can be estimated.

\subsection{Estimation of parameter sensitivities}
\label{sec:sens_estimator}
We now turn out attention to \textbf{Task 2} (see \eqref{estimation_task_sensitivity}), which concerns the estimation of parameter sensitivities when the propensity functions of the SRN depend on a finite set of parameters $\Theta$. As stated in Section \ref{prelim:expr_param_sens}, the main difficulty in using the integral representation~\eqref{sensitivity_koopman_repr} for parameter sensitivity estimation lies in computing the differences in the action of the Koopman operator \eqref{koopman_differences_path} across all states visited by the CTMC trajectory $\left(X^{(\Theta)}_x(t)\right)_{t \geq 0}$.  Applying our approximation~\eqref{approx_koopman_operator1}, these differences can be written as  
\begin{align*}
\Delta_k \mathcal{K}^{(\Theta)}_{t-s} f \left( X^{(\Theta)}_x(s) \right) &\approx \Delta_k \bar{\mathcal{K}}^{(\Theta)}_{t-s} f \left( X^{(\Theta)}_x(s) \right) = \sum_{j=1}^J \Delta_k \bar{\alpha}^{(\Theta)}_j(f, X^{(\Theta)}_x(s) ) \, e^{ - \bar{\sigma}_j (t-s)}.
\end{align*}
Substituting this into~\eqref{sensitivity_koopman_repr} yields  
\begin{align}
\label{spectral_sens_formula_1}
\mathcal{S}^{(\Theta)}_{t,\theta} f(x) 
& \approx \sum_{k=1}^K  \sum_{j=1}^J  \int_0^t \E\left[ \frac{\partial \lambda_k}{\partial \theta}(X^{(\Theta)}_x(s), \Theta)  \Delta_k \bar{\alpha}^{(\Theta)}_j(f, X^{(\Theta)}_x(s) ) \right] e^{ -\bar{\sigma}_j (t-s)} \, ds \notag  \\
& =  \sum_{j=1}^J  \int_0^t \left[ \bar{\mathcal{K}}^{(\Theta)}_s g_{\theta,j, f} (x) \right]   e^{ -\bar{\sigma}_j (t-s)} \, ds,
\end{align}
where  
\begin{align}
\label{defn_g_sensitivity}
g_{\theta,j, f} (y) = \sum_{k=1}^K  \frac{\partial \lambda_k}{\partial \theta}(y, \Theta)  \Delta_k \bar{\alpha}^{(\Theta)}_j(f, y ).
\end{align}

Let $\mathcal{F} = \{f_1,\dots,f_F\}$ be the set of observable functions for which the approximate Koopman operator of the form~\eqref{approx_koopman_operator1} has been estimated.  
Denote by $\mathbf{1}$ the constant function equal to $1$ for every $x$.  
Since $\bar{\mathcal{K}}^{(\Theta)}_t \mathbf{1} = \mathbf{1}$, it follows that for each $n = 1,\dots,F$ we have  
\begin{align*}
\bar{\mathcal{K}}^{(\Theta)}_t \bar{f}_n(y) \approx \sum_{\ell=1}^J  \bar{\alpha}^{(\Theta)}_\ell(f_n,y) \, e^{ -\bar{\sigma}_\ell t},
\end{align*}
where $\bar{f}_n(y) = f_n(y) -\E_\pi (f_n) \mathbf{1}$ is the centered version of $f_n$ with respect to its stationary expectation $\E_\pi(f_n)$.  
Define $\mathcal{L}(\bar{\mathcal{F}})$ as the linear span of the set $\bar{\mathcal{F}} = \{\mathbf{1},\bar{f}_1,\dots, \bar{f}_F\}$:  
\begin{align}
\label{defn_linear_span_bar_F}
\mathcal{L}(\bar{\mathcal{F}}) = \left\{ c_0 \mathbf{1} + \sum_{n=1}^F c_n \bar{f}_n(y) : c_0,\dots,c_F \in \mathbb{R} \right\}.
\end{align}
By the linearity of the Koopman operator, its approximate action on $\mathcal{L}(\bar{\mathcal{F}})$ is given by  
\begin{align}
\label{approx_koopman_linear_span}
\bar{\mathcal{K}}^{(\Theta)}_t \left(c_0 \mathbf{1} + \sum_{n=1}^F c_n \bar{f}_n(y) \right) \approx c_0 \mathbf{1}  + \sum_{n=1}^F c_n \bar{\mathcal{K}}^{(\Theta)}_t \bar{f}_n(y) = c_0 \mathbf{1} + \sum_{n=1}^F \sum_{\ell=1}^J  c_n  \bar{\alpha}^{(\Theta)}_\ell(f_n,y) \, e^{ -\bar{\sigma}_\ell t}.
\end{align}

We project each $g_{\theta,j, f}$ onto $\mathcal{L}(\bar{\mathcal{F}})$ in the Hilbert space $\mathcal{L}_2(\pi)$, obtaining coefficients $c^{(n)}_{\theta,j, f}$, $n=0,1,\dots,F$, such that  
\begin{align}
\label{projection_of_g_sens}
g_{\theta,j, f} (y)  \approx c^{(0)}_{\theta,j, f} \mathbf{1} + \sum_{n =1}^F  c^{(n)}_{\theta,j, f} \bar{f}_n(y).
\end{align}
Consequently, the Koopman image of $g_{\theta,j, f}$ can be approximated by  
\begin{align*}
\bar{\mathcal{K}}^{(\Theta)}_t g_{\theta,j, f} (y) \approx 
 c^{(0)}_{\theta,j, f} \mathbf{1} + \sum_{n =1}^F \sum_{\ell=1}^J  c^{(n)}_{\theta,j, f}    \bar{\alpha}^{(\Theta)}_\ell(f_n,y) \, e^{ -\bar{\sigma}_\ell t}.
\end{align*}
Substituting this into~\eqref{spectral_sens_formula_1} gives  
\begin{align*}
\mathcal{S}^{(\Theta)}_{t,\theta} f(x)  & \approx  \sum_{j=1}^J  \int_0^t \left[  c^{(0)}_{\theta,j, f} + \sum_{n =1}^F  \sum_{\ell=1}^J   c^{(n)}_{\theta,j, f}  \bar{\alpha}^{(\Theta)}_\ell(f_n,x) \, e^{ -\bar{\sigma}_\ell s} \right] e^{ -\bar{\sigma}_j (t-s)} \, ds.
\end{align*}
The time integral can now be computed \emph{explicitly}, leading---after simplification---to the approximate parameter sensitivity  
\begin{align}
\label{main_sensitivity_expression}
\mathcal{S}^{(\Theta)}_{t,\theta} f(x)   & \approx  \sum_{j=1}^J  \left[ \frac{ \bar{c}^{(0)}_{\theta, j, f} }{ \hat{\lambda}_j } \left( 1 -  e^{ -  \bar{\sigma}_j t}\right) + \sum_{n =1}^F  \sum_{\ell=1, \ell \neq j}^J   \bar{c}^{(n)}_{\theta,j, f}  \bar{\alpha}^{(\Theta)}_\ell(f_n,x) \left(  \frac{e^{ -\bar{\sigma}_\ell t } - e^{ -\bar{\sigma}_j t }  }{\bar{\sigma}_j - \bar{\sigma}_\ell}\right)  \right. \notag \\
&\quad \left. 
+ \sum_{n =1}^F   \bar{c}^{(n)}_{\theta,j, f}  \bar{\alpha}^{(\Theta)}_j(f_n,x) \, t \, e^{ -\bar{\sigma}_j t } \right].
\end{align}

As with the decay modes, the projection coefficients $\{c^{(n)}_{\theta,j, f} : n=0,1,\dots,F\}$ are independent of the initial state and can therefore be precomputed once for each $g_{\theta,j, f}$.  
For a new initial state $x$, only the linear coefficients $\bar{\alpha}^{(\Theta)}_j(f_n,x)$ need to be estimated in order to recover the full sensitivity dynamics $t \mapsto \mathcal{S}^{(\Theta)}_{t,\theta} f(x)$, thereby completing \textbf{Task 2}.

\subsection{Estimation of CSD}
\label{sec:csd_estimator}

We now turn to \textbf{Task~3} (see~\eqref{estimation_task_csd}), which concerns the estimation of cross-spectral densities (CSDs).  
Let $x \in \mathcal{E}$ be the initial state of the CTMC describing the SRN dynamics, and let $f_1, f_2 \in \mathcal{F}$ be two observable functions. As before, we define their centered versions by $\bar{f}_j(y) = f_j(y) -\E_\pi (f_j) \mathbf{1}$ for $j=1,2$.  
Our objective is to estimate the CSD for the two real-valued stochastic trajectories $\left(\bar{f}_1(X_x(t))\right)_{t \geq 0}$ and $\left(\bar{f}_2(X_x(t))\right)_{t \geq 0}$ over the time interval $[0,T]$, as defined in~\eqref{csd_definition}. We work with centered trajectories for CSD estimation because then the CSD remains well-defined as $T \to \infty$ and it captures the oscillatory behavior around the stationary mean.

Applying formula~\eqref{csd_in_terms_Koopman_operator} gives  
\begin{align}
\label{csd_approx_koopman_action_0}
\textnormal{CSD}_{\bar{f}_1, \bar{f}_2} (\omega, x, T) 
&= \frac{1}{T} \int_0^T \mathbb{E} \left[ \bar{f}_2(X_x(u)) \int_u^T 
\mathcal{K}_{t-u} \bar{f}_1(X_x(u)) \, e^{ - i \omega (t-u)} \, dt \right] du \\
&\quad + \frac{1}{T} \int_0^T \mathbb{E} \left[ \bar{f}_1(X_x(u)) \int_u^T \mathcal{K}_{t-u}
\bar{f}_2(X_x(u)) \, e^{  i \omega (t-u)} \, dt \right] du. \notag
\end{align}
The approximate action of the Koopman operator is  
\begin{align}
\label{csd_approx_koopman_action}
\bar{\mathcal{K}}_t \bar{f}_n(y) \approx \sum_{j=1}^J  \bar{\alpha}_j(f_n,y)  e^{ -\bar{\sigma}_j t},
\end{align}
which upon substitution in \eqref{csd_approx_koopman_action_0} yields
\begin{align*}
\textnormal{CSD}_{\bar{f}_1, \bar{f}_2} (\omega, x, T) 
&\approx  \frac{1}{T} \sum_{j=1}^J \int_0^T \mathbb{E} \left[ \bar{f}_2(X_x(u))
\bar{\alpha}_j(f_1,X_x(u)) \right]
\left( \int_u^T e^{ -\bar{\sigma}_j (t-u)} e^{ - i \omega (t-u)}  dt\right)  du \\
&\quad + \frac{1}{T} \sum_{j=1}^J  \int_0^T \mathbb{E} \left[ \bar{f}_1(X_x(u)) \bar{\alpha}_j(f_2,X_x(u)) \right] \left( \int_u^T e^{ -\bar{\sigma}_j (t-u)}  e^{  i \omega (t-u)} dt \right) du.
\end{align*}
Let $g_{f_1,f_2,j}(x) = \bar{f}_1(x) \bar{\alpha}_j(f_2, x)$ and $g_{f_2,f_1,j}(x) = \bar{f}_2(x) \bar{\alpha}_j(f_1, x)$, and define  
\begin{align*}
h_j(t,\omega) = \int_0^t e^{ -(\bar{\sigma}_j - i \omega )u  }   dt = \frac{1 -  e^{-(\bar{\sigma}_j - i \omega) t }}{\bar{\sigma}_j - i \omega}.
\end{align*}
Then the approximate CSD becomes  
\begin{align}
\label{approx_csd_koop_1}
\textnormal{CSD}_{\bar{f}_1, \bar{f}_2} (\omega, x, T) 
&\approx  \frac{1}{T} \sum_{j=1}^J \int_0^T \left[ \bar{\mathcal{K}}_t g_{f_2,f_1,j}(x)h_j(T-t, -\omega) + \bar{\mathcal{K}}_t g_{f_1,f_2,j}(x) h_j(T-t, \omega)\right]  dt.
\end{align}
As in Section~\ref{sec:sens_estimator}, we project $g_{f_1,f_2,j}$ and $g_{f_2,f_1,j}$ onto the linear span $\mathcal{L}(\bar{\mathcal{F}})$ to obtain coefficients $c^{(n)}_{f_1,f_2,j}$ and $c^{(n)}_{f_2,f_1,j}$ such that  
\begin{align}
\label{csd_g_func_projection}
g_{f_1,f_2,j}(y) \approx c^{(0)}_{f_1,f_2,j} \mathbf{1} + \sum_{n=1}^F c^{(n)}_{f_1,f_2,j} \bar{f}_n(y)\quad \textnormal{and} \quad
g_{f_2,f_1,j}(y) \approx c^{(0)}_{f_2,f_1,j} \mathbf{1} + \sum_{n=1}^F c^{(n)}_{f_2,f_1,j} \bar{f}_n(y).
\end{align}
Using \eqref{csd_approx_koopman_action}, their Koopman images are approximated as  
\begin{align*}
\bar{\mathcal{K}} _t g_{f_1,f_2,j}(y) &\approx  c^{(0)}_{f_1,f_2,j} \mathbf{1} + \sum_{n=1}^F \sum_{\ell=1}^J c^{(n)}_{f_1,f_2,j}\bar{\alpha}_\ell(f_n,y)  e^{ -\bar{\sigma}_\ell t}, \\
\bar{\mathcal{K}} _t g_{f_2,f_1,j}(y) &\approx  c^{(0)}_{f_2,f_1,j} \mathbf{1} + \sum_{n=1}^F \sum_{\ell=1}^J c^{(n)}_{f_2,f_1,j}\bar{\alpha}_\ell(f_n,y)  e^{ -\bar{\sigma}_\ell t}.
\end{align*}
Substituting these expressions into~\eqref{approx_csd_koop_1} and defining  
\begin{align*}
\kappa_{\bar{\sigma}_\ell, \bar{\sigma}_j} (\omega,T) = \int_0^T e^{ - \bar{\sigma}_\ell t} h_j (T-t, \omega)dt = \frac{1 -e^{ - \bar{\sigma}_\ell T} }{ \bar{\sigma}_\ell (\bar{\sigma}_j - i \omega)} - \frac{e^{ - \bar{\sigma}_\ell T} - e^{ - (\bar{\sigma}_j - i \omega) T} }{(\bar{\sigma}_j - i \omega) ( \bar{\sigma}_j -\bar{\sigma}_\ell - i \omega)},
\end{align*}
we obtain the simplified CSD formula  
\begin{align}
\label{approx_csd_koop_final_expression}
\textnormal{CSD}_{\bar{f}_1, \bar{f}_2} (\omega, x, T) 
&\approx  \frac{1}{T} \sum_{j=1}^J  c^{(0)}_{f_1,f_2,j} \kappa_{0, \bar{\sigma}_j} (\omega,T) +\frac{1}{T} \sum_{j=1}^J \sum_{n=1}^F \sum_{\ell=1}^J c^{(n)}_{f_1,f_2,j}\bar{\alpha}_\ell(f_n,y) \kappa_{\bar{\sigma}_\ell, \bar{\sigma}_j} (\omega,T) \\
&\quad + \frac{1}{T} \sum_{j=1}^J  c^{(0)}_{f_2,f_1,j} \kappa_{0, \bar{\sigma}_j} (\omega,T) +\frac{1}{T} \sum_{j=1}^J \sum_{n=1}^F \sum_{\ell=1}^J c^{(n)}_{f_2,f_1,j}\bar{\alpha}_\ell(f_n,y) \kappa_{\bar{\sigma}_\ell, \bar{\sigma}_j} (-\omega,T). \notag
\end{align}

Thus, by employing our approximate Koopman operator, we eliminate the double time integral in~\eqref{csd_approx_koopman_action_0} and obtain a closed-form analytical expression for the CSD as a function of frequency $\omega$ and terminal time $T$, thereby completing \textbf{Task~3}.  
As in Section~\ref{sec:sens_estimator}, the projection coefficients $\{c^{(n)}_{f_1,f_2,j} : n=0,1,\dots,F\}$ are independent of the initial state and can be precomputed once for each $g_{f_1,f_2,j}$.  
For any new initial state $x$, the CSD expression~\eqref{approx_csd_koop_final_expression} can be evaluated directly after estimating the linear coefficients $\bar{\alpha}_j(f_n,x)$.

\section{SKA: Implementation Details}
\label{SKA_details}

In this section, we present our methodology for constructing an approximate Koopman operator for stochastic reaction networks (SRNs) and for employing it for the estimation of parameter sensitivities and cross-spectral densities. We refer to this approach as SKA, an acronym for \texttt{S}tochastic \texttt{K}oopman \texttt{A}pproximation. A complete computational implementation of the method is available in our public GitHub repository: \url{https://github.com/ankitgupta83/SpectralKoopman.git}. This implementation is primarily written in \texttt{Python}, with selected \texttt{C} routines incorporated to accelerate certain estimations.

Suppose we are given a stochastic reaction network (SRN) with propensity functions $\lambda_k(x,\Theta)$ and stoichiometric vectors, and let $\mathcal{F}$ denote a finite set of observable functions on which the action of the Koopman operator is to be approximated. As outlined in Section~\ref{sec:intro}, SKA performs this task (see \textbf{Task~1} in \eqref{estimation_task1}) through a two-tiered procedure:  
\begin{enumerate}
    \item \textbf{Tier 1: Estimation of state-independent quantities.}  
    The decay modes and stationary expectations (i.e.\ $\E_\pi(f)$ for each $f\in\mathcal{F}$) are estimated. These quantities are independent of the initial state and therefore need to be computed only once for a given SRN. Reliable estimation requires multiple batches of stochastic trajectories from diverse initial states, which our implementation accelerates through parallelized execution on a graphical processing unit (GPU), whenever available.  
    \item \textbf{Tier 2: Construction of the approximate Koopman operator.}  
    For each initial state of interest, the approximate Koopman operator is constructed by estimating the linear coefficients in \eqref{approx_koopman_operator1}. This step is considerably less demanding, requiring only a small number of stochastic simulations that can be efficiently performed on a standard CPU.  
\end{enumerate}

We now discuss both these tiers in greater detail.

\subsection{Estimation of state-independent quantities}
\label{sec:est_state_independent}
To estimate the decay modes, we need to solve the optimization problem \eqref{cumulative_cost_fn}, which involves the unknown stationary distribution $\pi$. Another issue is that the state space $\mathcal{E}$ of the SRN is typically infinite, rendering the infinite sum in the cost function (see \eqref{const_fn_single_f_s}) computationally intractable. To overcome these difficulties, we construct a crude approximation of the stationary distribution, denoted by $\hat{\pi}$. Specifically, we simulate a small number of trajectories over a sufficiently long time horizon $[0,T]$ with $T \gg 1$, and then apply a basic clustering procedure to the terminal states. This yields $n_c$ distinct representative states $x_1,\dots,x_{n_c}$ together with associated probabilities $\hat{\pi}_1,\dots,\hat{\pi}_{n_c}$, such that
\begin{align}
\label{defn_approx_pi}
\hat{\pi}(x) = \sum_{n=1}^{n_c} \hat{\pi}_n \, \ind_{\{x_n\}}(x),
\end{align}
where $\ind_{\{x_n\}}(x)$ denotes the indicator function.\footnote{The indicator function is defined as
\[
\ind_{\{x\}}(y) = \begin{cases}
1 & \textnormal{if } y = x, \\
0 & \textnormal{otherwise}.
\end{cases}
\]}
With this approximation in place, any sum of the form $\sum_{x \in \mathcal{E}} A(x)\pi(x)$ can be replaced by the finite sum $\sum_{n=1}^{n_c} A(x_n)\hat{\pi}_n$, and the cost function \eqref{const_fn_single_f_s} can be approximated accordingly. We note here that a crude approximation of the stationary distribution is sufficient for our purpose, since it is never used directly for estimation. Instead, $\hat{\pi}$ serves only to approximate the norm in $\mathcal{L}_2(\pi)$ with respect to which the normalized error in \eqref{const_fn_single_f_s} is minimized.

We now fix a finite set of frequency values $\mathbb{S}$ and specify $J_{\textnormal{max}}$ as the maximum number of decay modes to be considered. For each $j \in \{1,\dots,J_{\textnormal{max}}\}$, $s \in \mathbb{S}$, $f \in \mathcal{F}$, and $x \in \{x_1,\dots,x_{n_c}\}$, we estimate the values of the iterated resolvents $\mathcal{R}^j_{s} f(x)$ via a Monte Carlo estimator based on the expression \eqref{resolvent_iterates_formula} (see the Appendix, Section \ref{supp_sec:monte_carlo_estimators}). For this purpose, with each initial state $x_n$ we need to simulate multiple CTMC trajectories over a sufficiently long time interval $[0,T]$, which can be computationally very demanding for larger SRNs. However this procedure is inherently parallelizable and can be efficiently accelerated with a GPU, as mentioned above. The CTMC trajectories generated for estimating the iterated resolvent can also be used for estimating the stationary expectations $\E_\pi(f)$ for each $f\in \mathcal{F}$ (see the Appendix, Section \ref{supp_sec:monte_carlo_estimators}).

Once the iterated resolvents and stationary expectations have been computed, the cost function $\mathcal{C}_J(\beta)$ (see \eqref{cumulative_cost_fn}) can be evaluated, and the optimization problem \eqref{defn_convex_optimization} is solved to obtain both the optimal value $C_J^*$ and the corresponding optimizer $\beta^*_J = (\beta^*_{J1}, \dots, \beta^*_{JJ})$. The approximate decay modes $\bar{\sigma}_1,\dots,\bar{\sigma}_J$ are then identified as the negated roots of the polynomial \eqref{defn_poly_beta}. Beginning with $J=1$, we increment $J$ iteratively, repeating this procedure until either $C_J^*$ falls below the prescribed tolerance $\epsilon_{\textrm{tol}}$, or 
until $J$ attains the upper bound $J_{\textnormal{max}}$. In the latter case, we select the value of $J$ that minimizes $C_J^*$. Furthermore, we impose the requirement that, for the selected value of $J$, all associated decay modes possess strictly positive real parts. In this way, both the number of decay modes and their numerical values are determined in a well-posed manner.

\subsection{Construction of the approximate Koopman operator}
\label{subsec:koopman_contruction}
Suppose we have identified $J$ decay modes $\bar{\sigma}_1,\dots, \bar{\sigma}_L$ for our SRN, along with the stationary expectations $\E_\pi(f)$ for each $f\in \mathcal{F}$. Then, for any given initial state $x \in \mathcal{E}$, we can construct the approximate Koopman operator~\eqref{approx_koopman_operator1} by estimating the linear coefficients $\bar{\alpha}_j(f,x)$ for each $j=1,\dots,J$ and $f \in \mathcal{F}$, collectively. We now elaborate on this procedure.

As discussed in Section \ref{num_coeff_identification}, we assign an order $J_\ell$ to each frequency $s_\ell$ in the finite set $\mathbb{S} = \{s_1,\dots,s_L\}$, and define the $J_{\textnormal{tot}} \times J$ matrix $\mathbf{A}_{\mathbb{S}}$ and the $J_{\textnormal{tot}} \times 1$ vector $\mathbf{R}_{\mathbb{S}}(f,x)$ by \eqref{over_linear_system_single_f}, where $J_{\textnormal{tot}} = \sum_{\ell=1}^L J_\ell$. Rather than solving the linear system \eqref{over_linear_system_single_f} separately for each $f \in \mathcal{F} = \{f_1,\dots,f_F\}$, to obtain the $J \times 1$ coefficient vector $\boldsymbol{\alpha}(f,x) = (\bar{\alpha}_1(f,x),\dots,\bar{\alpha}_J(f,x))$, we combine these systems into the single linear system
\begin{align}
\label{coeff_linear_system_all_f}
\mathbf{A}_{\mathbb{S}} \boldsymbol{\alpha}(\mathcal{F}, x) = \mathbf{R}_{\mathbb{S}}(x),
\end{align}
where $\boldsymbol{\alpha}(\mathcal{F}, x) = [\boldsymbol{\alpha}(f_1,x),\ \dots,\ \boldsymbol{\alpha}(f_F,x)]$ is the $J \times F$ matrix of unknown coefficients, and $\mathbf{R}_{\mathbb{S}}(x)$ is the $J_{\textnormal{tot}} \times F$ matrix obtained by horizontally stacking the vectors $\mathbf{R}_{\mathbb{S}}(f,x)$ for each $f \in \mathcal{F}$, i.e.\
\[
\mathbf{R}_{\mathbb{S}}(x) = [\mathbf{R}_{\mathbb{S}}(f_1,x),\ \dots,\ \mathbf{R}_{\mathbb{S}}(f_F,x)].
\]

To construct the matrix $\mathbf{R}_{\mathbb{S}}(x)$, we estimate differences of iterated resolvents of the form $\mathcal{R}_{s}^{m-1} f(x) - \mathcal{R}_{s}^{m} f(x)$ using Monte Carlo estimators we describe in Section \ref{supp_sec:monte_carlo_estimators} of the Appendix. For robustness, we ensure that the linear system \eqref{coeff_linear_system_all_f} is overdetermined (i.e.\ $J_{\textnormal{tot}} \geq J$) and solve it using the method of least squares. This yields not only an estimate of the coefficient matrix $\boldsymbol{\alpha}(\mathcal{F}, x)$ but also its $(JF) \times (JF)$ covariance matrix. The covariance matrix, in turn, enables the computation of standard deviations around our estimates of $\bar{\mathcal{K}}_t f(x)$ for each $t \geq 0$, thereby quantifying the associated uncertainty. Moreover, since both the sensitivity estimate (see \eqref{main_sensitivity_expression}) and the CSD estimate (see \eqref{approx_csd_koop_final_expression}) depend linearly on these coefficients, their standard deviations can likewise be derived from this covariance matrix.

\subsection{Estimation of parameter sensitivities and spectral densities}
\label{subsec:estimation_sens_spectral}
As discussed in Section~\ref{sec:sens_estimator}, once we have an approximate Koopman operator for each $f \in \mathcal{F} = \{f_1,\dots,f_F\}$, we also obtain one for the entire linear span $\mathcal{L}(\bar{\mathcal{F}})$ defined in \eqref{defn_linear_span_bar_F}. We then project each function $g_{\theta,j,f}$, defined in \eqref{defn_g_sensitivity}, onto the linear space $\mathcal{L}(\bar{\mathcal{F}})$ using weighted least squares, where the weights are determined by the approximate stationary distribution $\hat{\pi}$ (see \eqref{defn_approx_pi}). This yields the coefficients $\{c^{(n)}_{\theta,j,f} : n=0,1,\dots,F\}$ for each $g_{\theta,j,f}$ (see \eqref{projection_of_g_sens}), which are subsequently used for sensitivity estimation via \eqref{main_sensitivity_expression}. Similarly, for the estimation of spectral densities, we project the functions $g_{f_1,f_2,j}$ and $g_{f_2,f_1,j}$ (see Section~\ref{sec:csd_estimator}) onto $\mathcal{L}(\bar{\mathcal{F}})$ to obtain coefficients $\{c^{(n)}_{f_1,f_2,j} : n=0,1,\dots,F\}$ and $\{c^{(n)}_{f_2,f_1,j} : n=0,1,\dots,F\}$ such that \eqref{csd_g_func_projection} holds. These coefficients are then employed to estimate CSD via \eqref{approx_csd_koop_final_expression}. 

In both cases, the projection coefficients are independent of the initial state and need to be estimated only once. Thereafter, for any given initial state $x \in \mathcal{E}$, it suffices to estimate the coefficient matrix $\boldsymbol{\alpha}(\mathcal{F}, x)$ (see Section~\ref{subsec:koopman_contruction}), and then apply either \eqref{main_sensitivity_expression} or \eqref{approx_csd_koop_final_expression} to obtain the parameter sensitivities or the CSD corresponding to that state.

Observe that in order to estimate the projection coefficients, we must evaluate the projected functions at all the states $x$ lying in the support of the approximate stationary distribution $\hat{\pi}$. For CSD estimation, the projected functions take the form $g_{f_1,f_2,j}(x) = (f_1(x) - \E_{\pi}(f_1))\,\bar{\alpha}_j(f_2,x)$ and so they can be computed directly from all the linear coefficients $\boldsymbol{\alpha}(\mathcal{F}, x)$, estimated using the procedure described in Section \ref{subsec:koopman_contruction}. In contrast, sensitivity estimation involves projected functions $g_{\theta,j,f}$ (see~\eqref{defn_g_sensitivity}) that depend on difference of linear coefficients $\Delta_k \alpha_j(f,x) =\alpha_j(f, x + \zeta_k) - \alpha_j(f,x)$, where $\zeta_k$ is the stoichiometric vector corresponding to reaction $k$. Rather than estimating $\alpha_j(f, x+\zeta_k)$ and $\alpha_j(f,x)$ separately and subtracting, it is advantageous to estimate the difference $\Delta_k \alpha_j(f,x)$ directly, exploiting the fact that the matrix $\mathbf{A}_{\mathbb{S}}$ in the linear system \eqref{coeff_linear_system_all_f} is independent of the state $x$. Consequently, the same linear system remains valid when $\mathbf{R}_{\mathbb{S}}(x)$ and $\boldsymbol{\alpha}(\mathcal{F}, x)$ are replaced by their respective differences $\Delta_k \mathbf{R}_{\mathbb{S}}(x)$ and $\Delta_k \boldsymbol{\alpha}(\mathcal{F}, x)$. Solving this modified system provides estimates of all coefficient differences $\Delta_k \boldsymbol{\alpha}(\mathcal{F}, x)$ simultaneously. To estimate $\Delta_k \mathbf{R}_{\mathbb{S}}(x)$ accurately, we employ a coupled Monte Carlo estimator (see Section \ref{supp_sec:monte_carlo_estimators} of the Appendix), based on the CTMC coupling introduced in \cite{anderson2012efficient} and the random time–change representation \cite{ethier2009markov}.

\section{Examples}
\label{sec:examples}

In this section we illustrate the performance of our SKA method on several examples of SRNs modelling biological processes. In all cases, the observable class $\mathcal{F}$ consists of all first- and second-order (non-centred) moments of the $D$ species $\mathbf{X}_1,\dots,\mathbf{X}_D$
\begin{align}
\label{first_two_moments}
\mathcal{F} \;=\; \{x_j: j=1,\dots,D\} \,\cup\, \{x_j x_k : j,k=1,\dots,D,\; k \geq j\},
\end{align}
yielding $D(D+3)/2$ distinct functions. The set of frequency values is taken to be $\mathbb{S}=\{0.25,0.5,0.75,1.0\}$, and the maximum number of decay modes is fixed at $J_{\textnormal{max}}=8$. All required quantities are estimated from CTMC trajectories of the SRN simulated over the time horizon $[0,T]$ with $T=100$. To approximate the stationary distribution $\hat{\pi}$ (see \eqref{defn_approx_pi}), we select $n_c=50$ representative states obtained by clustering the terminal states of $1000$ SSA-generated trajectories on $[0,T]$. With each representative state $x$ as the initial condition, approximately $20,000$ CTMC trajectories are simulated to estimate the iterated resolvents $\mathcal{R}^j_s f(x)$ for $j=1,\dots,J_{\textnormal{max}}$, $s\in\mathbb{S}$, $f\in\mathcal{F}$, $x\in\{x_1,\dots,x_{n_c}\}$, as well as the stationary expectations $\E_\pi(f)$ for $f\in\mathcal{F}$. Once these quantities are obtained, the decay modes are identified following Section \ref{sec:est_state_independent}, with tolerance $\epsilon_{\textrm{tol}}=0.01$. For the construction of the approximate Koopman operator at a given initial state $x$, we simulate 100 CTMC trajectories and compute the linear coefficients in \eqref{approx_koopman_operator1} as described in Section \ref{subsec:koopman_contruction}, setting $J_\ell=2$ for each $s_\ell\in\mathbb{S}$. For every $f\in\mathcal{F}$ we then evaluate the error defined in~\eqref{defn_e_f_x}, and, with threshold $\epsilon_{\textrm{basis}}=0.1$, we define the set of basis functions $\mathcal{B}\subseteq\mathcal{F}$ as those for which the normalized error magnitude is below $\epsilon_{\textrm{basis}}$. The approximate Koopman operator is obtained directly for $f\in\mathcal{B}$ via the linear coefficients, and for $f\notin\mathcal{B}$ by projection onto the linear span of $\mathcal{B}$ in $\mathcal{L}_2(\hat{\pi})$, as discussed in Section \ref{subsec:estimation_sens_spectral}. We compare this with the estimate obtained directly using 1000 SSA-generated CTMC trajectories. 

For each SRN we consider, we also apply SKA to estimate all the parameter sensitivities and the cross-spectral densities between every pair of observable functions in $\mathcal{F}$, as described in Section \ref{subsec:estimation_sens_spectral}. The estimated sensitivities are compared against the \emph{coupled finite-difference (CFD)} method developed in \cite{anderson2012efficient}, while the spectral densities are obtained directly by applying the DFT to the time-discretized SSA trajectories. Additional details of the implementation are provided in Section \ref{supp:sensitivity_csd_impl_details} of the Appendix.

The large-scale simulations required to estimate iterated resolvents and stationary expectations were executed on a single NVIDIA RTX 4500 Ada Generation GPU, while the remaining computations were performed on a MacBook Pro with a 2.4\,GHz 8-core Intel Core i9 processor; the run-times for each example are reported in Tables \ref{table_computational_times} and \ref{table_task_times}.

\begin{table}[!htbp]
\begin{center} 
\begin{tabular}{||c |  c | c | c  ||}
\hline 
Network   & Decay Mode Estimation &  Sensitivity Coeffs. & CSD Coeffs. \\
 \hline  \hline
 Self-reg. Gene. Ex. &   175 & 0.78 & 9.5  \\  
 \hline
 Repressilator    & 210 & 4.5 & 13.4  \\  
\hline
Const. Gene. Ex.  + rAIF   & 560 & 155 & 52.5  \\ 
 \hline
Const. Gene. Ex.  + sAIF   & 600 & 103.5 & 55.0  \\ 
 \hline
\end{tabular}  
\end{center}
\caption{Computational times (in seconds) for estimating the state-independent quantities in SKA. The reported time for decay mode estimation includes not only the computation of the iterated resolvents and stationary expectations, but also the subsequent convex optimization steps (see Section \ref{sec:est_state_independent}). Sensitivity and CSD coefficients are defined in Section \ref{subsec:estimation_sens_spectral}.}
\label{table_computational_times}
\end{table}

\begin{table}[!htbp]
\begin{center} 
\begin{tabular}{||c | c | c | c | c ||}
\hline 
Network  & Task  & Method &  Computational times & Speedup factor \\
\hline
  \multirow{6}{*}{Self-reg. Gene. Ex.} &   \multirow{2}{*}{1} & SSA & 0.53 & \multirow{2}{*}{2.6}  \\ 
   &   & SKA & 0.2 &   \\  
 \cline{2-5} 
   &  \multirow{2}{*}{2}  & CFD & 8.0 &  \multirow{2}{*}{40}  \\
    &   & SKA & 0.2 &   \\
     \cline{2-5} 
        &  \multirow{2}{*}{3}  & SSA-DFT & 6.5 &  \multirow{2}{*}{24}  \\
    &   & SKA & 0.27 &   \\
     \cline{2-5} 
\hline
  \multirow{6}{*}{Repressilator} &   \multirow{2}{*}{1} & SSA & 0.84 & \multirow{2}{*}{3}  \\ 
   &   & SKA & 0.27 &   \\  
 \cline{2-5} 
   &  \multirow{2}{*}{2}  & CFD & 15.9 &  \multirow{2}{*}{44}  \\
    &   & SKA & 0.36 &   \\
     \cline{2-5} 
        &  \multirow{2}{*}{3}  & SSA-DFT & 10.2 &  \multirow{2}{*}{31}  \\
    &   & SKA & 0.32 &   \\
     \cline{2-5} 
\hline
     
\hline
  \multirow{6}{*}{Const. Gene. Ex. + rAIF} &   \multirow{2}{*}{1} & SSA & 1.9 & \multirow{2}{*}{2.4}  \\ 
   &   & SKA & 0.8 &   \\  
 \cline{2-5} 
   &  \multirow{2}{*}{2}  & CFD & 151.6 &  \multirow{2}{*}{56}  \\
    &   & SKA & 2.7 &   \\
     \cline{2-5} 
        &  \multirow{2}{*}{3}  & SSA-DFT & 20.5 &  \multirow{2}{*}{17.6}  \\
    &   & SKA & 1.2 &   \\
     \cline{2-5} 
\hline

  \multirow{6}{*}{Const. Gene. Ex. + sAIF} &   \multirow{2}{*}{1} & SSA & 2.9 & \multirow{2}{*}{2.8}  \\ 
   &   & SKA & 1.04 &   \\  
 \cline{2-5} 
   &  \multirow{2}{*}{2}  & CFD & 180.0 &  \multirow{2}{*}{45}  \\
    &   & SKA & 4.0 &   \\
     \cline{2-5} 
        &  \multirow{2}{*}{3}  & SSA-DFT & 28.8 &  \multirow{2}{*}{18}  \\
    &   & SKA & 1.6 &   \\
     \cline{2-5} 
\hline
\end{tabular}  
\end{center}
\caption{Computational times (in seconds) for three estimation tasks—Task 1 (\eqref{estimation_task1}, Koopman operator estimation), Task 2 (\eqref{estimation_task_sensitivity}, parameter sensitivity estimation), and Task 3 (\eqref{estimation_task_csd}, cross-spectral density estimation)—evaluated for a fixed initial state $x$. The reported speedup factor is defined as the ratio of the computational time of the comparison method (e.g., SSA for Task 1) to that of SKA.}
\label{table_task_times}
\end{table}

\subsection{Self-regulatory gene-expression network}
\label{main:self_ge_network}

\begin{figure}[!htbp]
  \centering
 \includegraphics[width=0.85\textwidth]{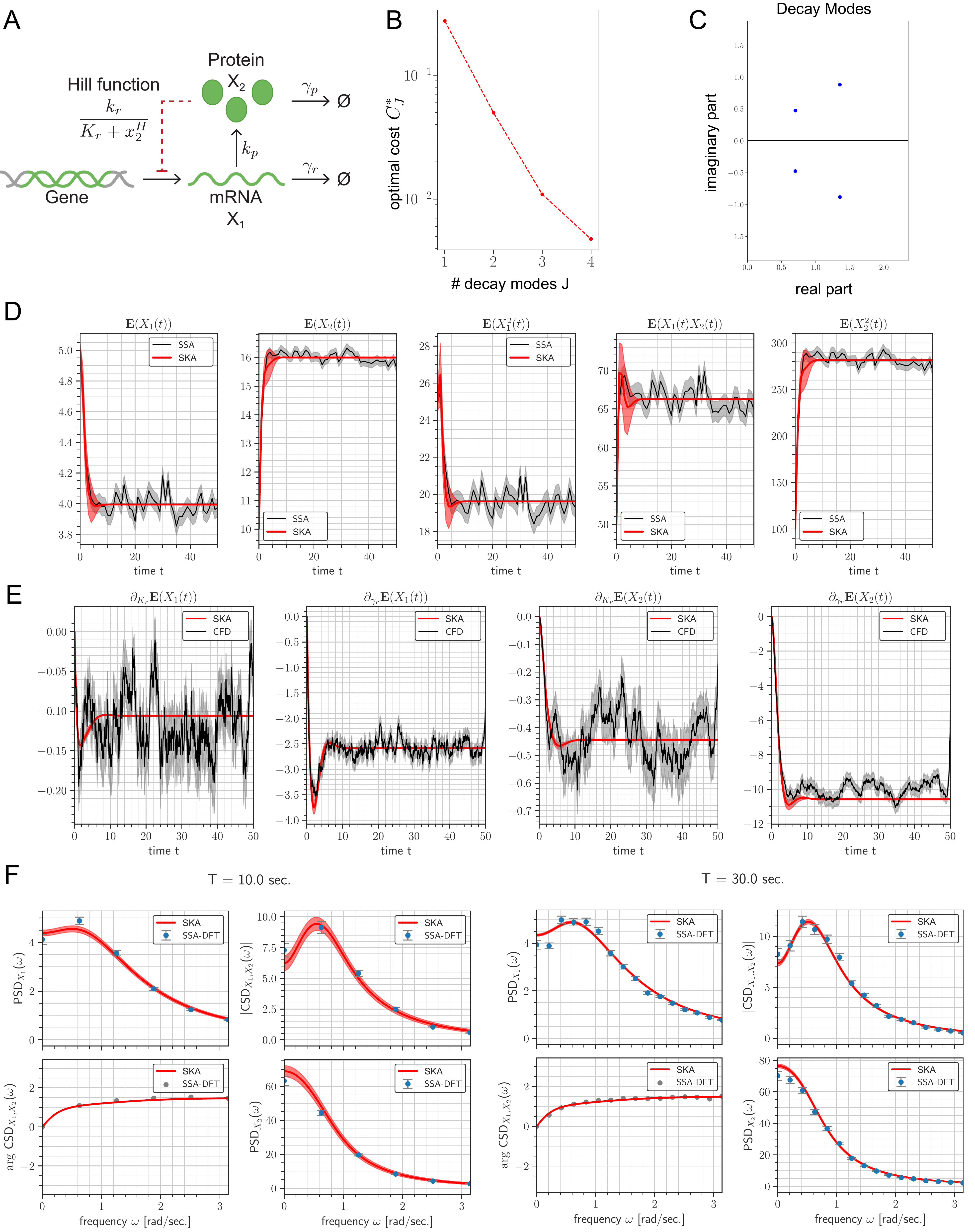}
\caption{ \small {\bf Analysis of the self-regulatory gene-expression network.} 
(A) Schematic of the network. (B) Optimal cost $C_J^*$ of the convex optimization problem \eqref{defn_convex_optimization} as a function of the number of decay modes $J$ (vertical axis on a logarithmic scale). (C) Identified decay modes on the complex plane, all with positive real parts and appearing in conjugate pairs. (D) Koopman trajectories for all observable functions in $\mathcal{F}$ (i.e., the first two moments \eqref{first_two_moments}) generated by SKA, compared with SSA for an arbitrary initial state $x=(x_1,x_2) = (5, 10)$. (E) Parameter sensitivity estimates for selected observable functions with respect to a subset of model parameters, compared with CFD (same initial state as in panel~D). (F) Cross-spectral density estimates for selected observable functions over the interval $[0,T]$ for two choices of $T$, compared with SSA–DFT (same initial state as in panel~D). The full evolution of SKA-estimated CSDs from $T=0$ to $T=30$ seconds is provided in Supplementary Movie 1. When $f_1=f_2$, $\textnormal{CSD}_{f_1,f_2}(\omega,x,T)$ reduces to the real-valued power spectral density $\textnormal{PSD}_{f_1}(\omega,x,T)$; when $f_1 \neq f_2$, the magnitude and argument of $\textnormal{CSD}_{f_1,f_2}(\omega,x,T)$ are plotted separately. In each of the plots, the solid curve represents the mean while the shaded region represents the symmetric one standard deviation interval around the mean.}
\label{main_fig_self_reg_gene_ex}
\end{figure}

As a first example, we consider the self-regulatory gene-expression network depicted in Figure \ref{main_fig_self_reg_gene_ex}(A). The network comprises two molecular species: mRNA ($\mathbf{X}_1$) and protein ($\mathbf{X}_2$). Both species degrade spontaneously, while $\mathbf{X}_1$ translates into $\mathbf{X}_2$, which in turn represses transcription of $\mathbf{X}_1$ through an inhibitory Hill function (see Section~\ref{supp:self_ge_network} of the Appendix for full details).

Figure \ref{main_fig_self_reg_gene_ex}(B) shows the decay of the optimal cost $C_J^*$ in the convex optimization problem \eqref{defn_convex_optimization} as the number of decay modes $J$ increases. The nearly linear trend on the logarithmic scale indicates that $C_J^*$ decreases approximately exponentially with $J$. The four identified decay modes are plotted in Figure \ref{main_fig_self_reg_gene_ex}(C). For a randomly chosen initial state, the approximate Koopman operator \eqref{approx_koopman_operator1} constructed by SKA is used to compute the dynamics of the first two moments. As shown in Figure \ref{main_fig_self_reg_gene_ex}(D), these trajectories closely match those obtained via SSA, while being smoother and exhibiting lower variance, despite SSA relying on ten times more CTMC simulations and requiring 2.6 times more computational time (see Table \ref{table_task_times}).  

In Figure \ref{main_fig_self_reg_gene_ex}(E), we report parameter sensitivity estimates obtained with SKA and compare them against the coupled finite-difference (CFD) method. Here, the computational advantages of SKA become more pronounced: it is approximately forty times faster (Table \ref{table_task_times}) and produces sensitivity estimates with substantially lower variance over time. Finally, in Figure \ref{main_fig_self_reg_gene_ex}(F), we apply SKA to estimate cross-spectral densities at two terminal times and benchmark against the SSA–DFT method. Not only is SKA twenty-three times faster (Table \ref{table_task_times}), it also yields a smooth, low-variance mapping between frequency and spectral density, in contrast to SSA–DFT, which produces estimates only at discrete frequency points. The full evolution of the SKA-estimated CSDs with the terminal time is provided in Supplementary Movie 1.

{\bf Inference of initial state distribution:}
We next demonstrate how SKA can be leveraged to infer the distribution of initial states in a cell population from single-cell measurements obtained with technologies like Flow Cytometry. Consider a population of cells, each governed by the self-regulatory gene-expression network described above, where the initial state $x=(x_1,x_2)$ of each cell is drawn from a product Poisson distribution $\mathcal{P}_{\hat{\eta}}(x)$ with an \emph{unknown} parameter vector $\hat{\eta}=(\hat{\eta}_1,\hat{\eta}_2)$, i.e.,
\begin{align}
\label{prod_poisson_distr}
\mathcal{P}_{\hat{\eta}}(x) = \frac{e^{-(\hat{\eta}_1 + \hat{\eta}_2)} \, \hat{\eta}_1^{x_1} \hat{\eta}_2^{x_2}}{x_1! \, x_2!}.
\end{align}
Suppose we have empirical measurements of the protein copy-number distribution at times $t_1,\dots,t_N$. Our objective is to use these data to infer the true parameter vector $\hat{\eta}$ characterizing the initial state distribution.

\begin{figure}[!htbp]
  \centering
 \includegraphics[width=1.0\textwidth]{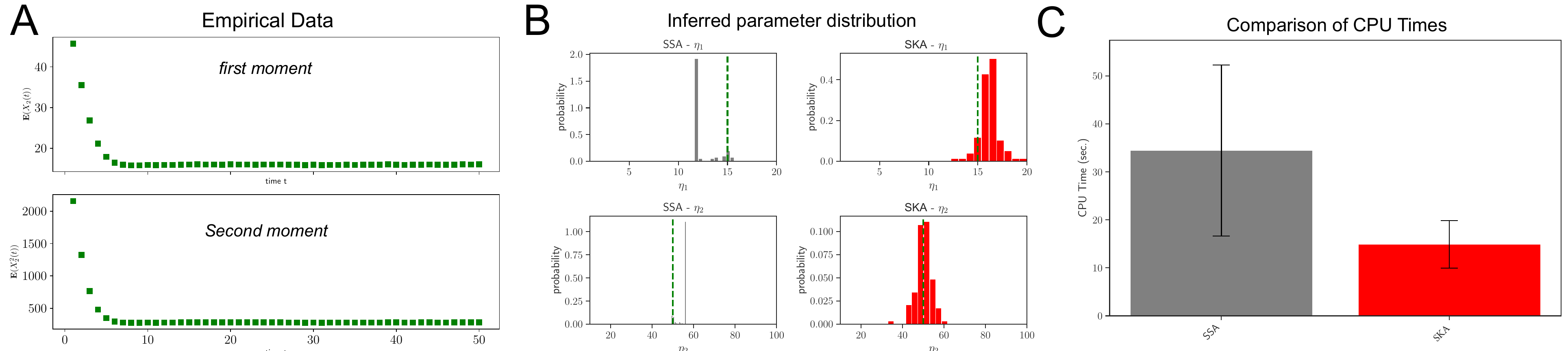}
\caption{\small {\bf Inference of the initial state distribution for the self-regulatory gene-expression network.} 
(A) First two moments of the protein copy numbers computed from simulated empirical data, with each CTMC initialized from the product Poisson distribution \eqref{prod_poisson_distr} with the true parameter vector $\hat{\eta} = (15, 50)$. (B) Marginal distributions of the inferred parameter vectors obtained from the SSA- and SKA-based approaches, each initialized from 100 randomly sampled parameter vectors. (C) Mean and standard deviation of the computational times required by the two approaches over 100 optimization runs.}
\label{fig_self_reg_gene_ex_inference}
\end{figure}

A natural approach is to first compute from the empirical data the first two moments of the protein copy number at each $t_n$, and then estimate $\eta$ by matching these to the corresponding moments predicted by the SRN model when the initial distribution is $\mathcal{P}_\eta(x)$. Let $f_1(x)=x_2$ and $f_2(x)=x_2^2$ denote the observables corresponding to the first and second moments of protein $\mathbf{X}_2$, and let $\hat{f}_1(t_n)$ and $\hat{f}_2(t_n)$ denote their empirical values at time $t_n$. Let $(X_x(t))_{t \geq 0} = (X_{x,1}(t), X_{x,2}(t))_{t \geq 0}$ denote the CTMC representing the SRN dynamics with initial state $x$. Then, the first two moments of the protein copy number are given by
\begin{align}
\label{self_reg_moment}
\E_{\mathcal{P}_\eta(x)}\!\left( X_{x,2}^k(t_n)\right) := \sum_{x \in \Z_{\geq 0}^2} \E\!\left(X_{x,2}^k(t_n)\right) \mathcal{P}_\eta(x),
\end{align}
for $k=1,2$, and their gradients $\nabla_\eta \E_{\mathcal{P}_\eta(x)}\!\left( X_{x,2}^k(t_n)\right)$ can be written as
\begin{align}
\nabla_\eta  \E_{\mathcal{P}_\eta(x)}\!\left( X_{x,2}^k(t_n)\right) = \sum_{x \in \Z_{\geq 0}^2} \E\left(X_{x,2}^k(t_n)\right) \nabla_\eta \mathcal{P}_\eta(x) = \E_{\mathcal{P}_\eta(x)} \left(\E\!\left(X_{x,2}^k(t_n)\right) \nabla_\eta \log \mathcal{P}_\eta(x) \right).
\end{align}
Thus, both the moments and their gradients can be estimated by sampling initial states from $\mathcal{P}_\eta$, simulating the CTMC over the interval $[t_1,t_N]$ using SSA for each sampled state $x$, and then averaging the values $X_{x,2}^k(t_n)$ and $\left(X_{x,2}^k(t_n)\nabla_\eta \log \mathcal{P}_\eta(x)\right)$, respectively.

The parameter vector $\eta$ can then be inferred by minimizing the cost function
\begin{align}
\label{inference_cost_function_direct}
C_{\textnormal{direct}}(\eta) = \frac{1}{2} \sum_{n=1}^N \sum_{k=1}^2 
\left(
   \frac{\E_\eta(X_2^k(t_n)) - \hat{f}_k(t_n)}{\E_\pi(f_k)}
\right)^2,
\end{align}
where normalization by $\E_\pi(f_k)$ ensures comparability across moments. As this cost function is non-convex in $\eta$, its minimization requires an iterative procedure involving evaluations of both the cost and its gradient with respect to $\eta$. These can be computed using the sampling-based estimates of $\E_{\mathcal{P}_\eta(x)}\!\left( X_{x,2}^k(t_n)\right)$ and $\nabla_\eta  \E_{\mathcal{P}_\eta(x)}\!\left( X_{x,2}^k(t_n)\right)$ described above.

However, the complexity of the non-convex cost function \eqref{inference_cost_function_direct} scales linearly with the number of measurement time points $N$, which can be very large in Flow Cytometry or other single-cell datasets. This makes the optimization problem computationally demanding. We now show how SKA circumvents this challenge by first extracting the linear coefficients of the Koopman operator \eqref{approx_koopman_operator1} from the empirical data, and then inferring $\eta$ by matching these coefficients with those of the Koopman operator for our SRN model. We assume that the state-independent decay modes $\bar{\sigma}_j$ and stationary expectations $\E_\pi(f_k)$ have been pre-estimated and are known.

Using the approximate Koopman operator \eqref{approx_koopman_operator1}, we can write
\begin{align}
\label{linear_coeff_reln_inference}
\E_{\mathcal{P}_\eta(x)}\!\left( X_{x,2}^k(t_n)\right)  \approx \E_\pi(f_k) + \sum_{j=1}^J \bar{\alpha}_j(f_k, \eta) e^{-\bar{\sigma}_j t_n},
\qquad k=1,2,\; n=1,\dots,N,
\end{align}
where $\bar{\alpha}_j(f_k,\eta)=\E_{\mathcal{P}_\eta(x)}\!\left(\bar{\alpha}_j(f_k,x)\right)$. Since the coefficients $\bar{\alpha}_j(f_k,\eta)$ appear linearly, they can be estimated from empirical data by substituting $\E_{\mathcal{P}_\eta(x)}\!\left( X_{x,2}^k(t_n)\right)$ with $\hat{f}_k(t_n)$ and performing linear regression to obtain estimates $\hat{\alpha}_{jk}$ for $k=1,2$ and $j=1,\dots,J$. The parameter vector $\eta$ can then be inferred by minimizing the SKA-based cost function
\begin{align}
\label{inference_cost_function_SKA}
C_{\textnormal{SKA}}(\eta) = \frac{1}{2} \sum_{j=1}^J \sum_{k=1}^2 
\left(
   \frac{\bar{\alpha}_j(f_k,\eta)-\hat{\alpha}_{jk}}{\E_\pi(f_k)}
\right)^2.
\end{align}
Unlike \eqref{inference_cost_function_direct}, the complexity of \eqref{inference_cost_function_SKA} scales with the number of decay modes $J$, which is typically much smaller than $N$, thereby making this SKA-based approach computationally more tractable. 

To minimize the cost function \eqref{inference_cost_function_SKA}, we must estimate the coefficients $\bar{\alpha}_j(f_k,\eta)$ and their gradients
\[
\nabla_\eta \bar{\alpha}_j(f_k,\eta) = \E_{\mathcal{P}_\eta(x)}\!\left( \bar{\alpha}_j(f_k,x)\,\nabla_\eta \log \mathcal{P}_\eta(x)\right).
\]
These can be obtained by fitting \eqref{linear_coeff_reln_inference} and its gradient using linear regression, where\\ $\E_{\mathcal{P}_\eta(x)}\left( X_{x,2}^k(t_n)\right)$ and $\nabla_\eta \E_{\mathcal{P}_\eta(x)}\!\left( X_{x,2}^k(t_n)\right)$ are estimated with the sampling-based approach described above. Crucially, this regression-based strategy does not require using the original measurement times $t_1,\dots,t_N$ at which empirical data are available; instead, one can freely choose a smaller set of time points spanning a shorter interval, thereby further reducing the computational cost of the SKA-based inference procedure.

In Figure \ref{fig_self_reg_gene_ex_inference}, we compared the direct SSA-based and SKA-based inference approaches using synthetic data from 10{,}000 single-cell trajectories of the self-regulatory gene-expression network. Each CTMC was initialized from the product Poisson distribution $\mathcal{P}_{\hat{\eta}}(x)$ with parameter $\hat{\eta} = (15, 50)$, and $N = 50$ measurement times were uniformly spaced from $t_1 = 1$ to $t_{50} = 50$. The SSA-based approach minimized the cost function \eqref{inference_cost_function_direct} using the iterative L-BFGS optimization algorithm, with the first two moments and their gradients for each $\eta$ estimated via 10{,}000 SSA-generated CTMC simulations over $[1,50]$. The SKA-based approach minimized the cost function \eqref{inference_cost_function_SKA} using the same algorithm, with the linear coefficients and their gradients for each $\eta$ estimated from 10{,}000 CTMC trajectories over the shorter interval $[0,5]$. Figure~\ref{fig_self_reg_gene_ex_inference}(B) shows the marginal probability distributions of the components of $\eta = (\eta_1,\eta_2)$ obtained from 100 randomly sampled initial parameter vectors in $[1,20]\times[10,100]$, where the SKA-based approach produced a sharper distribution with its peak closer to the true parameter values than the SSA-based approach. In the SSA-based approach, many optimization runs became trapped in local minima, producing a sharp peak in the histogram away from the true parameter value. Figure \ref{fig_self_reg_gene_ex_inference}(C) reports the computational times of the 100 optimization runs, showing that SKA was at least twice more efficient, due to the lower complexity of its cost function.

\subsection{\emph{Repressilator} network}
\label{main:repressilator_network}

\begin{figure}[!htbp]
  \centering
 \includegraphics[width=0.95\textwidth]{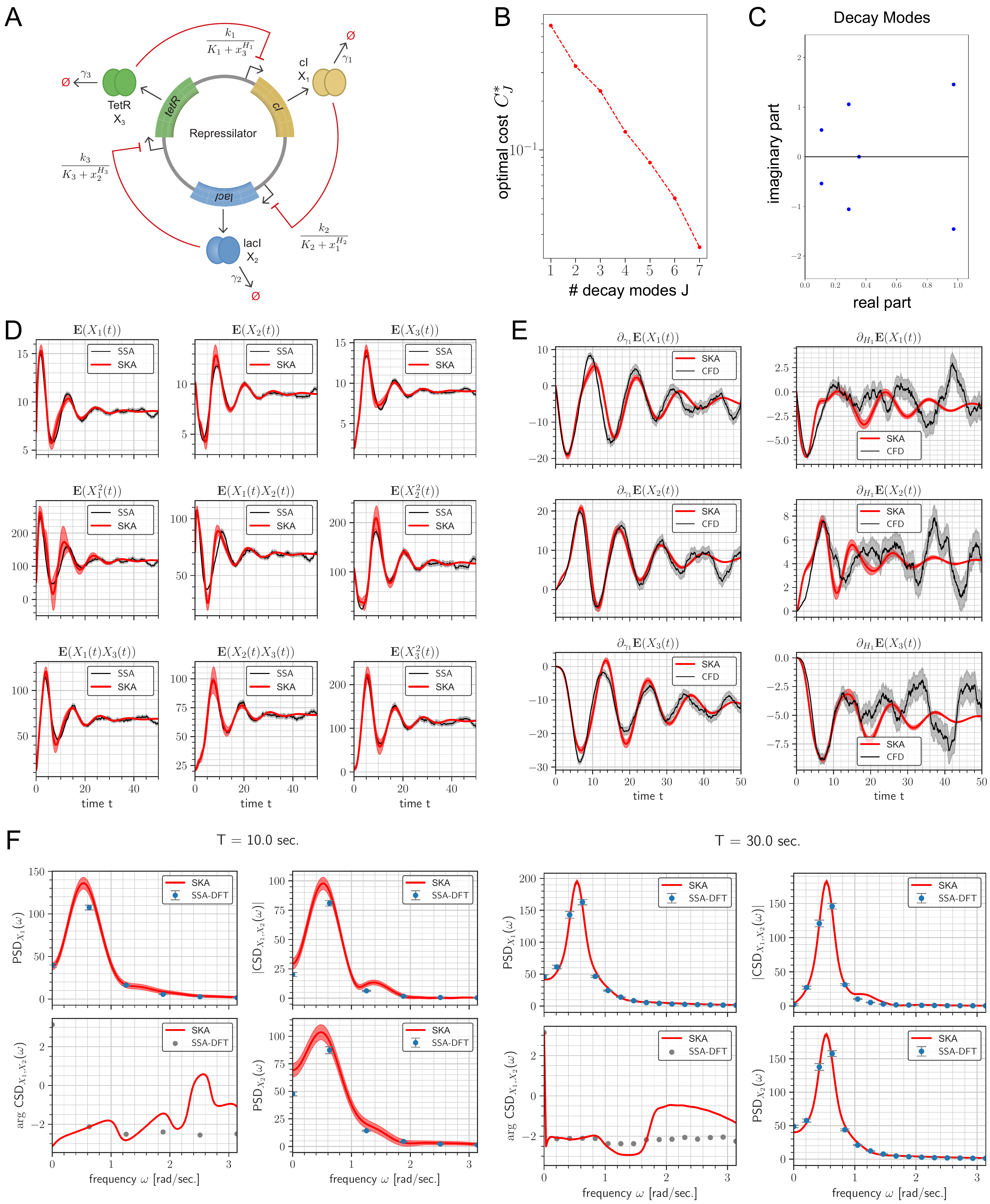}
\caption{{\bf Analysis of the \emph{repressilator} network.} 
(A) Schematic of the network. Panels (B–F) correspond to the same analyses described in Figure \ref{main_fig_self_reg_gene_ex}, but here computed with the arbitrary initial state $x = (x_1,x_2,x_3) = (7, 10, 2)$. Corresponding to panel (F), the full evolution of SKA-estimated CSDs from $T=0$ to $T=30$ seconds is provided in Supplementary Movie 2.
}
\label{main_fig_repress}
\end{figure}

The \emph{repressilator} \cite{elowitz2000synthetic}, the first synthetic genetic oscillator, consists of three genes that repress one another in a cyclic fashion (Figure~\ref{main_fig_repress}(A)). These genes produce the proteins cI ($\mathbf{X}_1$), TetR ($\mathbf{X}_2$), and LacI ($\mathbf{X}_3$), which mutually repress transcription through inhibitory Hill functions (see Section~\ref{supp:repressilator_network} of the Appendix for full details).  

Figure \ref{main_fig_repress}(B) shows the decay of the optimal cost $C_J^*$ as a function of the number of decay modes $J$, which decreases nearly exponentially. The seven identified decay modes are plotted in Figure \ref{main_fig_repress}(C). For an arbitrary initial state, we then apply SKA to estimate (i) the dynamics of the first two moments, (ii) the time evolution of parameter sensitivities, and (iii) cross-spectral densities. The results are compared with SSA, CFD, and SSA–DFT in Figures \ref{main_fig_repress}(D–F), respectively. Across all tasks, SKA consistently provides accurate, low-variance, and smooth estimates. Moreover, it achieves substantial computational gains: being approximately three times faster than SSA for moment dynamics, forty-four times faster than CFD for parameter sensitivities, and thirty-one times faster than SSA–DFT for spectral densities (see Table \ref{table_computational_times}).

\begin{figure}[!htbp]
  \centering
 \includegraphics[width=0.95\textwidth]{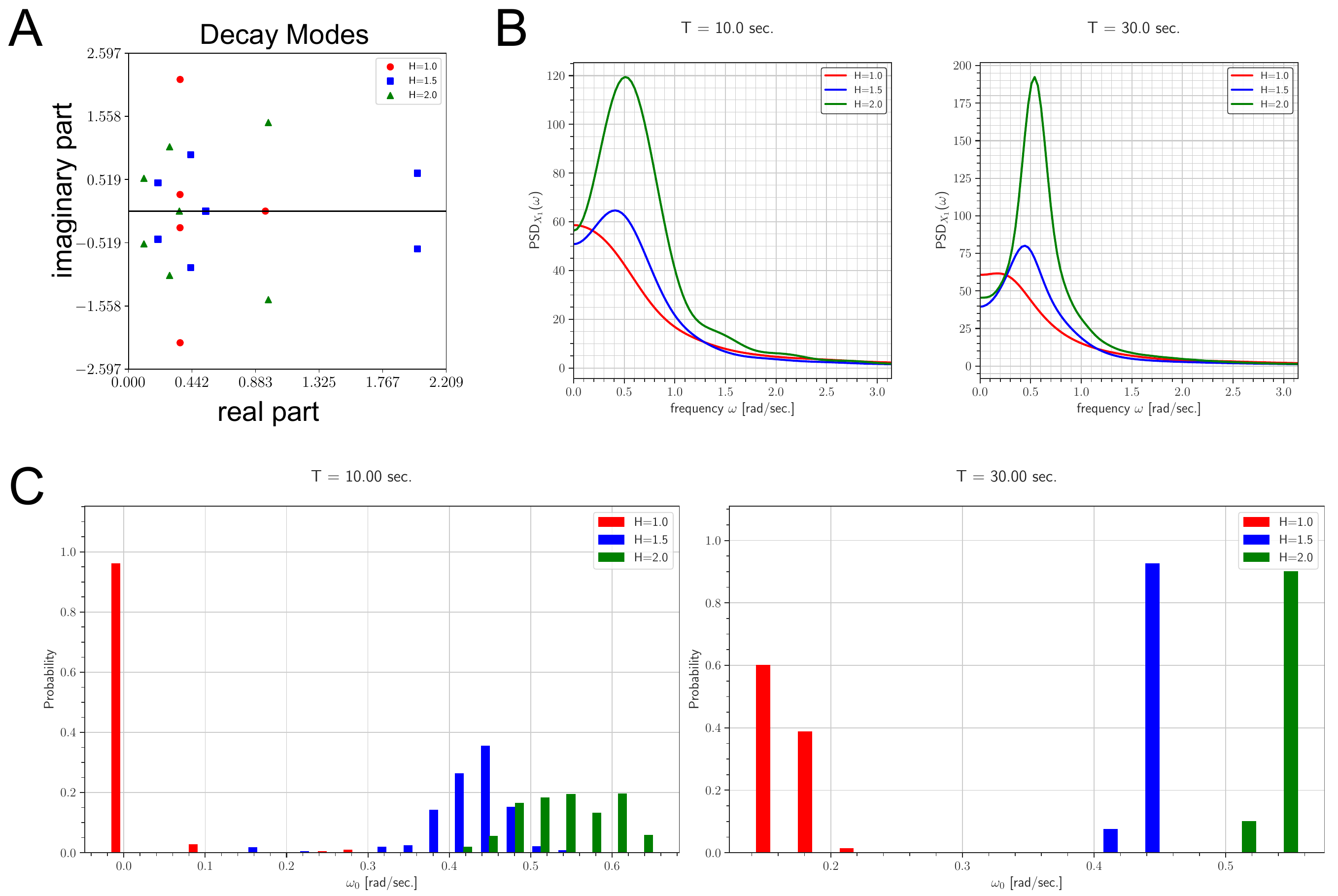}
\caption{{\bf Effect of cooperativity on the \emph{repressilator} network.} We consider three repressilator networks with all Hill coefficients in the repression mechanism set to $H=1.0$, $1.5$, and $2.0$. 
(A) Decay modes of the three networks. (B) $\mathcal{L}_2(\hat{\pi})$ norm of the PSD of the $\mathbf{X}_1$ copy-number trajectory for the three networks, shown for two terminal times $T$, where $\hat{\pi}$ denotes the approximate stationary distribution. The full temporal dynamics from $T=0$ to $30$ seconds are provided in Supplementary Movie 5. (C) Probability distribution of the peak frequency ($\omega_0$), assuming the initial state is distributed according to $\hat{\pi}$, plotted at the two terminal times $T$. The full temporal dynamics from $T=0$ to $30$ seconds are provided in Supplementary Movie 6.}
\label{main_fig_repress_comparison}
\end{figure}

{\bf Evaluating the effect of repression cooperativity:}
The single-cell trajectories of the repressilator exhibit oscillatory behavior, as evidenced by the sharp peaks in the cross-spectral densities in Figure~\ref{main_fig_repress}(F). These plots reveal a distinct peak frequency, $\omega_0$, at which the oscillations are most pronounced. However, when oscillations are averaged across a population of cells, the repressilator loses coherence over time due to intrinsic noise in the dynamics, as indicated by the moment dynamics in Figure~\ref{main_fig_repress}(D). This deleterious effect of intrinsic noise is exacerbated by reduced cooperativity in the repression mechanism, encoded by the Hill coefficients $H_j$ for $j=1,2,3$ (see Figure~\ref{main_fig_repress}(A)). To study this effect quantitatively, we apply SKA to three repressilator networks with all Hill coefficients set to $H=1.0$, $1.5$, and $2.0$ (the last corresponding to the network analyzed in Figure \ref{main_fig_repress}). The results are summarized in Figure \ref{main_fig_repress_comparison}. As cooperativity $H$ decreases, the leading decay modes shift further to the right, indicating greater stability and diminished oscillatory power, independent of the initial state. This trend is consistent with the PSD plots in Figure~\ref{main_fig_repress_comparison}(B) (see also 
Supplementary Movie 5). Figure~\ref{main_fig_repress_comparison}(C) shows the probability distribution of the peak frequency at two terminal times (full evolution shown in 
Supplementary Movie 6), revealing a tendency for the peak frequency to increase with $H$. Notably, SKA enables tractable estimation of PSDs across initial states, and the smooth curves it produces facilitate comparisons between networks and analysis of peak locations.

\subsection{Constitutive gene-expression with the Antithetic Integral Feedback (AIF) controller} \label{main:aif}

The \emph{antithetic integral feedback} (AIF) controller, proposed in \cite{briat2016antithetic}, is an \emph{in vivo} bio-molecular control motif that ensures \emph{robust perfect adaptation} (RPA) in stochastic intracellular networks. RPA guarantees that the expected copy number of an output species, $\mathbf{X}_\ell$, converges to a prescribed set-point despite constant perturbations to network parameters. Strikingly, any bio-molecular controller achieving RPA in the stochastic regime must embed the AIF motif \cite{gupta2019universal}, establishing its universality. This highlights the importance of analyzing closed-loop stochastic reaction networks (SRNs) formed by coupling the AIF controller to target systems \cite{gupta2018antithetic, olsman2019architectural, kell2023noise}, and in this section we demonstrate how SKA enables such analysis.

\begin{figure}[!htbp]
  \centering
 \includegraphics[width=0.93\textwidth]{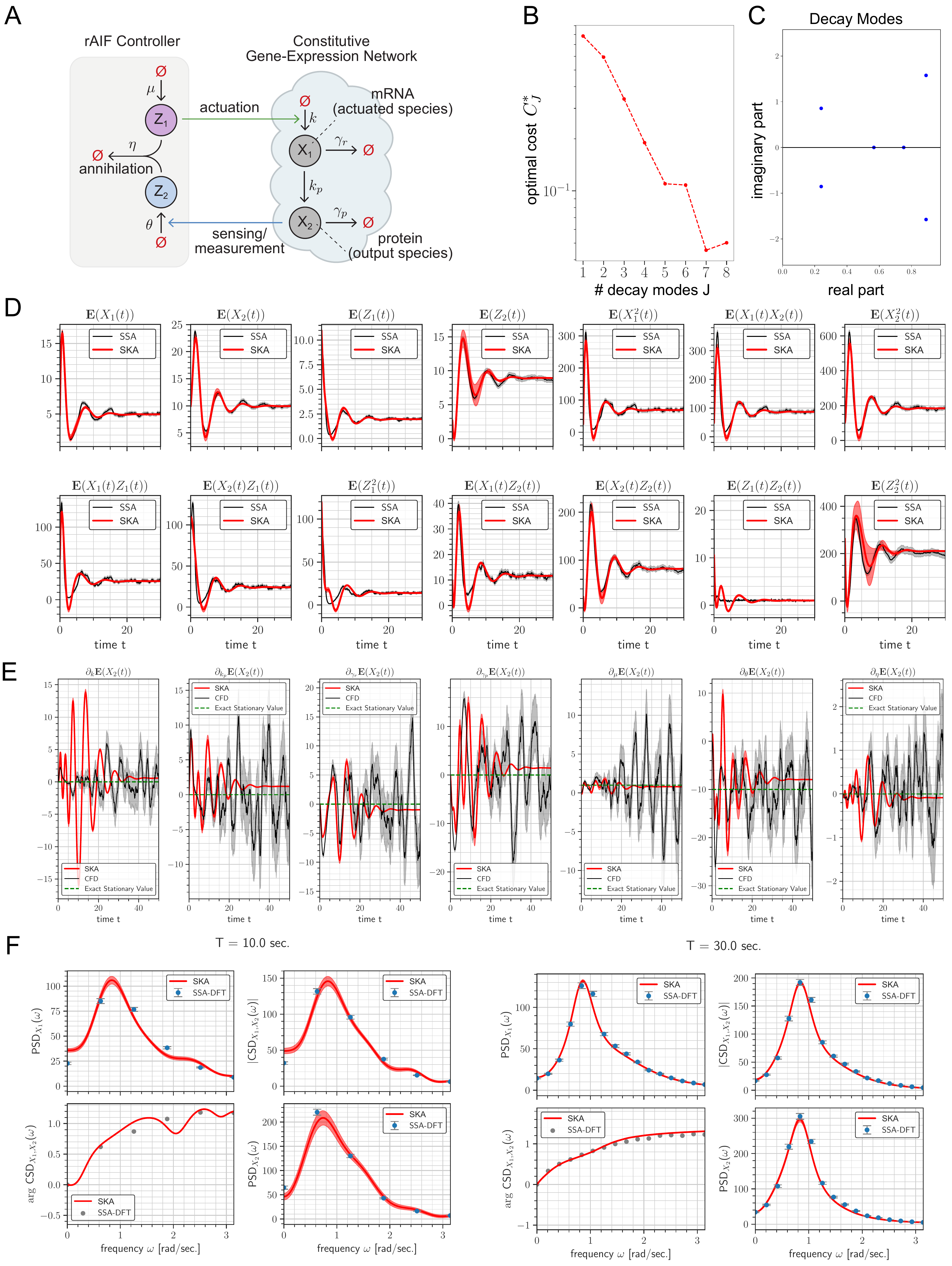}
\caption{{\bf Analysis of the closed-loop network consisting of constitutive gene-expression with the reference-based Antithetic Integral Feedback (rAIF) controller.} (A) Schematic of the network. Panels (B–F) correspond to the same analyses shown in Figure~\ref{main_fig_self_reg_gene_ex}, here evaluated for the arbitrary initial state $x = (x_1,x_2,z_1, z_2) = (5, 10, 11, 2)$. Corresponding to panel (F), the full evolution of SKA-estimated CSDs from $T=0$ to $T=30$ seconds is provided in Supplementary Movie 3.
}
\label{main_fig_raif}
\end{figure}

The AIF controller consists of two species: $\mathbf{Z}_1$ (reference) and $\mathbf{Z}_2$ (sensor), which mutually inactivate each other through annihilation. The production rate of $\mathbf{Z}_1$ is constitutive, with rate constant $\mu$, while the production of $\mathbf{Z}_2$ is catalyzed by the output species $\mathbf{X}_\ell$ with rate constant $\theta$. The AIF achieves RPA by ensuring that the expected copy number of $\mathbf{X}_\ell$ converges with time to the ratio $\mu/\theta$, regardless of other parameter values in either the controller or the controlled network. We consider two versions of the AIF proposed in \cite{briat2016antithetic}, which differ in their \emph{actuation} mechanisms. In the \emph{reference-based} AIF (rAIF), actuation is mediated by the reference species $\mathbf{Z}_1$, which catalyses the production of a network species $\mathbf{X}_1$. In the \emph{sensor-based} AIF (sAIF), actuation is mediated by the sensor species $\mathbf{Z}_2$, which represses the production of $\mathbf{X}_1$ (see \cite{kell2023noise, filo2023hidden}). As the controlled network, we adopt the constitutive gene-expression model (see Section~\ref{ex:cons_gene_ex}), where the mRNA $\mathbf{X}_1$ serves as the actuated species and the protein $\mathbf{X}_2$ as the output species. The closed-loop networks with rAIF and sAIF are shown in Figures~\ref{main_fig_raif}--\ref{main_fig_saif}(A) and described in detail in Sections~\ref{supp:sec_raif}--\ref{supp:sec_saif} of the Appendix. As discussed therein, we choose parameters for both controllers such that their actuation rates and actuation gains are approximately matched at the steady-state, ensuring a fair comparison of their performance.

In Figures \ref{main_fig_raif}--\ref{main_fig_saif}(B--F), we analyze the two closed-loop networks in the same way as the previous examples. We present the optimal costs, identified decay modes, and, for a representative initial state, SKA-estimated dynamics of the first two moments, parameter sensitivities, and cross-spectral densities. These results are benchmarked against SSA, CFD, and SSA-DFT. Across all tasks, SKA provides accurate estimates while offering substantially greater computational efficiency (Table \ref{table_computational_times}). 

\begin{figure}[!htbp]
  \centering
 \includegraphics[width=0.93\textwidth]{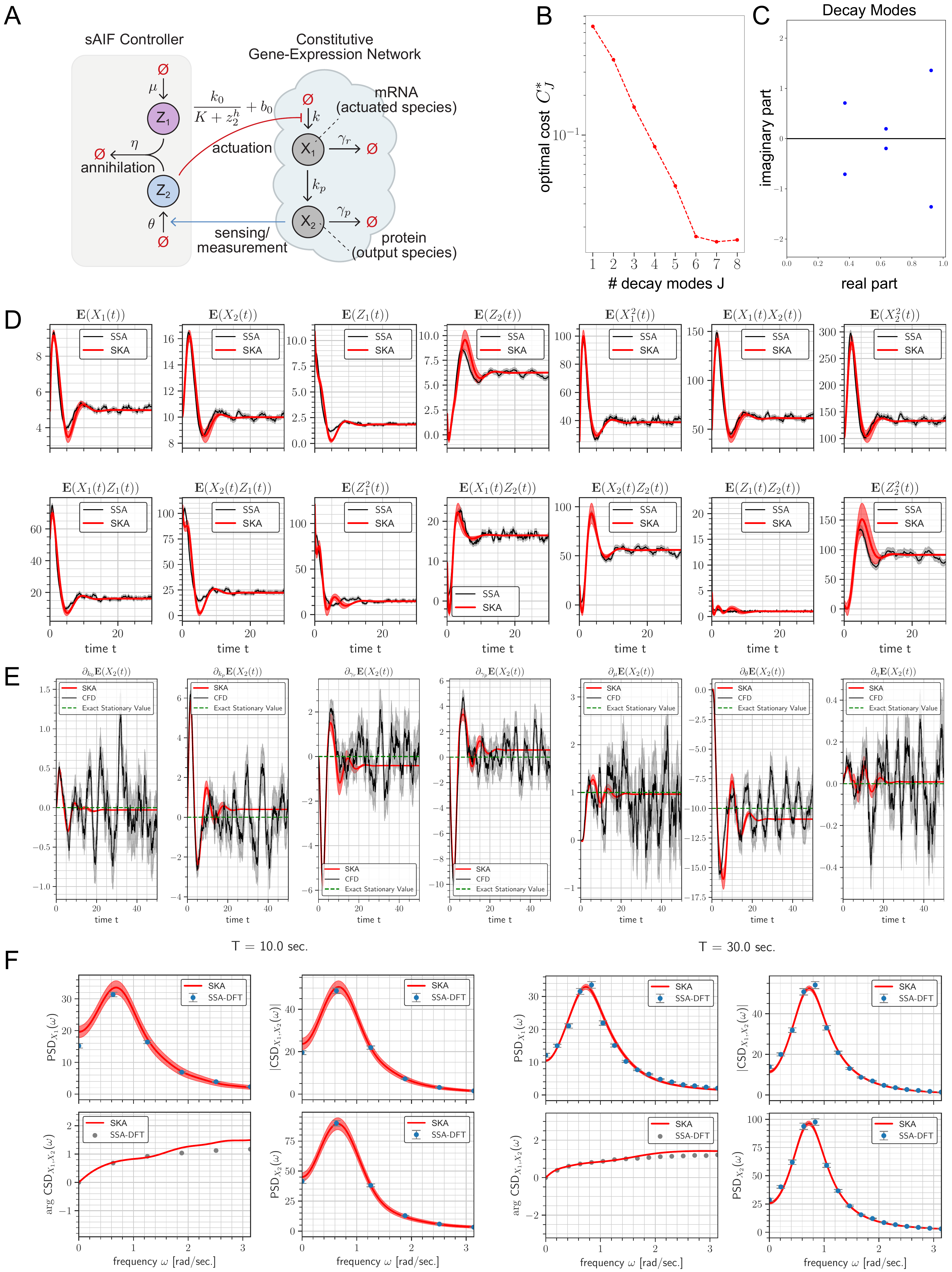}
\caption{{\bf Analysis of the closed-loop network consisting of constitutive gene-expression with the sensor-based Antithetic Integral Feedback (sAIF) controller.} (A) Schematic of the network. Panels (B–F) correspond to the same analyses shown in Figure \ref{main_fig_self_reg_gene_ex}, here evaluated for the arbitrary initial state $x = (x_1,x_2,z_1, z_2) = (5, 10, 11, 2)$. Corresponding to panel (F), the full evolution of SKA-estimated CSDs from $T=0$ to $T=30$ seconds is provided in Supplementary Movie 4.
}
\label{main_fig_saif}
\end{figure}

Since the steady-state expected copy number of the output species satisfies $\lim_{t \to \infty}\mathbb{E}(X_2(t))= \mu/\theta$ for both networks, the sensitivity of this quantity with respect to all network parameters can be computed analytically at steady state. These analytical values are shown as dashed green lines in Figures \ref{main_fig_raif}--\ref{main_fig_saif}(E). The results demonstrate that while SKA estimates converge over time to values close to the analytical benchmark, CFD frequently produces estimates that oscillate around the exact value with high variance, highlighting the superior performance of SKA. We note that the difference between the estimated sensitivity and the exact steady-state value can be interpreted as the ``RPA error'' incurred by an AIF controller, and that a superior controller should exhibit lower RPA error not only at steady state but also transiently.

\begin{figure}[h!]
  \centering
 \includegraphics[width=0.96\textwidth]{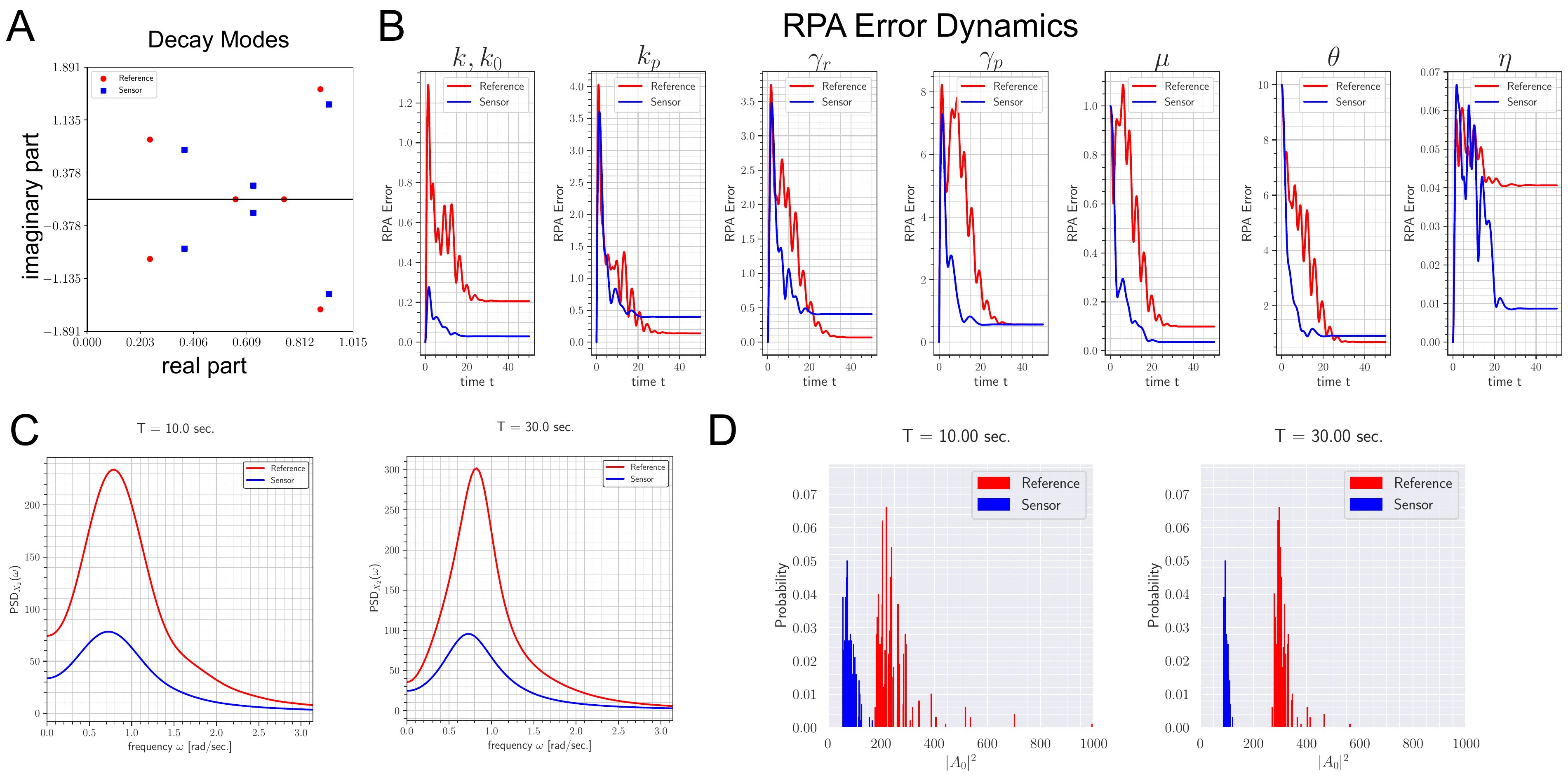}
\caption{{\bf Comparison of closed-loop networks with reference-based AIF (rAIF) and sensor-based AIF (sAIF).} (A) Decay modes of the two networks. (B) Dynamics of the RPA error, defined as the $\mathcal{L}_2(\hat{\pi})$ norm of the difference between the estimated sensitivity at any time $t$ and the exact steady-state value, where $\hat{\pi}$ denotes the approximate stationary distribution. (C) $\mathcal{L}_2(\hat{\pi})$ norm of the PSD of the output species copy-number trajectory for both controllers, shown for two terminal times $T$. The full temporal dynamics from $T=0$ to $30$ seconds are provided in Supplementary Movie 7. (D) Probability distribution of the peak amplitude ($|A_0|^2$), assuming the initial state is distributed according to $\hat{\pi}$, plotted at the two terminal times $T$. The full temporal dynamics from $T=0$ to $30$ seconds are provided in Supplementary Movie 8.
}
\label{main_fig_raif_vs_saif}
\end{figure}

{\bf Comparison between the rAIF and the sAIF controller:} In Figure \ref{main_fig_raif_vs_saif}, we use SKA to systematically compare the performance of the two controller topologies in achieving RPA for the constitutive gene-expression network. As shown in Figure~\ref{main_fig_raif_vs_saif}(A), the leading decay modes for sAIF are shifted further to the right relative to rAIF, indicating greater stability and reduced oscillatory tendency, independent of the initial state. This trend is consistent with the PSD plots in Figure \ref{main_fig_raif_vs_saif}(C) and the probability distribution of peak amplitudes in Figure~\ref{main_fig_raif_vs_saif}(D). While these plots correspond to two specific terminal times, the full evolution is shown in Supplementary Movies 7 and 8. The dynamics of the RPA error in Figure~\ref{main_fig_raif_vs_saif}(B) likewise reveal that sAIF incurs lower error over time. To ensure that our comparison is independent of the initial state, in Figures~\ref{main_fig_raif_vs_saif}(B--C) we estimated the relevant quantities across all initial states in the support of the approximate stationary distribution $\hat{\pi}$ and averaged with respect to the $\mathcal{L}_2(\hat{\pi})$ norm. Performing such computations over a large collection of initial states is far more tractable with SKA than with CFD or SSA-DFT. Overall, our results demonstrate that sAIF is a superior controller compared to rAIF, as it provides greater dynamical stability, lower RPA error, and smaller oscillation amplitudes across time and initial states. These findings are in line with those recently reported in \cite{filo2023hidden}.

\section{Conclusion}
\label{sec:conclusions}

The increasing availability of single-cell measurement technologies is transforming our ability to interrogate the dynamics of intracellular biochemical networks at unprecedented resolution. These dynamics are inherently stochastic, driven by the low copy numbers of many molecular species, and are naturally modeled as stochastic reaction networks (SRNs) evolving as continuous-time Markov chains (CTMCs). While the growing accessibility of single-cell data offers the opportunity to more accurately resolve SRN models, doing so remains computationally challenging. Our goal in this work was to develop a powerful computational framework to aid in this effort.

We introduced SKA (Stochastic Koopman Approximation), a method for estimating the stochastic Koopman operator associated with an SRN, which characterizes how the expectations of observable functions evolve over time as a function of the initial state. Exploiting the compactness of this operator, SKA constructs its spectral decomposition, explicitly identifying the contributions of its eigenmodes. The method proceeds in two stages: first, large-scale simulations are used to extract state-independent spectral properties of the operator; second, these are combined with linear regression techniques to efficiently estimate the state-dependent coefficients for any given initial condition. We develop this approach in the frequency-domain approach by leveraging properties of the resolvent operator of the underlying CTMC dynamics, enabling not only the estimation of expected observable function dynamics but also the dynamics of their parameter sensitivities and the cross-spectral densities of SRN trajectories.

Through diverse biological case studies, we demonstrated that SKA offers substantial computational speedups over existing approaches, particularly in the tasks of estimating parameter sensitivities and cross-spectral densities. We further showed that it can be used to infer initial state distributions from single-cell data, analyze synthetic oscillators \cite{elowitz2000synthetic}, and compare competing \emph{cybergenetic} controllers \cite{briat2016antithetic}, thereby highlighting its broad applicability.

The main drawback of SKA is that it relies on stochastic simulations and thus inherits their computational cost. This cost is particularly acute for multiscale networks, but this can be somewhat mitigated by integrating variance-reduction and approximate simulation approaches such as $\tau$-leaping \cite{cao2006efficient}, multilevel Monte Carlo schemes \cite{anderson2012multilevel}, or model reduction strategies \cite{kang2013separation,gupta2015adaptive}. In addition, because SKA yields parameter sensitivities, it could naturally be extended into a full parameter inference framework \cite{munsky2009listening}, akin to our demonstration of initial-state inference. Finally, an especially promising direction is to integrate the spectral expansion provided by SKA with machine-learning approaches such as DeepCME \cite{gupta2021deepcme}, thereby combining interpretability with scalability. We are actively pursuing this direction (see \cite{Badolle2025Interpretable}) and anticipate that such a hybrid framework will further enhance our ability to computationally analyze SRNs at scale and infer them from single-cell experimental data.

 \section*{Acknowledgments}
This work was supported by the Swiss National Science Foundation (SNSF) Advanced Grant: Theory and Design of Advanced Genetically Engineered Control Systems (grant number 216505). The authors also gratefully acknowledge Dr. Stephanie Aoki for her assistance in improving the figures.


\newpage

\textwidth=6.5true in
\textheight=9true in
\topmargin-0.5true in
\oddsidemargin=-0.25true in
\allowdisplaybreaks

\setcounter{section}{0}
\setcounter{equation}{0}
\setcounter{figure}{0}
\setcounter{table}{0}
\renewcommand{\thesection}{A\arabic{section}}
\renewcommand{\thetable}{A\arabic{table}} 
\renewcommand{\thefigure}{A\arabic{figure}}
\renewcommand{\theequation}{A\arabic{equation}}
\renewcommand{\figurename}{Appendix Figure}
\renewcommand{\tablename}{Appendix Table} 
\renewcommand{\contentsname}{Appendix}

\def \non{{\nonumber}}
\def \hat{\widehat}
\def \tilde{\widetilde}
\def \bar{\overline}
\newcommand{\ag}[1]{{\color{blue}#1}}

\def\ind{{\mathchoice {\rm 1\mskip-4mu l} {\rm 1\mskip-4mu l}
{\rm 1\mskip-4.5mu l} {\rm 1\mskip-5mu l}}}





\newpage

\begin{center}
{\Large {\bf Appendix}}
\end{center}

\section{Compactness of the Koopman operator}
\label{supp:compactness_of_the_koopman_operator}

Consider a simple birth--death process in which a single species $\mathbf{X}$ undergoes two reactions
\begin{align*}
\emptyset \stackrel{k}{\longrightarrow} \mathbf{X} \stackrel{\gamma} {\longrightarrow} \emptyset ,
\end{align*}
with propensity functions defined by mass-action kinetics (see \eqref{main:massactionkinetics}) and positive rate constants indicated above the reaction arrows. When the state (i.e., the number of molecules of $\mathbf{X}$) is $x$, the birth reaction occurs at a constant rate $k$, thereby increasing the state to $(x+1)$, while the death reaction occurs at rate $\gamma x$, decreasing the state to $(x-1)$. Modeling this system as a Stochastic Reaction Network (SRN), we obtain a continuous-time Markov chain (CTMC) whose generator is
\begin{align*}
\mathbb{A} f(x) = k (f(x+1) - f(x)) +\gamma x (f(x-1) - f(x)).
\end{align*}
This CTMC is exponentially ergodic with state space $\mathcal{E}$ consisting of all nonnegative integers, and stationary distribution $\pi$ given by the Poisson distribution with rate $k/\gamma$. The evolution of the probability distribution is governed by the Chemical Master Equation (CME) (see \eqref{cme_as_linear_system} in the main text), whose evolution operator $\mathbf{A}$ is the adjoint of the generator $\mathbb{A}$.  

The Finite State Projection (FSP) method approximates solutions of the CME by truncating the state space and thereby replacing $\mathbf{A}$ with a finite-dimensional operator. However, this approach suffers from fundamental limitations (see Section \ref{sec:intro} in the main text), primarily because $\mathbf{A}$ is not compact and thus cannot be accurately approximated by finite-dimensional operators. To see this lack of compactness in the present birth--death example, note that the eigenvalues of $\mathbf{A}$ (or equivalently of $\mathbb{A}$) are given by
\begin{align*}
\{ -n \gamma: n=0,1,2,\dots \}
\end{align*}
by the Karlin--McGregor spectral representation (see Chapter 3 in \cite{schoutens2012stochastic}). These eigenvalues diverge to $-\infty$, rather than accumulating at $0$, as required for compactness \cite{kato2013perturbation}. By contrast, the eigenvalues of the associated Koopman operator $\mathcal{K}_t$ are
\begin{align*}
\{ e^{-n \gamma t}: n=0,1,2,\dots \},
\end{align*}
which do accumulate at $0$ for any $t>0$, demonstrating that $\mathcal{K}_t$ is compact. This compactness underpins the approximation method for Koopman operators developed in this paper.  

The spectral representation result referenced above (see Chapter 3 in \cite{schoutens2012stochastic}) extends to more general classes of birth--death processes, suggesting that analogous compactness arguments apply more broadly. Furthermore, sufficient conditions for exponential ergodicity (one of our key assumptions) of general multi-species SRNs have been established by interpreting them as generalized cascades of birth--death networks (see \cite{gupta2014scalable, gupta2018computational}). In particular, the drift condition in \cite{gupta2014scalable}, expressed via a linear Foster--Lyapunov function $V(x)$, is satisfied by many SRN models of interest in systems and synthetic biology. This condition not only establishes exponential ergodicity but also implies that the Koopman operator is \emph{quasi-compact} (see \cite{herve2014approximating} and references therein). Quasi-compactness guarantees that the Koopman operator is well-approximated by a compact operator plus a contraction operator whose effect decays rapidly with time.  

It is important to note that for certain SRNs the Koopman operator may in fact be non-compact. In such cases, the approximation method developed here may not provide an accurate representation of the operator, as its validity fundamentally relies on compactness. Crucially, however, this limitation can be systematically diagnosed: one monitors the optimal value $C^*_J$ obtained from solving the convex optimization problem \eqref{defn_convex_optimization} as the number of decay modes $J$ increases. If the Koopman operator is non-compact, this will be reflected in the absence of a systematic decay of $C^*_J$ with increasing $J$. Thus, the method not only provides accurate approximations when compactness or quasi-compactness holds, but also offers a built-in mechanism for detecting when these properties fail.  

\section{Properties of the Resolvent operator}
\label{supp:resolvent_properties}

In the next proposition we collect various properties of the resolvent operator which are useful for the development of our method.

\begin{proposition}
\label{properties_of_the_resolvent_operator}
Let $\mathcal{R}_s f(x)$ be the resolvent operator defined by \eqref{defn_resolvent_operator} in the main text. Then we have the following:
\begin{itemize}
\item[(A)] The map $s \mapsto \mathcal{R}_s f(x)$ is complex analytic for $s \in \C_+$. 
\item[(B)] Final Value Theorem
$$  \lim_{s \to 0} \mathcal{R}_s f (x) =  \E_\pi(f).$$
\item[(C)] Iterates of the resolvent are given by
\begin{align*}
\mathcal{R}^{m}_s f (x) = \left\{ 
\begin{array}{cc}
f(x) & \quad \textnormal{if} \quad m = 0 \\
\frac{s^m}{(m-1)!} \int_0^\infty t^{m-1} e^{- s t} \mathcal{K}_{t} f(x)dt  & \quad \textnormal{for} \quad m = 1,2,\dots 
\end{array}\right.
\end{align*}
\item[(D)] Derivatives of the resolvent satisfy
\begin{align*}
\frac{\partial^m }{ \partial s^m}\mathcal{R}_s f (x) = \frac{(-1)^{m-1} m! }{s^m}  \left( \mathcal{R}^m_s f(x) -  \mathcal{R}^{m+1}_s f(x) \right),
\end{align*}
for any $m=1,2,\dots$.
\end{itemize}
\end{proposition}
\begin{proof}
Proofs of parts (A) and (B) are provided in Chapters 4 and 5 of \cite{engel2000one}. We note that our definition of the resolvent operator includes an additional multiplicative factor of \( s \), compared to the standard definition commonly found in the literature, including in~\cite{engel2000one}.

We now prove part (C). Due to the Markov property, the Koopman operator satisfies the semigroup property
\begin{align*}
\mathcal{K}_{t_1+t_2} f(x) =  \mathcal{K}_{ t_1}\left[ \mathcal{K}_{ t_2} f \right](x), \quad \textnormal{for any} \quad t_1,t_2 \geq 0.
\end{align*}
As the resolvent map is complex-analytic, it suffices to prove the relation in part (C) over the positive real line (i.e. $\mathbb{R}_+$), for it to hold over the whole positive half-plane $\mathbb{C}_+$. For a $s\in \mathbb{R}_+$, if we take $m=1$, then $\mathcal{R}^m_s f $ can be viewed as the expected value of the Koopman operator, applied on function $f$ at an independent random time $\tau_s$ which is exponentially distributed with rate $s$. Therefore for $m=2$, 
\begin{align*}
\mathcal{R}^{2}_s f (x) =\E \left( \mathcal{K}_{ \tau^{(1)}_s} \mathcal{R}_s f(x) \right) = \E \left( \mathcal{K}_{ \tau^{(1)}_s}\E \left( \mathcal{K}_{ \tau^{(2)}_s} f(x)\right)  \right), 
\end{align*}
where $\tau^{(1)}_s$ and $\tau^{(2)}_s$ are independent exponential random variables with rate $s$. Utilizing the independence of these two exponential random variables, and the semigroup property of the Koopman operator we can see that 
\begin{align*}
\mathcal{R}^{2}_s f (x) =\E \left( \mathcal{K}_{ \tau_s} f(x) \right), 
\end{align*}
where $\tau_s:= \tau^{(1)}_s + \tau^{(2)}_2$ is now a Gamma distributed random variable with rate parameter $s$ and shape parameter $m=2$, i.e.\ the p.d.f. of $\tau_s$ is given by
\begin{align*}
f_{ \tau_s} (t) = \frac{s^m}{\Gamma(m)} t^{m-1} e^{- s t} \qquad \textnormal{for} \qquad t \geq 0,
\end{align*}
where $\Gamma(m) = (m-1)!$ is the usual Gamma function. This can be generalized to any positive integer $m$ to obtain the formula in part (C) for $\mathcal{R}^{m}_s f (x)$.

We now prove part (D) by mathematical induction. Note that for
\begin{align*}
\frac{\partial}{\partial s} \mathcal{R}_s f(x) = \frac{\partial}{\partial s} \int_0^\infty s e^{-s t} \mathcal{K}_t f(x) = \int_0^\infty  e^{-s t} \mathcal{K}_t f(x) - s \int_0^\infty t e^{-s t} \mathcal{K}_t f(x).
\end{align*}
The first term of the right-hand side is simply $s^{-1} \mathcal{R}_s f(x)$, while the second term is $s^{-1} \mathcal{R}^2_s f(x)$ due to part (C). This proves that the relation in part (D) holds for $m=1$. We now assume that part (D) holds for $m$ and prove that it will continue to hold for $m+1$. For this, note that
\begin{align}
\label{iterative_derivative_proof1}
\frac{\partial^{m+1}}{\partial s^{m+1}} \mathcal{R}_s f(x) & = \frac{\partial}{\partial s} \left[ \frac{\partial^{m}}{\partial s^{m}} \mathcal{R}_s f(x) \right]  = (-1)^{m-1} \frac{\partial}{\partial s} \left[ \frac{ m! }{s^m}  \left( \mathcal{R}^m_s f(x) -  \mathcal{R}^{m+1}_s f(x) \right) \right]. 
\end{align}
However from part (C) we know that
\begin{align*}
\frac{m! }{s^m}  \left( \mathcal{R}^m_s f(x) -  \mathcal{R}^{m+1}_s f(x) \right) = m \int_0^\infty t^{m-1} e^{- s t} \mathcal{K}_{t} f(x)dt - s \int_0^\infty t^{m} e^{- s t} \mathcal{K}_{t} f(x)dt
\end{align*}
and hence, again using part (C)
\begin{align*}
 \frac{\partial}{\partial s} \left[ \frac{ m! }{s^m}  \left( \mathcal{R}^m_s f(x) -  \mathcal{R}^{m+1}_s f(x) \right) \right] &= 
-(m+1) \int_0^\infty t^{m} e^{- s t} \mathcal{K}_{t} f(x)dt
+ s\int_0^\infty t^{m+1} e^{- s t} \mathcal{K}_{t} f(x)dt \\
& = -\frac{(m+1)!}{s^{m+1}} \mathcal{R}^{m+1}_s f(x) 
+ \frac{(m+1)!}{s^{m+1}} \mathcal{R}^{m+2}_s f(x). 
\end{align*}
Plugging this in \eqref{iterative_derivative_proof1} proves (D) for $m+1$ and this completes the proof of this proposition.
\end{proof}

\section{Frequency Domain Characterization of the Exact Finite Representation of the Koopman operator}
\label{supp:sec_frequency_domain_characterization}
\begin{proof}[Proof of Theorem \ref{thm_1}]
Suppose that \eqref{spectral_expansion} holds exactly. Note that for any $\sigma \in \C_+$ and  $m=1,2,\dots$
\begin{align*}
\frac{s^m}{(m-1)!} \int_0^\infty t^{m-1} e^{- s t} e^{-\sigma t}dt = \frac{s^m}{(m-1)!} \int_0^\infty t^{m-1} e^{- (s +\sigma) t}dt = \left( \frac{s}{ s + \sigma} \right)^m.
\end{align*}
Therefore using the expression for the resolvent iterates given in Proposition \ref{properties_of_the_resolvent_operator}(C), we obtain
\begin{align*}
\mathcal{R}^{m}_s f(x) = \E_\pi(f) +  \sum_{j=1}^J \alpha_j(f,x) \left(\frac{s}{s+\sigma_j} \right)^m
\end{align*}
for $m=0,1,2,\dots$. Set $\sigma_0 = 0$ and $\alpha_0(f,x) = \E_\pi(f)$. Then we obtain the following system of $(J+1)$ linear equations
\begin{align}
\label{proof_linear_system1}
\mathcal{R}^{m}_s f(x) = \sum_{j=0}^J \alpha_j(f,x) \left(\frac{s}{s+\sigma_j} \right)^m, \quad \textnormal{for} \quad m=0,1,\dots,J.  
\end{align}
Let ${\bf V}$ be the $(J+1) \times (J+1)$ \emph{Vandermonde} matrix given by
\begin{align*}
\mathbf{V} = V \left( \frac{s}{s+\sigma_0},\dots, \frac{s}{s+\sigma_J} \right)
\end{align*}
where
\begin{align}
\label{supp_vandermonde_defnition}
V(x_0,\dots,x_{J}) = \begin{bmatrix}
1 & x_0 & x_0^2 & \cdots & x_0^{J} \\
1 & x_1 & x_1^2 & \cdots & x_1^{J} \\
1 & x_2 & x_2^2 & \cdots & x_2^{J} \\
\vdots & \vdots & \vdots & \ddots & \vdots \\
1 & x_J & x_J^2 & \cdots & x_J^{J}.
\end{bmatrix}
\end{align}
We can write the system of linear equations \eqref{proof_linear_system1} as
\begin{align*}
\mathcal{R}^{m}_s f(x) = \sum_{j=0}^J \alpha_j(f,x) \mathbf{V}_{jm}, \quad \textnormal{for} \quad m=0,1,\dots,J.  
\end{align*}
Let $\mathbf{R}(f,x)$ and $\boldsymbol{\alpha}(f,x) $ be $(J+1) \times 1$ column vectors whose $j$-th component is $\mathcal{R}^{j}_s f (x)$ and $\alpha_{j}(f,x)$ respectively. Then the last linear system can be written succinctly as
 \begin{align*}
\mathbf{R}(f,x)  & = {\bf V}^T \boldsymbol{\alpha}(f,x) ,
\end{align*}
which implies that
 \begin{align*}
\boldsymbol{\alpha}(f,x)  = \left( {\bf V}^T \right)^{-1}  \mathbf{R}(f,x).
\end{align*}
We know from interpolation theory that $\left( {\bf V}^T \right)^{-1}  = \mathbf{L}(s)$, where $\mathbf{L}(s)$ is the $(J+1) \times (J+1)$ \emph{Lagrange} matrix whose each row $\mathbf{L}_j(s)$ for $j=0,1,\dots, J$ consists of the coefficients of the Lagrange interpolation polynomial
\begin{align*}
\sum_{k=0}^J \mathbf{L}_{jk}(s) x^{k} = \prod_{k=0, k \neq j}^{J} \left( \frac{x -x_k}{x_j -x_k}\right)
\end{align*}
where each $x_k = s/(s+\sigma_k)$. This implies that each $\alpha_j(f,x)$ can be expressed as
\begin{align}
\label{formula_for_alpha_f_x}
\alpha_j(f,x) =  \sum_{k=0}^J \mathbf{L}_{jk}(s) \mathcal{R}^{k}_s f(x) = 
\prod_{k=0, k \neq j}^J \left( \frac{ \mathcal{R}_s - \frac{s}{s+\sigma_k} \mathbf{I} }{ \frac{s}{s+\sigma_j} - \frac{s}{s+\sigma_k} }  \right) f(x).
\end{align}
In particular, for $j=0$ we know that
\begin{align}
\label{thm_pf_reln_sigma_0}
\E_\pi(f) = \alpha_0(f,x) =  \prod_{k=1}^J \left( \frac{ \mathcal{R}_s - \frac{s}{s+\sigma_k} \mathbf{I} }{ 1 - \frac{s}{s+\sigma_k} }  \right) f(x)
\end{align}
Noting that
\begin{align*}
\frac{ \mathcal{R}_s - \frac{s}{s+\sigma_k} \mathbf{I} }{ 1 - \frac{s}{s+\sigma_k} } = \frac{ s (\mathcal{R}_s - \mathbf{I} )+\sigma_k \mathcal{R}_s }{ \sigma_k }
\end{align*}
proves that \eqref{resol_cond1} holds, thereby proving the ``if" part of the theorem.

We now prove the ``only if" part of the theorem which is more demanding. Define a function
\begin{align*}
G_s(x) = \prod_{j=1}^J \left( s( \mathcal{R}_s -\mathbf{I}) +\sigma_j \mathcal{R}_s \right)  f(x) -\E_\pi(f) \prod_{j=1}^J \sigma_j 
\end{align*}
which is complex analytic on $\C_+$ as a function of $s$. By condition \eqref{resol_cond1} we know that there exists a $s^* \in C_+$ such that $G_{s^*}(x) = 0$ for all $x\in \mathcal{E}$. Applying the product rule, the first order derivative of $G_s(x)$ with respect to $s$ is given by
\begin{align}
\label{G_derivative_expr}
\frac{\partial}{\partial s}G_s(x) = \sum_{j=1}^J \left[\prod_{k=1, k \neq j}^J \left( s( \mathcal{R}_s -\mathbf{I}) +\sigma_k \mathcal{R}_s \right)  \right] \left( \mathcal{R}_s - \mathbf{I} +(s + \sigma_j) \frac{\partial }{\partial s} \mathcal{R}_s \right) f(x).
\end{align}
However we know from Proposition \ref{properties_of_the_resolvent_operator}(D) that 
\begin{align*}
\frac{\partial  }{ \partial s }\mathcal{R}_s f (x) = \frac{1}{s} \left( \mathcal{R}_s - \mathcal{R}^2_s\right) f(x)  = -\frac{1}{s} \mathcal{R}_s ( \mathcal{R}_s - \mathbf{I})f(x).
\end{align*}
This shows that
\begin{align*}
\left( \mathcal{R}_s - \mathbf{I} +(s + \sigma_j) \frac{\partial }{\partial s} \mathcal{R}_s \right) f(x) &= \left( \mathcal{R}_s - \mathbf{I}  - \left( \frac{s + \sigma_j}{s} \right) \mathcal{R}_s ( \mathcal{R}_s - \mathbf{I}) \right) f(x) \\
& =  \left(   \left( \frac{s + \sigma_j}{s} \right) \mathcal{R}_s - \mathbf{I} \right) (  \mathbf{I} - \mathcal{R}_s) f(x) \\
& = \frac{1}{s} \left( s (\mathcal{R}_s - \mathbf{I}) + \sigma_j \mathcal{R}_s \right) (  \mathbf{I} - \mathcal{R}_s) f(x).
\end{align*}
Plugging this in \eqref{G_derivative_expr} we get
\begin{align*}
\frac{\partial}{\partial s}G_s(x) & = \sum_{j=1}^J \left[\prod_{k=1, k \neq j}^J \left( s( \mathcal{R}_s -\mathbf{I}) +\sigma_k \mathcal{R}_s \right)  \right] \frac{1}{s} \left( s (\mathcal{R}_s - \mathbf{I}) + \sigma_j \mathcal{R}_s \right) (  \mathbf{I} - \mathcal{R}_s) f(x) \\
& = \sum_{j=1}^J \left[\prod_{k=1}^J \left( s( \mathcal{R}_s -\mathbf{I}) +\sigma_k \mathcal{R}_s \right)  \right] \frac{1}{s}  (  \mathbf{I} - \mathcal{R}_s) f(x) \\
& = \frac{J}{s} (  \mathbf{I} - \mathcal{R}_s)\left[\prod_{k=1}^J \left( s( \mathcal{R}_s -\mathbf{I}) +\sigma_k \mathcal{R}_s \right)  \right]f(x) \\
& =  \frac{J}{s} (  \mathbf{I} - \mathcal{R}_s) \left( G_s(x) + \E_\pi(f) \prod_{j=1}^J \sigma_j  \right).
\end{align*}
However since $ \E_\pi(f) \prod_{j=1}^J \sigma_j $ is a constant, its image under the operator $(  \mathbf{I} - \mathcal{R}_s)$ is the zero function. Hence we obtain the following expression for the first order derivative of $G_s(x)$
\begin{align}
\label{g_der_first_order_expression}
\frac{\partial}{\partial s}G_s(x)  = \frac{J}{s} (  \mathbf{I} - \mathcal{R}_s) G_s(x),
\end{align}
which shows that the first order derivative of $G_s(x)$ is expressible in terms of $G_s(x)$. Therefore if $G_{s^*}(x) =0$ for all $x\in \mathcal{E}$ then it implies that the first order derivative
$$\frac{\partial}{\partial s}G_s(x) \Big \vert_{s = s^*} = 0 \quad \textnormal{for all} \quad x\in \mathcal{E}.$$
This argument can be extended to show that all the high-order derivatives are identically zero as well, which then implies by the Identity Theorem for complex-analytic functions \cite{stein2010complex} that for all $s \in \C_+$ we have $G_s(x) = 0$ for all $x\in \mathcal{E}$. This proves that if \eqref{resol_cond1} holds for one $s \in \C_+$ then it will hold for each $s \in  \C_+$. 

Now let us fix some $s\in \C_+$ for each $j=1,\dots,J$ suppose $\alpha_j(f,x,s)$ is defined by the right-hand side of \eqref{coeff_alphas} (with $\sigma_0 = 0$ as before). First thing to note that $\alpha_j(f,x,s)$ can also be equivalently defined by the right-hand side of \eqref{formula_for_alpha_f_x}. This is because
\begin{align*}
\prod_{k=0, k \neq j}^J \left( \frac{ \mathcal{R}_s - \frac{s}{s+\sigma_k} \mathbf{I} }{ \frac{s}{s+\sigma_j} - \frac{s}{s+\sigma_k} }  \right) f(x) & =
\left[ \prod_{k=1, k \neq j}^J \left( \frac{ \mathcal{R}_s - \frac{s}{s+\sigma_k} \mathbf{I} }{ \frac{s}{s+\sigma_j} - \frac{s}{s+\sigma_k} }  \right) \right] \left( \frac{\mathcal{R}_s  - \mathbf{I}}{\frac{s}{s+\sigma_j} - 1} \right)f (x) \\
&= \left( \frac{s+\sigma_j}{\sigma_j} \right) \left[ \prod_{k=1, k \neq j}^J \left( \frac{ \mathcal{R}_s - \frac{s}{s+\sigma_k} \mathbf{I} }{ \frac{s}{s+\sigma_j} - \frac{s}{s+\sigma_k} }  \right) \right] (\mathbf{I} -\mathcal{R}_s )f(x) \\
& = \left( \frac{s+\sigma_j}{\sigma_j} \right) \sum_{k=1}^J \mathbf{L}_{jk}(s) \mathcal{R}^{k-1}_s (\mathbf{I} -\mathcal{R}_s )f(x) \\
&  = \left( \frac{s+\sigma_j}{\sigma_j} \right) \sum_{k=1}^J \mathbf{L}_{jk}(s) (\mathcal{R}^{k-1}_s  -\mathcal{R}^k_s )f(x) \\
& = \alpha_j(f,x,s)
\end{align*}
where $\mathbf{L(s)}$ refers in this calculation to the $J \times J$ Lagrange interpolation matrix satisfying \eqref{lagrange_interpolation_poly} with interpolation points $x_j=s/(s+\sigma_j)$ for $j=1,\dots, J$.

Note that using the formula \eqref{formula_for_alpha_f_x} we see that 
\begin{align*}
\mathcal{R}_s \alpha_j(f,x,s)  & = \prod_{k=0, k \neq j}^J \left( \frac{ \mathcal{R}_s - \frac{s}{s+\sigma_k} \mathbf{I} }{ \frac{s}{s+\sigma_j} - \frac{s}{s+\sigma_k} }  \right) \mathcal{R}_s f(x) \\
& = \prod_{k=0, k \neq j}^J \left( \frac{ \mathcal{R}_s - \frac{s}{s+\sigma_k} \mathbf{I} }{ \frac{s}{s+\sigma_j} - \frac{s}{s+\sigma_k} }  \right) \left(  \mathcal{R}_s  -  \frac{s}{s+\sigma_j} \mathbf{I} + \frac{s}{s+\sigma_j} \mathbf{I} \right) f(x) \\
& = \frac{ \left( \mathcal{R}_s - \mathbf{I}\right)\prod_{k=1}^J  \left( \mathcal{R}_s - \frac{s}{s+\sigma_k} \mathbf{I} \right) f(x)}{\prod_{k=0, k \neq j}^J  \left( \frac{s}{s+\sigma_j} - \frac{s}{s+\sigma_k} \right) }   +   \frac{s}{s+\sigma_j} \alpha_j(f,x, s).
\end{align*}
Since $G_s(x) = 0$ for all $x\in \mathcal{E}$, the map $x\mapsto \prod_{k=1}^J  \left( \mathcal{R}_s - \frac{s}{s+\sigma_k} \mathbf{I} \right) f(x)$ is a constant and hence its image under the operator $\left( \mathcal{R}_s - \mathbf{I}\right)$ is just $\mathbf{0}$, showing that $\alpha_j(f,x,s)$ is in fact an eigenfunction for the resolvent operator $\mathcal{R}_s$ corresponding to the eigenvalue $s/(s+\sigma_j)$.

The argument we made above can be extended to prove that each $\alpha_j(f,x,s)$, defined by the right-hand side of \eqref{coeff_alphas}, does not depend on $s$, under condition \eqref{resol_cond1}. To see this note that from the right-hand side of expression \eqref{formula_for_alpha_f_x}, each $\alpha_j(f,x,s)$ can be expressed as
\begin{align*}
\alpha_j(f,x,s) = c_j \left( \frac{s+\sigma_j}{s} \right)^J
h_j(s,x) .
\end{align*}
where $c_j = \left(\prod_{k=0, k \neq j}^J (\sigma_k - \sigma_j) \right)^{-1} $ is a constant and 
\begin{align*}
h_j(s,x) = \prod_{k=0, k \neq j}^J \left( s( \mathcal{R}_s - \mathbf{I}) + \sigma_k \mathcal{R}_s  \right) f(x).
\end{align*}
We see from a calculation very similar to the one we did above that
\begin{align*}
\frac{\partial}{\partial s} h_j(s,x) = \frac{J}{s}  \left( \mathbf{I} - \mathcal{R}_s \right) h_j(s,x),
\end{align*}
and hence using the product rule we obtain
\begin{align*}
\frac{\partial}{\partial s} \alpha_j(f,x,s) &=- c_j \left( \frac{s+\sigma_j}{s} \right)^J \frac{J}{s} \left( \frac{\sigma_j}{s+\sigma_j} \right) h_j(s,x) + c_j \left( \frac{s+\sigma_j}{s} \right)^J
\frac{J}{s}  \left( \mathbf{I} - \mathcal{R}_s \right) h_j(s,x) \\
& = -\frac{J}{s} \left( \frac{\sigma_j}{s+\sigma_j} \mathbf{I} -\mathbf{I} + \mathcal{R}_s  \right) \alpha_j(f, x, s) \\
& = - \frac{J}{s} \left( \mathcal{R}_s - \frac{s}{s+\sigma_j} \mathbf{I}   \right) \alpha_j(f,x,s) \\
& = 0
\end{align*}
because $\alpha_j(f,x,s)$ is an eigenfunction for $\mathcal{R}_s$ corresponding to the eigenvalue $s/(s+\sigma_j)$, and so
\begin{align}
\label{proof_eigv_reln_alpha_j_x_s}
\left( \mathcal{R}_s - \frac{s}{s+\sigma_j} \mathbf{I}   \right) \alpha_j(f,x,s) = 0 \quad \textnormal{for all} \quad x \in \mathcal{E}.
\end{align}
This argument can be extended to show that all higher-order derivatives of $\alpha_j(f,x,s)$ with respect to $s$ are identically $0$ which proves that for each $j=1,\dots,J$, the map $s \mapsto \alpha_j(f,x,s)$ is a constant. Henceforth, we drop the frequency argument $s$, and write each $\alpha_j(f,x,s)$ as $\alpha_j(f,x)$. Now \eqref{proof_eigv_reln_alpha_j_x_s} proves \eqref{eigenfunction_reln_phi_j}.

Observe that if we sum the rows of the $J \times J$ Lagrange matrix $\mathbf{L}_{jk}(s)$ in expression \eqref{coeff_alphas}, then we obtain a $1\times J$ vector whose first component $1$ and the rest are all zeros. Therefore from \eqref{coeff_alphas} we get
\begin{align*}
\sum_{j=1}^J \left( \frac{\sigma_j}{s+\sigma_j} \right) \alpha_j(f,x) = \sum_{k=1}^J \sum_{j=1}^J  \mathbf{L}_{jk} (s) \left(  \mathcal{R}^{k-1}_s -   \mathcal{R}^{k}_s \right) f(x) = \left(  \mathbf{I} -   \mathcal{R}_s \right) f(x),
\end{align*}
which upon rearranging gives us
\begin{align*}
\mathcal{R}_s f(x) = f(x) - \sum_{j=1}^J \alpha_j(f,x) + \sum_{j=1}^J \left( \frac{s}{s+\sigma_j} \right) \alpha_j(f,x).
\end{align*}
From the Final Value Theorem (see Proposition \ref{properties_of_the_resolvent_operator}) we know that $\lim_{s \to 0} \mathcal{R}_s f(x) = \E_\pi(f)$ and hence 
\begin{align*}
f(x) - \sum_{j=1}^J \alpha_j(f,x) =\E_\pi(f) \quad \textnormal{for all} \quad x \in \mathcal{E}.
\end{align*}
Therefore the last expression for $\mathcal{R}_s f(x)$ simplifies to
\begin{align*}
\mathcal{R}_s f(x) = \E_\pi(f)+ \sum_{j=1}^J \left( \frac{s}{s+\sigma_j} \right) \alpha_j(f,x),
\end{align*}
which proves that \eqref{spectral_expansion} holds exactly for all $x\in \mathcal{E}$ and all $t \geq 0$. This completes the proof of this theorem.
\end{proof}

\section{Supporting Lemmas}
\label{supp:supporting_lemmas}

\begin{lemma}
\label{supp_lemm:error_function_form}
Keeping the notation as in Section \ref{decay_mode_identification} of the main text, the ``error" $E_{f,J}(x,s)$ defined by \eqref{decay_mode_Estimation_error_fn1} can be expressed as \eqref{decay_mode_Estimation_error_fn2}. 
\end{lemma}
\begin{proof}
Recall from Section \ref{decay_mode_identification} that $\beta_1,\dots, \beta_J$ are the coefficients of the polynomial
\begin{align}
\label{supp:beta_poly}
1 + \sum_{j=1}^J \beta_j s^j = \frac{\prod_{j=1}^{J} (s + \sigma_j)}{ \prod_{j=1}^J \sigma_j}. 
\end{align}
Let $\mathcal{R}^{-1}_s$ denote the inverse of the resolvent operator and let $\mathcal{R}^{-j}_s$ denote the $j$-th iterate of this inverse. Then, using \eqref{supp:beta_poly}, we can write $E_{f,J}(x,s)$ as
\begin{align*}
E_{f,J}(x,s)& =\mathcal{R}^J_s \left( \frac{ \prod_{j=1}^J \left( s( \mathcal{R}_s -\mathbf{I})\mathcal{R}^{-1}_s  +\sigma_j \right) }{ \prod_{j=1}^J \sigma_j}\right) f(x) - \E_\pi(f) \\
& = \mathcal{R}^J_s \left(\mathbf{I} + \sum_{j=1}^J \beta_j s^j ( \mathcal{R}_s -\mathbf{I})^j\mathcal{R}^{-j}_s \right) f(x) - \E_\pi(f) \\
& = \mathcal{R}^J_s f(x) + \sum_{j=1}^J s^j ( \mathcal{R}_s -\mathbf{I})^j\mathcal{R}^{J-j}_s f(x) - \E_\pi(f).
\end{align*}
Applying the binomial expansion, we see that $( \mathcal{R}_s -\mathbf{I})^j\mathcal{R}^{J-j}_s f(x)$ is exactly the coefficient defined by \eqref{error_fn_coefficient_defn} in the main text. This completes the proof of this lemma.
\end{proof}

\begin{lemma}
\label{supp_lem:linear_coeff}
Keeping the same notation as in Section \ref{num_coeff_identification} of the main text, the first $J-1$ derivatives of $\mathcal{R}_s f(x)$ and $\bar{\mathcal{R}}_s f(x)$ coincide at $s = \bar{s}$ 
\begin{align}
\label{supp:derivatives_match_condition}
\frac{\partial^m}{\partial s^m} \mathcal{R}_s f(x) \Big|_{s = \bar{s}} 
= \frac{\partial^m}{\partial s^m} \bar{\mathcal{R}}_s f(x) \Big|_{s = \bar{s}}, 
\quad \text{for each } m = 1, \dots, J-1.
\end{align}
Furthermore, we have the relation
\begin{align}
\label{supp:resolvent_error_relation}
\mathcal{E}_{\bar{s}} f(x):=\mathcal{R}_{\bar{s}} f(x) - \bar{\mathcal{R}}_{\bar{s}} f(x) 
= f(x) - \mathbb{E}_\pi(f) - \sum_{j=1}^J \bar{\alpha}_j(f,x),
\end{align}
where $\bar{\alpha}_1(f,x), \dots, \bar{\alpha}_J(f,x)$ are the approximate linear coefficients estimated via \eqref{coeff_alphas_2}. Moreover, if $C^*_J$ is the optimal value of the optimization problem 
\eqref{defn_convex_optimization} in the main text, and if $\bar{s} \in \mathbb{S}$ as well as $f \in \mathcal{F}$, 
then we must have
\begin{align}
\label{lemma_reln_optimal_cost_error}
\frac{\| \mathcal{E}_{\bar{s}} f(\cdot)\|_{\mathcal{L}_2(\pi)} }{ \|f\|_{\mathcal{L}_2(\pi)} } 
\leq C^*_J.
\end{align}
\end{lemma}
\begin{proof}
Note that the approximate resolvent $\bar{\mathcal{R}}_s f(x)$ is defined by \eqref{defn_resolvent_operator_approximate} in the main text. Noting that
\begin{align*}
\frac{\partial^m}{ \partial s^m} \left( \frac{s}{s+\sigma}\right) = (-1)^{m-1} m! \frac{\sigma}{(s+\sigma)^{m+1}},
\end{align*}
and using \eqref{coeff_alphas_2}, we see that for each $m = 1, \dots, J-1$
\begin{align}
\label{supp_deriv_match_proof}
\frac{\partial^m}{\partial s^m} \bar{\mathcal{R}}_s f(x) \Big|_{s = \bar{s}} & =(-1)^{m-1} m! \sum_{j=1}^J \left( \frac{ \bar{\alpha}_j(f,x) \bar{\sigma}_j}{ \bar{s} + \bar{\sigma}_j}\right) \frac{1}{(\bar{s}+\bar{\sigma}_j)^{m}} \notag \\
& = (-1)^{m-1} m! \sum_{j=1}^J 
 \sum_{k=1}^J \frac{1}{(\bar{s}+\bar{\sigma}_j)^{m}} \mathbf{L}_{jk} (\bar{s}) \left(  \mathcal{R}^{k-1}_{\bar{s}} -   \mathcal{R}^{k}_{\bar{s}} \right) f(x) \notag \\
 & = \frac{(-1)^{m-1} m! }{\bar{s}^m}
 \sum_{k=1}^J \left[ \sum_{j=1}^J \left( \frac{\bar{s}}{\bar{s}+\bar{\sigma}_j} \right)^m \mathbf{L}_{jk} (\bar{s}) \right] \left(  \mathcal{R}^{k-1}_{\bar{s}} -   \mathcal{R}^{k}_{\bar{s}} \right) f(x).
\end{align}
If $\mathbf{V}$ is the $J \times J$ Vandermonde matrix (see \eqref{supp_vandermonde_defnition})
\begin{align*}
\mathbf{V} = V \left( \frac{\bar{s}}{\bar{s}+\bar{\sigma}_1},\dots, \frac{\bar{s}}{\bar{s}+\bar{\sigma}_J} \right)
\end{align*}
then we know that $\mathbf{V}^T \mathbf{L}(\bar{s}) = \mathbf{I}$ (the $J \times J$ identity matrix). Therefore 
\begin{align*}
\sum_{j=1}^J \left( \frac{\bar{s}}{\bar{s}+\bar{\sigma}_j} \right)^m \mathbf{L}_{jk} (\bar{s}) = \sum_{j=1}^J \mathbf{V}^T_{(m+1) j} \mathbf{L}_{jk} (\bar{s}) = \delta_{(m+1)k} 
\end{align*}
where $\delta_{ij}$ is the Kronecker delta function. Plugging this in \eqref{supp_deriv_match_proof} we obtain
\begin{align*}
\frac{\partial^m}{\partial s^m} \bar{\mathcal{R}}_s f(x) \Big|_{s = \bar{s}} = \frac{(-1)^{m-1} m!}{\bar{s}^m}
 \left(  \mathcal{R}^{m}_{\bar{s}} -   \mathcal{R}^{m+1}_{\bar{s}} \right) f(x),
\end{align*}
which proves \eqref{supp:derivatives_match_condition} due to Proposition \ref{properties_of_the_resolvent_operator}(D).

Similarly, one can see that 
\begin{align}
\label{suppl_s11_proof_1}
\sum_{j=1}^J \left( \frac{ \bar{\alpha}_j(f,x) \bar{\sigma}_j}{ \bar{s} + \bar{\sigma}_j}\right) & = 
 \sum_{k=1}^J \sum_{j=1}^J  \mathbf{L}_{jk} (\bar{s}) \left(  \mathcal{R}^{k-1}_{\bar{s}} -   \mathcal{R}^{k}_{\bar{s}} \right) f(x) \notag \\
 & =  \sum_{k=1}^J \sum_{j=1}^J \mathbf{V}^T_{1j} \mathbf{L}_{jk} (\bar{s}) \left(  \mathcal{R}^{k-1}_{\bar{s}} -   \mathcal{R}^{k}_{\bar{s}} \right) f(x) \notag \\
 & = \sum_{k=1}^J \delta_{1k} \left(  \mathcal{R}^{k-1}_{\bar{s}} -   \mathcal{R}^{k}_{\bar{s}} \right) f(x) \notag \\
 & = \left(  \mathbf{I} -   \mathcal{R}_{\bar{s}} \right)  f(x).
\end{align}
However 
\begin{align*}
\sum_{j=1}^J \left( \frac{ \bar{\alpha}_j(f,x) \bar{\sigma}_j}{ \bar{s} + \bar{\sigma}_j}\right) &= \sum_{j=1}^J  \bar{\alpha}_j(f,x) - \sum_{j=1}^J  \bar{\alpha}_j(f,x) \left( \frac{ s}{ \bar{s} + \bar{\sigma}_j}\right) \\
& = \sum_{j=1}^J  \bar{\alpha}_j(f,x) +\E_\pi(f) - \bar{\mathcal{R}}_{\bar{s}} f(x). 
\end{align*}
Equating this expression with \eqref{suppl_s11_proof_1} proves \eqref{supp:resolvent_error_relation}.

Finally, to prove \eqref{lemma_reln_optimal_cost_error} it suffices to show that $\mathcal{E}_{\bar{s}} f(x)$ is identical to $E_{f,J}(x,\bar{s})$ defined by \eqref{decay_mode_Estimation_error_fn1} in the main text, i.e.
\begin{align}
\label{decay_mode_Estimation_error_fn1_supp}
\mathcal{E}_{\bar{s}} f(x) =\left( \frac{ \prod_{j=1}^J \left( \bar{s}( \mathcal{R}_{\bar{s}} -\mathbf{I}) +\bar{\sigma}_j \mathcal{R}_{\bar{s}} \right) }{ \prod_{j=1}^J \bar{\sigma}_j}\right) f(x) - \E_\pi(f).
\end{align}
As shown in the proof of Theorem \ref{thm_1}, letting $\bar{\sigma}_0 = 0$, we can express each $\bar{\alpha}_j(f,x)$ as
\begin{align}
\label{supp_lem_proof_alternative_alpha_j}
\bar{\alpha}_j(f,x)  = \sum_{k=0}^J \mathbf{L}_{jk}(\bar{s}) \mathcal{R}^{k}_{\bar{s}} f(x) =
\prod_{k=0, k \neq j}^J \left( \frac{ \mathcal{R}_{\bar{s}} - \frac{\bar{s}}{\bar{s}+\bar{\sigma}_k} \mathbf{I} }{ \frac{\bar{s}}{\bar{s}+\bar{\sigma}_j} - \frac{\bar{s}}{\bar{s}+\bar{\sigma}_k} }  \right) f(x),\quad \textnormal{for} \quad j=0,1,\dots,J,
\end{align}
where $\mathbf{L}(s)$ is the $(J+1) \times (J+1)$ \emph{Lagrange} matrix whose each row $\mathbf{L}_j(s)$ for $j=0,1,\dots, J$ consists of the coefficients of the Lagrange interpolation polynomial
\begin{align*}
\sum_{k=0}^J \mathbf{L}_{jk}(s) x^{k} = \prod_{k=0, k \neq j}^{J} \left( \frac{x -x_k}{x_j -x_k}\right)
\end{align*}
with $x_k = s/(s+\sigma_k)$. Since the rows of matrix $\mathbf{L}(s)$ sum up to a vector whose first component is $1$ and the rest are all zeros we have
\begin{align*}
\bar{\alpha}_0(f,x) + \sum_{j=1}^J \bar{\alpha}_j(f,x)  = \sum_{j=0}^J \bar{\alpha}_j(f,x) = \sum_{k=0}^J \sum_{j=0}^J \mathbf{L}_{jk}(\bar{s}) \mathcal{R}^{k}_{\bar{s}} f(x) = \mathcal{R}^{0}_{\bar{s}} f(x) =f(x),
\end{align*}
and hence
\begin{align*}
\bar{\alpha}_0(f,x) = f(x) - \sum_{j=1}^J \bar{\alpha}_j(f,x).
\end{align*}
Equation this with $\bar{\alpha}_0(f,x)$ we get from \eqref{supp_lem_proof_alternative_alpha_j} we obtain
\begin{align*}
 f(x) - \sum_{j=1}^J \bar{\alpha}_j(f,x) = \bar{\alpha}_0(f,x) & = \prod_{k=1}^J \left( \frac{ \mathcal{R}_{\bar{s}} - \frac{\bar{s}}{\bar{s}+\bar{\sigma}_k} \mathbf{I} }{ 1 - \frac{\bar{s}}{\bar{s}+\bar{\sigma}_k} }  \right) f(x) \\
 &= \left( \frac{ \prod_{j=1}^J \left( \bar{s}( \mathcal{R}_{\bar{s}} -\mathbf{I}) +\bar{\sigma}_j \mathcal{R}_{\bar{s}} \right) }{ \prod_{j=1}^J \bar{\sigma}_j}\right) f(x),
\end{align*}
which shows that
\begin{align*}
\mathcal{E}_{\bar{s}} f(x) = f(x) - \mathbb{E}_\pi(f) - \sum_{j=1}^J \bar{\alpha}_j(f,x) = \left( \frac{ \prod_{j=1}^J \left( \bar{s}( \mathcal{R}_{\bar{s}} -\mathbf{I}) +\bar{\sigma}_j \mathcal{R}_{\bar{s}} \right) }{ \prod_{j=1}^J \bar{\sigma}_j}\right) f(x) - \mathbb{E}_\pi(f).
\end{align*}
This proves \eqref{decay_mode_Estimation_error_fn1_supp} and completes the proof of this lemma.
\end{proof}

\section{Monte Carlo Estimators}
\label{supp_sec:monte_carlo_estimators}

In this section we present the Monte Carlo (MC) estimators for the quantities required by the SKA method. All these estimators rely on simulating CTMC trajectories of an SRN, which can be generated using Gillespie's Stochastic Simulation Algorithm (SSA) \cite{gillespie1977exact} or its variants. For example, if $(X^{(1)}_x(t))_{t\geq 0},\dots,(X^{(N)}_x(t))_{t\geq 0}$ denote $N$ such trajectories starting from an initial state $x \in \mathcal{E}$, then a simple MC estimator for the Koopman operator is given by
\begin{align}
\label{mc_koop_1}
\hat{\mathcal{K}_t}f(x) = \frac{1}{N} \sum_{n=1}^N f(X^{(n)}_x(t)) 
\quad \textnormal{for each} \quad t \geq 0 \quad \textnormal{and} \quad f \in \mathcal{F}.
\end{align}
While this MC estimator is straightforward to implement, it suffers from several drawbacks discussed in Section \ref{sec:intro}. In particular, its computational cost grows linearly with $N$, whereas its standard error decreases only at the rate $1/\sqrt{N}$. Moreover, it yields estimates of the Koopman operator only at the finitely many time points where \eqref{mc_koop_1} is evaluated; it does not provide an analytic mapping $t \mapsto \mathcal{K}_t f(x)$ valid for all $t \in [0,\infty)$.The most significant drawback, however, is that for each new initial state $x$, the MC estimator relies exclusively on a fresh batch of CTMC trajectories generated with that initial state, while the estimations obtained for previous initial states provide no useful information. This renders the estimation of $\mathcal{K}_t f(x)$ over a range of initial states computationally infeasible. Motivated by these issues, we develop our SKA method that approximately analytically constructs the map $t \mapsto \mathcal{K}_t f(x)$, by applying the spectral expansion \eqref{spectral_expansion}. In comparison to the MC estimator, our method is more efficient is providing estimates of this map for several initial states.

Note that since our SRN is ergodic, for any \( f \in \mathcal{F} \), its expectation under the stationary distribution \( \pi \) coincides with the long-term time average of the stochastic process \( (f(X_x(t)))_{t \geq 0} \), irrespective of the initial state \( x \) (see~\cite{gupta2014scalable}), i.e.,
\begin{align*}
\E_\pi(f) = \lim_{T \to \infty} \frac{1}{T} \int_0^T f(X_x(t))dt \quad \textnormal{almost surely}.
\end{align*}
Taking expectations both sides and noting that $\E(f(X_x(t))) = \mathcal{K}_t f(x)$ we get.
\begin{align}
\label{stationary_expectation_expected_formula}
\E_\pi(f) = \lim_{T \to \infty} \frac{1}{T} \int_0^T \mathcal{K}_t f(x)dt.
\end{align}

We first develop Monte Carlo (MC) estimators for the quantities required by \textrm{SKA} to estimate both the state-independent decay modes and the stationary expectations (see Section~\ref{sec:est_state_independent} in the main text). Let \( \mathcal{X} = \{x_1, \dots, x_{n_c}\} \) denote the support of the approximate stationary distribution \( \hat{\pi} \), defined in \eqref{defn_approx_pi} of the main text. For each initial state \( x \in \mathcal{X} \), we generate \( N \) independent CTMC trajectories \( (X^{(1)}_x(t))_{t \geq 0}, \dots, (X^{(N)}_x(t))_{t \geq 0} \) over the time interval \([0, T]\), which we assume is sufficiently large for the process to effectively reach the stationary regime. Hence from \eqref{stationary_expectation_expected_formula}, we see that an accurate MC estimator for the stationary expectation is given by
\begin{align}
\label{ergodic_estimator}
\hat{\E}_\pi(f) =  \frac{1}{n_c T N} \sum_{x \in \mathcal{X}} \sum_{n=1}^N \int_0^T f(X^{(n)}_x(t))dt.
\end{align}
Using the same set of trajectories, we can construct an MC estimator for the iterated resolvent \( \mathcal{R}^m_s f(x) \) (as defined in~\eqref{resolvent_iterates_formula}) for any positive integer \( m \). We shall truncate the infinite time integer to a finite time integral over $[0,T]$, and assume that $\mathcal{K}_t f(x) \approx \E_\pi(f)$ for $t \geq T$. Noting that
\begin{align*}
\frac{s^m}{(m-1)!} \int_{a}^b t^{m-1} e^{-s t }dt = e^{-sa}\sum_{k=0}^{m-1}\frac{(sa)^k}{k!}
- e^{-sb}\sum_{k=0}^{m-1}\frac{(sb)^k}{k!},
\end{align*}
we obtain the following MC estimator
\begin{align}
\label{mc_estimatir_iter_resolvent1}
\hat{\mathcal{R}}^m_s f(x) = \frac{s^m}{(m-1)!} \frac{1} {N} \sum_{n=1}^N \int_0^T t^{m-1} e^{-s t } f(X^{(n)}_x(t))dt + e^{-s T} \sum_{k=0}^{m-1}\frac{(sT)^k}{k!}\E_\pi(f).
\end{align}

Observe that since the CTMC \( (X_x(t))_{t \geq 0} \) is a pure jump process, it remains constant between jump times. This allows us to evaluate the time integral in~\eqref{ergodic_estimator} \emph{exactly} as
\begin{align*}
\int_0^T f(X_x(t))dt = f(x)\tau_1 + \sum_{r=1}^{R-1} f(X_x(\tau_r)) ( \tau_{r+1} -\tau_r) + f(X_x(\tau_R)) ( T -\tau_R),
\end{align*}
where \( \tau_r \) is the \( r \)-th jump time of the process \( (X_x(t))_{t \geq 0} \), and \( R \) denotes the last jump time before time \( T \) (i.e., \( \tau_R < T \) but \( \tau_{R+1} \geq T \)). In the same way, the time integral in~\eqref{mc_estimatir_iter_resolvent1}, when multiplied by the constant \( \frac{s^m}{(m-1)!} \), can also be evaluated \emph{exactly} as
\begin{align}
\label{mc_est_iter_resolvent_exact_time_integral}
\frac{s^m}{(m-1)!} \int_0^T t^{m-1} e^{-s t } f(X_x(t))dt &= f(x)\left(1 -  e^{-s \tau_1} \sum_{k=0}^{m-1} \frac{(s\tau_1)^k}{k!}\right) \\
&\quad + \sum_{r=1}^{R-1} f(X_x(\tau_r)) \left( e^{-s \tau_r}\sum_{k=0}^{m-1}\frac{(s\tau_r)^k}{k!}
- e^{-s\tau_{r+1}}\sum_{k=0}^{m-1}\frac{(s\tau_{r+1})^k}{k!} \right) \notag \\
&\quad +  f(X_x(\tau_R)) \left( e^{-s \tau_R}\sum_{k=0}^{m-1}\frac{(s\tau_R)^k}{k!}
- e^{-sT }\sum_{k=0}^{m-1}\frac{(sT)^k}{k!} \right). \notag
\end{align}
The simulations required for the above MC estimators can be computationally demanding, as they must be generated for multiple initial states, and with a large number of trajectories $N$ to ensure that the MC estimators have sufficiently low variance for reliable estimation of decay modes. To address this, we recommend leveraging the inherent parallelism of the estimation procedure and executing the simulation and integration steps on a GPU. Our implementation provides GPU code to facilitate efficient computation.

We now develop MC estimators for differences in resolvent iterates, i.e.\ $\mathcal{R}_{s}^{m-1} f(x) - \mathcal{R}_{s}^{m} f(x) $ that are needed for the estimation of linear coefficients in \eqref{approx_koopman_operator1} for any given initial state $x$ (see Section \ref{subsec:koopman_contruction} in the main text). From \eqref{mc_estimatir_iter_resolvent1}, it is immediate that a suitable MC estimator for this difference is given by
\begin{align}
\label{MC_estimator_resolvent_diff}
\hat{\mathcal{R}}_{s}^{m-1} f(x) - \hat{\mathcal{R}}_{s}^{m} f(x) = \begin{cases}
f(x) -  \frac{s}{N} \sum_{n=1}^N \int_0^T e^{- s t} f(X^{(n)}_x(t)) dt - e^{-s T} \E_\pi(f), & m = 1, \\
\frac{1}{N} \sum_{n=1}^N \int_0^T w_m(s, t) e^{- s t} f(X^{(n)}_x(t))dt - e^{-s T} \frac{(sT)^{m-1}}{(m-1)!} \E_\pi(f), & m \geq 2,
\end{cases} 
\end{align}
where $(X^{(1)}_x(t))_{t \geq 0}, \dots, (X^{(N)}_x(t))_{t \geq 0}$ are $N$ independent CTMC trajectories over the time interval $[0, T]$ and the function $w_m(s,t)$ is given by
\begin{align*}
w_m(s,t) = \left( \frac{s^{m-1}}{(m-2)!}  t^{m-2} - \frac{s^{m}}{(m-1)!} t^{m-1}\right).
\end{align*}
For the case $m=1$ the time integral in \eqref{MC_estimator_resolvent_diff} can be evaluated using \eqref{mc_est_iter_resolvent_exact_time_integral}, while for $m\geq 2$ the time integral in \eqref{MC_estimator_resolvent_diff} can be evaluated as
\begin{align*}
\int_0^T w_m(s,t) e^{- s t} f(X_x(t))dt & =f(x) e^{-s \tau_1} \frac{(s\tau_1)^{m-1}}{(m-1)!} \\
&\quad + \sum_{r=1}^{R-1} f(X_x(\tau_r)) \left( e^{-s\tau_{r+1}}\frac{(s\tau_{r+1})^{m-1}}{(m-1)!}  - e^{-s \tau_r}\frac{(s\tau_r)^{m-1}}{(m-1)!}
\right)  \\
&\quad +  f(X_x(\tau_R)) \left( 
 e^{-sT } \frac{(sT)^{m-1}}{(m-1)!} - e^{-s \tau_R} \frac{(s\tau_R)^{m-1}}{(m-1)!}\right).
\end{align*}

Recall from Section~\ref{subsec:estimation_sens_spectral} of the main text that, in order to estimate parameter sensitivities, it is necessary to compute the vector \(\Delta_k \mathbf{R}_{\mathbb{S}}(x)\), whose components are given by
\begin{align*}
D_{m,k}(x) 
&= \Delta_k \Bigl(\mathcal{R}_{s}^{m-1} f(x) - \mathcal{R}_{s}^{m} f(x)\Bigr)  \\
&= \mathcal{R}_{s}^{m-1} f(x+\zeta_k) - \mathcal{R}_{s}^{m} f(x+\zeta_k) - \mathcal{R}_{s}^{m-1} f(x) + \mathcal{R}_{s}^{m} f(x).
\end{align*}
We estimate this quantity by simulating two coupled processes \((X_x(t))_{t \geq 0}\) and \((X_{x+\zeta_k}(t))_{t \geq 0}\), following the coupling scheme introduced in~\cite{anderson2012efficient} and represented via random time–change formulations~\cite{ethier2009markov}:
\begin{align}
\label{suppl:rtc_formulation1}
X_x(t) &= x 
+ \sum_{\ell=1}^K Y^{(1)}_\ell\!\left( \int_0^t \lambda_{\ell, \textrm{min}}( X_x(s), X_{x+\zeta_k}(s)) \, ds \right) \zeta_k \notag  \\
&\quad + \sum_{\ell=1}^K Y^{(2)}_\ell\!\left( \int_0^t \Bigl( \lambda_\ell(X_x(s)) - \lambda_{\ell, \textrm{min}}( X_x(s), X_{x+\zeta_k}(s)) \Bigr) ds \right) \zeta_k, \\
\label{suppl:rtc_formulation2}
X_{x+\zeta_k}(t) &= x+\zeta_k 
+ \sum_{\ell=1}^K Y^{(1)}_\ell\!\left( \int_0^t \lambda_{\ell, \textrm{min}}( X_x(s), X_{x+\zeta_k}(s)) \, ds \right) \zeta_k \notag \\
&\quad + \sum_{\ell=1}^K Y^{(3)}_\ell\!\left( \int_0^t \Bigl( \lambda_\ell(X_{x+\zeta_k}(s)) - \lambda_{\ell, \textrm{min}}( X_x(s), X_{x+\zeta_k}(s)) \Bigr) ds \right) \zeta_k,
\end{align}
where \(Y^{(1)}_\ell, Y^{(2)}_\ell,\) and \(Y^{(3)}_\ell\) denote independent unit–rate Poisson processes for \(\ell=1,\dots,K\), and
\begin{align*}
\lambda_{\ell, \textrm{min}}\bigl( X_x(t), X_{x+\zeta_k}(t) \bigr) 
= \min\!\left\{ \lambda_\ell(X_x(t)), \, \lambda_\ell(X_{x+\zeta_k}(t)) \right\}.
\end{align*}
By simulating \(N\) such independent coupled trajectories,
\[
(X^{(1)}_x(t), X^{(1)}_{x+\zeta_k}(t))_{t \geq 0}, \;\dots,\; (X^{(N)}_x(t), X^{(N)}_{x+\zeta_k}(t))_{t \geq 0},
\]
and applying \eqref{MC_estimator_resolvent_diff}, we obtain the Monte Carlo estimator for \(D_{m,k}(x)\):
\begin{align}
\label{MC_estimator_resolvent_diff_diff}
\hat{D}_{m,k}(x) = 
\begin{cases}
f(x+\zeta_k) - f(x) - \dfrac{s}{N} \displaystyle\sum_{n=1}^N \int_0^T e^{- s t} \Bigl( f(X^{(n)}_{x+\zeta_k}(t)) - f(X^{(n)}_{x}(t)) \Bigr)\, dt, & m = 1, \\[1.25em]
\dfrac{1}{N} \displaystyle\sum_{n=1}^N \int_0^T w_m(s, t) e^{- s t} \Bigl( f(X^{(n)}_{x+\zeta_k}(t)) - f(X^{(n)}_{x}(t)) \Bigr)\, dt, & m \geq 2.
\end{cases} 
\end{align}
This estimator may also be applied when the two processes \((X^{(n)}_x(t))_{t \geq 0}\) and \((X^{(n)}_{x+\zeta_k}(t))_{t \geq 0}\) are simulated independently. However, the coupling described above substantially reduces its variance. An additional advantage of the coupling is that if, at some random time \(\tau^{(n)}\), the two processes coincide, i.e.,
\[
X^{(n)}_{x}(\tau^{(n)}) = X^{(n)}_{x+\zeta_k}(\tau^{(n)}),
\]
then they evolve together for all \(t \geq \tau^{(n)}\). As a consequence, in the estimator~\eqref{MC_estimator_resolvent_diff_diff}, the time integration interval can be shortened from \([0,T]\) to \([0,\tau^{(n)}]\) whenever \(\tau^{(n)} < T\), since the integrand vanishes on the remainder \([\tau^{(n)},T]\).

\section{Description of Reaction Networks}\label{supp:sec_reaction_networks}

In this section we provide details on the examples considered in the main paper to illustrate the developed SKA method. 

\subsection{Self-regulatory gene-expression network}
\label{supp:self_ge_network}

The self-regulatory gene-expression network (Figure \ref{main_fig_self_reg_gene_ex}(A)) involves two molecular species, the mRNA ($\mathbf{X}_1$) and the protein ($\mathbf{X}_2$), and is governed by the following four reactions:
\begin{align*}
& \emptyset  \stackrel{H_r(x_2)}{\longrightarrow} \mathbf{X}_1  && \textnormal{(mRNA transcription)} \\
& \mathbf{X}_1  \stackrel{k_p}{\longrightarrow} \mathbf{X}_1 + \mathbf{X}_2  && \textnormal{(protein translation)} \\
& \mathbf{X}_1  \stackrel{\gamma_r}{\longrightarrow} \emptyset  && \textnormal{(mRNA degradation)} \\
& \mathbf{X}_2  \stackrel{\gamma_p}{\longrightarrow} \emptyset  && \textnormal{(protein degradation)}.
\end{align*}
The last three reactions follow mass-action kinetics (see \eqref{main:massactionkinetics} in the main text). In contrast, the propensity of the transcription reaction is described by an inhibitory Hill function of the protein copy number $x_2$,
\begin{align*}
H_r(x_2) = \frac{k_r}{K_r + x_2^H},
\end{align*}
with Hill coefficient $H$. The nonlinearity introduced by this function prevents a direct solution for the time evolution of the first two moments due to the moment-closure problem~\cite{schnoerr2014validity}. Nevertheless, as illustrated in Figure \ref{main_fig_self_reg_gene_ex}(D) of the main text, SKA accurately approximates the dynamics of the first two moments, the associated cross spectral densities (CSDs), as well the parameter sensitivities, for any given initial state $x = (x_1, x_2)$. For the results presented, the following parameter values were used:
\begin{align*}
& k_{r} = 100.0\ \textnormal{sec.}^{-1}, \quad K_{r} = 10.0, \quad H = 1.0, \quad k_{p} = 2.0\ \textnormal{sec.}^{-1}, \\
& \qquad \gamma_r = 1.0\ \textnormal{sec.}^{-1}, \quad \gamma_p = 0.5\ \textnormal{sec.}^{-1}.
\end{align*}

\subsection{The \emph{Repressilator} network}
\label{supp:repressilator_network}

The \emph{repressilator}~\cite{elowitz2000synthetic} is a synthetic genetic regulatory network composed of three transcriptional repressors: \emph{tetR} from the Tn10 transposon, \emph{cI} from bacteriophage~$\lambda$, and \emph{lacI} from the lactose operon. These genes are arranged in a cyclic architecture, where each gene product represses the transcription of the next (Figure \ref{main_fig_repress}(A)). The corresponding proteins, TetR ($\mathbf{X}_1$), cI ($\mathbf{X}_2$), and LacI ($\mathbf{X}_3$), thus form a three-node negative feedback loop. As in the previous example, we describe the repression kinetics using inhibitory Hill functions of the form
\begin{align*}
H_j(x_k) = \frac{k_j}{K_j + x_k^{H_j}}, \quad j=1,2,3,
\end{align*}
which characterise the transcriptional repression governing the expression of protein $\mathbf{X}_j$ as a function of the copy number $x_k$ of its upstream repressor protein $\mathbf{X}_k$. Owing to the cyclic structure, we set $x_k = x_{j-1}$ for $j = 1,2,3$, with the convention $x_0 = x_3$. The overall \emph{repressilator} network consists of the three proteins and the following six reactions:
\begin{align*}
& \emptyset  \stackrel{H_1(x_3)}{\longrightarrow} \mathbf{X}_1  && \textnormal{(production of TetR)} \\
& \emptyset  \stackrel{H_2(x_1)}{\longrightarrow} \mathbf{X}_2  && \textnormal{(production of cI)} \\
& \emptyset  \stackrel{H_3(x_2)}{\longrightarrow} \mathbf{X}_3  && \textnormal{(production of LacI)} \\
& \mathbf{X}_1  \stackrel{\gamma_1}{\longrightarrow}   \emptyset && \textnormal{(degradation of TetR)} \\
& \mathbf{X}_2  \stackrel{\gamma_2}{\longrightarrow}   \emptyset && \textnormal{(degradation of cI)} \\
& \mathbf{X}_3  \stackrel{\gamma_3}{\longrightarrow}   \emptyset && \textnormal{(degradation of LacI)}.
\end{align*}

For our numerical experiments, we set the following parameter values:
\begin{align*}
k_j = 200.0\ \textnormal{sec.}^{-1}, \quad K_j = 10.0, \quad \gamma_j = 0.5\ \textnormal{sec.}^{-1}, \quad H_j = 2.0, 
\quad j=1,2,3.
\end{align*}
Here the Hill coefficients $n_j$ quantify the degree of cooperativity in the repressor–promoter binding interactions. To investigate how the oscillatory behavior depends on the cooperativity level, we additionally consider simulations with all Hill coefficients set to $H_j = 1.0$ and $H_j = 1.5$ (see Figure \ref{main_fig_repress_comparison}).

\subsection{Constitutive gene-expression with rAIF controller} \label{supp:sec_raif}

We begin by providing a description of the reference-based antithetic integral feedback (rAIF) controller. The controller consists of two bio-molecular species, $\mathbf{Z}_1$ and $\mathbf{Z}_2$, interacting through four elementary reactions governed by mass-action kinetics:
\begin{align}
\label{supp:reference}
& \emptyset  \stackrel{\mu}{\longrightarrow} {\bf Z_1}  && \textnormal{(reference)} \\
\label{supp:sense}
& {\bf X_\ell}  \stackrel{\theta}{\longrightarrow} {\bf X_\ell}  + {\bf Z_2} && \textnormal{(output sensing)}  \\
\label{supp:annihilation}
& {\bf Z_1} + {\bf Z_2}  \stackrel{\eta}{\longrightarrow}  \emptyset && \textnormal{(annihilation)}  \\
\label{supp:actuation}
& {\bf Z_1}  \stackrel{k}{\longrightarrow}  {\bf Z_1} + {\bf X_1} && \textnormal{(actuation)}.
\end{align}
Here, the reaction rate constants are indicated above the arrows, and ${\bf X_1}$ denotes the actuated species. Through downstream reactions and intermediates, ${\bf X_1}$ positively regulates the copy-number of the output species ${\bf X_\ell}$. The rAIF controller drives the mean level of ${\bf X_\ell}$ to the prescribed set point $\mu/\theta$, where $\mu$ is the production rate of ${\bf Z_1}$ (reaction \eqref{supp:reference}) and $\theta$ is the rate constant for output sensing (reaction \eqref{supp:sense}). Actuation occurs via production of ${\bf X_1}$ (reaction \eqref{supp:actuation}), while the feedback loop is closed by the bimolecular annihilation reaction \eqref{supp:annihilation}.  

In our illustrative example, the controlled network linking ${\bf X_1}$ and ${\bf X_\ell}$ is given by the constitutive gene-expression model introduced in Section~\ref{ex:cons_gene_ex}. Specifically, mRNA $\mathbf{X}_1$ serves as the actuated species in \eqref{supp:actuation} (transcription), while the protein $\mathbf{X}_2$ acts as the output species ($\ell=2$) in \eqref{supp:sense}. The associated reactions are:
\begin{align*}
& \mathbf{X}_1  \stackrel{k_p}{\longrightarrow} \mathbf{X}_1  + \mathbf{X}_2  && \textnormal{(translation)}  \\
& \mathbf{X}_1   \stackrel{\gamma_r}{\longrightarrow}  \emptyset  && \textnormal{(mRNA degradation)}  \\
& \mathbf{X}_2  \stackrel{\gamma_p}{\longrightarrow}   \emptyset && \textnormal{(protein degradation)}.
\end{align*}
We assume mass-action kinetics \eqref{main:massactionkinetics}, with rate constants specified above. A schematic of the rAIF controller connected to the gene-expression network is shown in Figure \ref{main_fig_raif}(A).  

The resulting closed-loop dynamics are described by the continuous-time Markov chain (CTMC) $(X(t), Z(t))_{t \geq 0}$, where $X(t) = (X_{1}(t), X_{2}(t))$ represents the state of the gene-expression network and $Z(t) = (Z_{1}(t), Z_{2}(t))$ the state of the controller. Due to the mass-action assumption, the first-moment dynamics satisfy:
\begin{align}
\label{first_moment_raif_eqns}
\frac{d}{dt} \E[X_1(t)] & =  k \E[Z_1(t)] - \gamma_r \E[X_1(t)], \\
\frac{d}{dt} \E[X_2(t)] & =  k_p \E[X_1(t)] - \gamma_p \E[X_2(t)], \notag \\
\frac{d}{dt} \E[Z_1(t)] & =  \mu  - \eta \E[Z_1(t) Z_2(t)], \notag \\
\frac{d}{dt} \E[Z_2(t)] & =  \theta \E[X_2(t)]  - \eta \E[Z_1(t) Z_2(t)]. \notag
\end{align}
Because of the nonlinearity introduced by the annihilation term $\E[Z_1(t) Z_2(t)]$, closed-form solutions to this system of ODEs are unavailable. Nevertheless, as shown in Figure \ref{main_fig_raif}(D) of the main text, SKA provides accurate approximations of both first- and second-moment dynamics for a given initial condition $x = (x_1, x_2, z_1, z_2)$. Moreover, SKA can compute cross-spectral densities (CSDs) and parameter sensitivities.

Setting the right-hand side of \eqref{first_moment_raif_eqns} equal to zero, we obtain the following steady-state expectations under the stationary distribution $\pi$:
\begin{align*}
\E_\pi(X_2) = \frac{\mu}{\theta}, \quad 
\E_\pi(X_1) = \frac{\mu \gamma_p}{\theta k_p}, \quad 
\E_\pi(Z_1) = \frac{\mu \gamma_p \gamma_r}{\theta k_p k}.
\end{align*}
In particular, the steady-state parameter sensitivities of $\E_\pi(X_2)$ with respect to a network parameter $\delta$ is given by
\begin{align*}
\partial_\delta \E_\pi(X_2) = \left\{ \begin{array}{cc}
\frac{1}{\theta} & \quad \textnormal{if} \quad \delta = \mu \\
-\frac{\mu}{\theta^2} & \quad \textnormal{if} \quad \delta = \theta \\
0 & \quad \textnormal{if} \quad \delta \in \{\eta, k, k_p,\gamma_r,\gamma_p\}.
\end{array} \right.
\end{align*}
These steady-state sensitivities are plotted in Figure \ref{main_fig_raif}(E), and provide a benchmark for evaluating the accuracy of our sensitivity estimates.  

For the numerical experiments, we fix the rAIF parameters as
\begin{align*}
\mu = 10.0 \ \textnormal{sec.}^{-1}, \quad \theta = 1.0 \ \textnormal{sec.}^{-1}, \quad \eta = 10.0 \ \textnormal{sec.}^{-1}, \quad k = 5.0 \ \textnormal{sec.}^{-1},
\end{align*}
and the gene-expression parameters as
\begin{align}
\label{antithetic_gene_ex_parameters}
k_p = 2.0 \ \textnormal{sec.}^{-1}, \quad \gamma_r = 2.0 \ \textnormal{sec.}^{-1}, \quad \gamma_p = 1.0 \ \textnormal{sec.}^{-1}.
\end{align}

\subsection{Constitutive gene-expression with sAIF controller}
\label{supp:sec_saif}

With the setup being identical to Section \ref{supp:sec_raif}, the sensor-based antithetic integral feedback (sAIF) controller modifies the rAIF motif by altering the actuation reaction \eqref{supp:actuation} to
\begin{align}
\label{supp:sensor_actuation_fn}
\emptyset \stackrel{H(z_2)}{\longrightarrow} \mathbf{X}_1,
\end{align}
where $H(z_2)$ is an inhibitory Hill function of the copy number $z_2$ of the controller species $\mathbf{Z}_2$, with a small basal production rate $b_0$:
\begin{align}
\label{supp:inh_hill_sensor_act}
H(z_2) = \frac{k_0}{ K + z^{h}_2} + b_0.
\end{align}
Thus, whereas in rAIF the controller actuates the process through $\mathbf{Z}_1$ catalyzing production of $\mathbf{X}_1$, in sAIF the actuation is mediated by $\mathbf{Z}_2$ repressing the production of $\mathbf{X}_1$. Despite this modification, the sAIF controller—like rAIF—drives the mean of ${\bf X_\ell}$ to the prescribed set point $\mu/\theta$, where $\mu$ is the production rate of ${\bf Z_1}$ (reaction \eqref{supp:reference}) and $\theta$ is the output sensing rate constant (reaction \eqref{supp:sense}).

The system of ODEs for the first-moment dynamics under sAIF is:
\begin{align}
\label{first_moment_saif_eqns}
\frac{d}{dt} \E[X_1(t)] & =   \E[ H(Z_2(t))] - \gamma_r \E[X_1(t)], \\
\frac{d}{dt} \E[X_2(t)] & =  k_p \E[X_1(t)] - \gamma_p \E[X_2(t)], \notag \\
\frac{d}{dt} \E[Z_1(t)] & =  \mu  - \eta \E[Z_1(t) Z_2(t)], \notag \\
\frac{d}{dt} \E[Z_2(t)] & =  \theta \E[X_2(t)]  - \eta \E[Z_1(t) Z_2(t)]. \notag
\end{align}
At stationarity, setting the right-hand sides equal to zero yields
\begin{align}
\label{supp_stationary_expectations_sAIF}
\E_\pi(X_2) = \frac{\mu}{\theta}, \quad 
\E_\pi(X_1) = \frac{\mu \gamma_p}{\theta k_p}, \quad 
\E_\pi(H(Z_2)) = \frac{\mu \gamma_p \gamma_r}{\theta k_p}.
\end{align}
Notably, the steady-state actuation rate $\E_\pi(H(Z_2))$ matches $k \E_\pi(Z_1)$ from rAIF, reflecting the identical controlled process. Moreover, the steady-state parameter sensitivities of $\E_\pi(X_2)$ with respect to a network parameter $\delta$ is given by
\begin{align*}
\partial_\delta \E_\pi(X_2) = \left\{ \begin{array}{cc}
\frac{1}{\theta} & \quad \textnormal{if} \quad \delta = \mu \\
-\frac{\mu}{\theta^2} & \quad \textnormal{if} \quad \delta = \theta \\
0 & \quad \textnormal{if} \quad \delta \in \{\eta, k_0, K, h, b_0, k_p,\gamma_r,\gamma_p\}.
\end{array} \right.
\end{align*}
Some of these steady-state sensitivities are plotted in Figure \ref{main_fig_saif}(E), and provide a benchmark for evaluating the accuracy of our sensitivity estimates.

To compare rAIF and sAIF performance, we must also align the \emph{actuation gains}. By definition, this gain is the magnitude of the sensitivity of the actuation rate with respect to the relevant controller species: $\mathbf{Z}_1$ for rAIF and $\mathbf{Z}_2$ for sAIF.  
For rAIF, the actuation gain is simply the mass-action constant $k$ in \eqref{supp:actuation}.  
For sAIF, the steady-state gain is
\begin{align*}
\E_\pi \big[ | H'(Z_2) | \big] 
&= \E_\pi\!\left[ \left| - \frac{k_0 h Z^{h-1}_2}{(K+ Z^h_2)^2} \right| \right] 
= \frac{h}{k_0} \E_\pi \Big[ \big( H(Z_2)-b_0 \big)^2 Z^{h-1}_2 \Big].
\end{align*}
Choosing Hill coefficient $h=1$ and approximating
\begin{align}
\label{sAIF_approx}
\E_\pi \big[ (H(Z_2)-b_0)^2 \big] \approx \big(\E_\pi[ H(Z_2) ] \big)^2,
\end{align}
we obtain, using \eqref{supp_stationary_expectations_sAIF}, the approximate steady-state gain
\begin{align*}
\E_\pi \big[ | H'(Z_2) | \big] \approx \frac{1}{k_0} \left( \frac{\mu \gamma_p \gamma_r}{\theta k_p} \right)^2.
\end{align*}
To enforce a fair comparison, we set this equal to the rAIF gain $k$, yielding the condition
\begin{align}
\label{sAIF_rAIF_fair_comparsison_cond}
k = \frac{1}{k_0} \left( \frac{\mu \gamma_p \gamma_r}{\theta k_p} \right)^2.
\end{align}

For numerical experiments, we retain the same gene-expression parameters as in \eqref{antithetic_gene_ex_parameters} and set
\begin{align*}
& \mu = 10.0 \ \textnormal{sec.}^{-1}, \quad \theta = 1.0 \ \textnormal{sec.}^{-1}, \quad \eta = 10.0 \ \textnormal{sec.}^{-1}, \quad b_0 = 2.0 \ \textnormal{sec.}^{-1}, \\
& h = 1.0, \quad K = 1.0, \quad k_0 = \frac{1}{k} \left( \frac{\mu \gamma_p \gamma_r}{\theta k_p} \right)^2 
= \frac{1}{5.0} \left( \frac{10.0 \times 1.0 \times 2.0}{1.0 \times 2.0} \right)^2 = 20.0 \ \textnormal{sec.}^{-1}.
\end{align*}
Our numerical results confirm that, with this parameter choice, the approximation \eqref{sAIF_approx} is closely satisfied.

\section{Implementation Details}
\label{supp:sensitivity_csd_impl_details}

\subsection*{Cross Spectral Density Estimation with SSA-DFT}

In the main text we applied our SKA method to estimate cross spectral densities (CSDs) for pairs of observables $f_1, f_2 \in \mathcal{F}$, and compared its performance against the SSA–DFT approach, which combines the Stochastic Simulation Algorithm (SSA) with the Discrete Fourier Transform (DFT). We now describe the SSA–DFT method and our implementation in detail.  

Fix a terminal time $T>0$, and let $(X_x(t))_{t \geq 0}$ denote a CTMC trajectory of the SRN generated by SSA from initial state $x$. Choose a uniform discretization of $[0,T]$ with step size $\delta = T/N$, yielding time points $t_j = j \delta$ for $j=0,\dots,N$. For each observable $f_k \in \mathcal{F}$ ($k=1,2$), we obtain the discrete time series
\[
\{ f^{(j)}_k : j=0,1,\dots,N \}, 
\quad \textnormal{where} \quad f^{(j)}_k = f_k(X_x(t_j)).
\]  
Applying the DFT to these sequences yields the one-sided Fourier transforms
\begin{align*}
\mathcal{F}_{f_k}(\omega_m) 
= \sqrt{\frac{\delta}{N}} \sum_{j=0}^{N-1} f^{(j)}_k \, e^{- i \omega_m j \delta},
\quad \omega_m = \frac{2 \pi m}{T}, 
\quad m=0,1,\dots,\tfrac{N}{2}.
\end{align*}
The CSD between $f_1$ and $f_2$ is then estimated by
\[
\mathcal{F}_{f_1}(\omega_m)\, \overline{\mathcal{F}_{f_2}(\omega_m)},
\]
averaged over an ensemble of $1000$ independent SSA trajectories. This procedure provides a numerical approximation of $\textnormal{CSD}_{f_1,f_2}(\omega,x,T)$ (see \eqref{csd_definition} in the main text) at the discrete set of frequencies $\{\omega_m : m=0,1,\dots,\tfrac{N}{2}\}$.

The SSA–DFT estimator inherits two inherent limitations: 1. it is biased due to the finite discretization step $\delta$, which constrains the frequency resolution, and 2. it can suffer from high variance unless a large ensemble of SSA trajectories is simulated, making it computationally expensive. By contrast, SKA circumvents both drawbacks: it provides CSD estimates directly in the frequency domain, without time discretization, and achieves reliable accuracy with a smaller sample size.

\subsection*{Parameter Sensitivity Estimation with CFD}

In the main paper we apply our SKA method to estimate parameter sensitivities with respect to all parameters of the SRN, and compare its performance against the coupled finite-difference (CFD) method of \cite{anderson2012efficient}, which is considered as one of the most reliable methods for sensitivity estimation for SRNs. We now describe the CFD method and our implementation in greater detail.  

Let the parameter vector be denoted by $\Theta = (\theta_1,\dots, \theta_p)$, and let $(X^{(\Theta)}_x(t))_{t \geq 0}$ denote the CTMC trajectories of the SRN with parameters $\Theta$ and initial state $x$. For a given observable $f$, the sensitivity with respect to parameter $\theta_j$ at time $t$ (see \eqref{theta_sensitivity} in the main text) is approximated in the CFD framework by the finite-difference
\begin{align*}
\mathcal{S}_{t,\theta_j}^{(\Theta)} f(x) 
:= \frac{\partial}{\partial \theta_j} \mathcal{K}^{(\Theta)}_t f(x) 
\;\approx\; \frac{\mathcal{K}^{(\Theta + h \mathbf{e}_j)}_t f(x) - \mathcal{K}^{(\Theta)}_t f(x)}{h} 
= \frac{\E\!\left[f(X^{(\Theta + h \mathbf{e}_j)}_x(t)) - f(X^{(\Theta)}_x(t)) \right]}{h},
\end{align*}
where $h > 0$ is the finite-difference step size, and $\mathbf{e}_j$ denotes the standard unit vector in the $j$-th coordinate. The processes $(X^{(\Theta + h \mathbf{e}_j)}_x(t))_{t \geq 0}$ and $(X^{(\Theta)}_x(t))_{t \geq 0}$ are coupled via their random time-change representations \eqref{suppl:rtc_formulation1}–\eqref{suppl:rtc_formulation2}, with identical initial states but distinct parameter values.  

Like our SKA-based approach, CFD is a biased estimator owing to the finite-difference approximation. Moreover, CFD suffers from an additional drawback: for each parameter $\theta_j$, an independent batch of coupled trajectories must be generated. This requirement becomes computationally prohibitive when the parameter dimension $p$ is large. In contrast, the computaional complexity of our SKA-based approach does not change with the parameter dimension.

To reduce the discretization bias in CFD, we employ a centered finite-difference scheme with step size $h=0.01$:
\begin{align*}
\mathcal{S}_{t,\theta_j}^{(\Theta)} f(x) 
\;\approx\; \frac{\E\!\left[f(X^{(\Theta + h \mathbf{e}_j)}_x(t)) - f(X^{(\Theta - h \mathbf{e}_j)}_x(t)) \right]}{2h}.
\end{align*}
The expectations are approximated using a standard Monte Carlo estimator, based on $10^4$ coupled trajectory pairs for each parameter over the prescribed time horizon.

\section{Supplementary Movies}

Description of the Supplementary Movies: \\
\begin{enumerate}
    \item \textbf{Supplementary Movie 1.} Evolution of the cross-spectral densities (CSDs) with the terminal time $T$ ranging from $0$ to $30$ seconds for the self-regulatory gene-expression network. Snapshots at $T=10$ and $T=30$ seconds are shown in Figure~\ref{main_fig_self_reg_gene_ex}(F) and compared with the SSA-DFT method.

    \item \textbf{Supplementary Movie 2.} Same as Supplementary Movie 1, but for the \emph{repressilator network}. Snapshots at $T=10$ and $T=30$ seconds are shown in Figure~\ref{main_fig_repress}(F) and compared with the SSA-DFT method.

    \item \textbf{Supplementary Movie 3.} Same as Supplementary Movie 1, but for the network comprising \emph{constitutive gene expression with the reference-based Antithetic Integral Feedback (rAIF) controller}. Snapshots at $T=10$ and $T=30$ seconds are shown in Figure~\ref{main_fig_raif}(F) and compared with the SSA-DFT method.

    \item \textbf{Supplementary Movie 4.} Same as Supplementary Movie 1, but for the network comprising \emph{constitutive gene expression with the sensor-based Antithetic Integral Feedback (sAIF) controller}. Snapshots at $T=10$ and $T=30$ seconds are shown in Figure~\ref{main_fig_saif}(F) and compared with the SSA-DFT method.

    \item \textbf{Supplementary Movie 5.} Evolution of the $\mathcal{L}_2(\hat{\pi})$ norm of the PSDs for the three \emph{repressilator networks} considered in Figure~\ref{main_fig_repress_comparison}, with terminal time $T$ ranging from $0$ to $30$ seconds. Snapshots at $T=10$ and $T=30$ seconds are shown in Figure~\ref{main_fig_repress_comparison}(B).

    \item \textbf{Supplementary Movie 6.} Evolution of the probability distribution of the peak frequency $\omega_0$, assuming the initial state is distributed according to $\hat{\pi}$, for the three \emph{repressilator networks} considered in Figure~\ref{main_fig_repress_comparison}, with terminal time $T$ ranging from $0$ to $30$ seconds. Snapshots at $T=10$ and $T=30$ seconds are shown in Figure~\ref{main_fig_repress_comparison}(C).

    \item \textbf{Supplementary Movie 7.} Evolution of the $\mathcal{L}_2(\hat{\pi})$ norm of the PSDs for the two AIF-based closed-loop networks considered in Figure~\ref{main_fig_raif_vs_saif}, with terminal time $T$ ranging from $0$ to $30$ seconds. Snapshots at $T=10$ and $T=30$ seconds are shown in Figure~\ref{main_fig_raif_vs_saif}(C).

    \item \textbf{Supplementary Movie 8.} Evolution of the probability distribution of the peak amplitude $|A_0|^2$, assuming the initial state is distributed according to $\hat{\pi}$, for the two AIF-based closed-loop networks considered in Figure~\ref{main_fig_raif_vs_saif}, with terminal time $T$ ranging from $0$ to $30$ seconds. Snapshots at $T=10$ and $T=30$ seconds are shown in Figure~\ref{main_fig_raif_vs_saif}(D).
\end{enumerate}

\newpage

\bibliographystyle{unsrt}
\bibliography{references}

\end{document}